\documentclass[prb,showpacs,preprintnumbers,amsfonts,amssymb,amsmath,floats,twocolumn,aps,superscriptaddress]{revtex4-1}

\usepackage{amsmath}
\usepackage{amsthm}
\usepackage{amssymb}
\usepackage{graphicx}%
\usepackage{bm}%
\usepackage{bbold}%
\usepackage{xcolor}%

\usepackage[pdftex]{hyperref} %
\definecolor{darkblue}{rgb}{0,0,.5} %
\definecolor{black}{rgb}{0,0,0} %
\hypersetup{
  colorlinks=true, breaklinks=true, linkcolor=darkblue,
  menucolor=darkblue,
  urlcolor=darkblue, citecolor=darkblue }

\def\tr{\operatorname{tr}}% trace
\def\Tr{\operatorname{Tr}}% generalized trace

\def\mc{\mathcal}
\def\bs{\boldsymbol}

\def\op{}% operators
\def\mat{\bs}% matrices
\def\fcl{\widehat}% functional
\def\gop{\mc}% grandcanoncial operators

\def\ord{\mc T}% (time) ordering operator

\def\cre#1#2{#1^\dagger_{#2}}%
\def\ann#1#2{#1^{\vphantom{\dagger}}_{#2}}%

\def\cc#1{\cre{c}{#1}}%
\def\ac#1{\ann{c}{#1}}%

\def\ra{\rightarrow}%
\def\com[#1,#2]{\left[#1,#2\right]}
\def\contcom[#1,#2]{\left[#1\stackrel{\circ}{,}#2\right]}

\def\ev<#1>{\left<#1\right>}

\def\bra<#1|{\left<#1\right|}
\def\ket|#1>{\left|#1\right>}
\def\braket<#1|#2|#3>{\langle#1|#2|#3\rangle}

\begin{document}
  
\title{Nonequilibrium self-energy functional theory}

\author{Felix Hofmann} \email{fhofmann@physik.uni-hamburg.de}
\affiliation{I. Institut f\"ur Theoretische Physik, Universit\"at
  Hamburg, Jungiusstra\ss{}e 9, 20355 Hamburg, Germany} %
\author{Martin  Eckstein} \affiliation{Max Planck Research Department for Structural
  Dynamics at the University of Hamburg, CFEL, Notkestra\ss{}e 85,
  22607 Hamburg, Germany} %
\author{Enrico Arrigoni}
\affiliation{Institute of Theoretical and Computational Physics,
Graz University of Technology, Petersgasse 16, 8010 Graz, Austria
}
\author{Michael Potthoff}
\affiliation{I. Institut f\"ur Theoretische Physik, Universit\"at
  Hamburg, Jungiusstra\ss{}e 9, 20355 Hamburg, Germany} %

\begin{abstract}
  The self-energy functional theory (SFT) is generalized to describe
  the real-time dynamics of correlated lattice-fermion models far from
  thermal equilibrium.  This is achieved by starting from a reformulation
  of the original equilibrium theory in terms of double-time Green's
  functions on the Keldysh-Matsubara contour.  With the help of a
  generalized Luttinger-Ward functional, we construct a functional
  $\fcl\Omega[\mat \Sigma]$ which is stationary at the physical
  (nonequilibrium) self-energy $\mat \Sigma$ and which yields the
  grand potential of the initial thermal state $\Omega$ at the
  physical point.  Non-perturbative approximations can be defined by
  specifying a reference system that serves to generate trial
  self-energies.  These self-energies are varied by varying the
  reference system's one-particle parameters on the Keldysh-Matsubara
  contour.  In case of thermal equilibrium, the new approach reduces
  to the conventional SFT.  Contrary to the equilibrium theory,
  however, ``unphysical'' variations, i.e., variations that are
  different on the upper and the lower branch of the Keldysh contour,
  must be considered to fix the time-dependence of the optimal
    physical parameters via the
  variational principle.  Functional derivatives in the
  nonequilibrium SFT Euler equation are carried out analytically to
  derive conditional equations for the variational parameters that are
  accessible to a numerical evaluation via a time-propagation scheme.
  Approximations constructed by means of the nonequilibrium SFT are
  shown to be inherently causal, internally consistent and to respect
  macroscopic conservation laws resulting from gauge symmetries of the
  Hamiltonian.
  This comprises the nonequilibrium dynamical mean-field theory but
  also dynamical-impurity and variational-cluster approximations that
  are specified by reference systems with a {\em finite} number of
  degrees of freedom.  In this way, non-perturbative and consistent
  approximations can be set up, the numerical evaluation of which is
  accessible to an exact-diagonalization approach.
\end{abstract}

% 71.10.-w Theories and models of many-electron systems 71.10.Fd
% Lattice fermion models (Hubbard model, etc.)  71.15.Qe Excited
% states: methodology 78.47.J- Ultrafast spectroscopy (<1 psec)
% 67.85.-d Ultracold gases, trapped gases

\maketitle

% ----------------------------------------------------
\section{Introduction}
\label{sec:introduction}

The development of new theoretical methods to study the real-time
dynamics of systems of strongly correlated fermions far from thermal
equilibrium has become more and more important recently.  Apart from
fundamental questions related, e.g., to the concept of thermalization,
\cite{kinoshita2006} to dynamical phase
transitions,\cite{mitra2006,diehl2010} and other open problems in
quantum statistics, \cite{leggett1987} this interest is to a large
extent triggered by the experimental progress which made it possible
to control microscopic degrees of freedom with high temporal
resolution.  Examples are given by femtosecond pump-probe spectroscopy
from transition-metal oxides \cite{iwai2003,perfetti2006,wall2009} or
by the dynamics of ultracold atomic gases trapped in optical
lattices. \cite{jaksch1998,bloch2008,strohmaier2010}

For correlated lattice-fermion models with local interactions, such as
the Hubbard model \cite{gutzwiller1963,hubbard1963,kanamori1963} as a
prototype, a conceptually appealing and pragmatic theoretical idea is
the mean-field approach. \cite{weiss1907} With the invention of
dynamical mean-field theory (DMFT)
\cite{metzner1989,georges1992,jarrell1992,georges1996} we have
\emph{the} optimal mean-field theory at hand that comprises a number
of important properties, including its non-perturbative character and
its internal consistency.  Those features are also shared by the
nonequilibrium (NE) generalization of the DMFT
\cite{freericks2006,schmidt2002} which has already been applied
successfully to a number of
problems. \cite{eckstein2009b,tsuji2009,eckstein2010,werner2012,amaricci2012}

On the operational level, DMFT (both for equilibrium and for
nonequilibrium) requires the computation of the fermion self-energy of
an effective impurity model with self-consistently determined
parameters.  For the equilibrium case, quantum Monte-Carlo (QMC)
techniques \cite{gull2011} nowadays represent a standard tool to treat
the many-body impurity problem efficiently and accurately.  Employing
exact diagonalization (ED) \cite{caffarel1994} as a ``solver''
represents a competitive alternative in case of single- and multi-band
\cite{liebsch2012} models.  It is easily implemented, computationally
efficient and highly accurate.  A disadvantage of the ED solver
consists in the essentially \emph{ad hoc} character of the
self-consistency condition that fixes the Weiss field.  This
originates from the impossibility to fit a continuous Weiss field with
any finite number of bath degrees of freedom, and it becomes a serious
problem, if, for reasons of limited computational resources, only a
small number of bath sites can be used in the effective impurity
model.

The \emph{ad hoc} character of the bath representation can lead to a
violation of thermodynamic consistency and conservation laws. This
problem could be solved within the framework of the self-energy
functional theory (SFT)
\cite{potthoff2003,potthoff2003c,potthoff2007,potthoff2012} where the
DMFT self-consistency condition is replaced by the condition for
stationarity of the system's grand potential with respect to the bath
parameters of the impurity or ``reference'' system.  Thereby the bath
parameters are efficiently determined by a physically meaningful and
unique procedure which provides consistent results for impurity models
with a few parameters only and recovers the full DMFT in the continuum
limit.  Very precise studies of phase diagrams have been done in this
way, see Refs.~\onlinecite{pozgajcic2004,eckstein2007} for example.

In the nonequilibrium case, the situation is more complicated:
QMC-based solvers have been employed successfully but suffer from a
severe sign (or phase) problem contrary to the equilibrium case where
the sign problem is absent or mild. \cite{gull2011,werner2009}
Simplified, e.g., perturbative approximations, such as the iterative
perturbation theory \cite{schmidt2002} the non-crossing approximation,
\cite{eckstein2010,werner2012} or simplified models, such as the
Falicov-Kimball model \cite{freericks2006,tsuji2009} have been
considered instead, as well as a nonequilibrium variant of the
dual-fermion approach.\cite{jung2012} For the study of steady-state
properties, a non-trivial extension of ED-based DMFT has been
suggested recently.\cite{arrigoni2012} The development of ED-based
impurity solvers to compute the \emph{real-time} evolution within DMFT
is more challenging, as it is by no means obvious how to fix the
time-dependent parameters to fit a given Weiss field, i.e., a given
non-homogeneous function of two time variables with certain analytical
properties. One indeed can find mapping strategies which are accurate
and systematic at short times, \cite{gramsch2013} but in general, and
in particular for the long-time limit, the reduction of the
Hamiltonian representation of the Weiss field to a \emph{small} number
of parameters remains somehow \emph{ad hoc}.

The goal of the present study is therefore to explore whether
non-perturbative and internally consistent approximations based on the
exact-diagonalization of a reference system with a finite (small)
number of bath sites can be formulated by means of a proper
generalization of the self-energy functional theory to the
nonequilibrium case.  Preceding attempts in this direction are not
satisfactory yet.  The nonequilibrium cluster-perturbation theory
\cite{potthoff2011c,balzer2012,jurgenowski2013} does make use of the
exact diagonalization of a finite reference system out of equilibrium
and provides the one-particle propagator for a nonequilibrium state of
the correlated lattice model.  However, the approach does not rely on
a variational principle at all and does not involve any
self-consistent or variational optimization of the parameters of the
reference system.  On the other hand, a self-consistent parameter
optimization is part of a similar ED-based cluster approach
\cite{knap2011,nuss2012} which has been formulated and applied to
study the steady state of an out-of-equilibrium correlated lattice
model.  Here a physically motivated self-consistency condition is used
which, however, is not yet shown to derive from a general variational
principle that also applies to the transient dynamics.

There are several problems that must be solved in order to construct a
nonequilibrium self-energy functional theory (NE-SFT): First, a
functional $\fcl\Omega[\mat \Sigma]$ of the double-time nonequilibrium
self-energy must be constructed formally and shown to be stationary at
the physical self-energy of the lattice model.  Ideally, the
functional, if evaluated at the physical self-energy, has a precise
physical meaning.  In the spirit of the equilibrium SFT, the
functional should be accessible to an exact numerical evaluation for
trial nonequilibrium self-energies generated by a reference system,
which typically consists of a small number of sites such that it is
tractable by exact-diagonalization techniques. Next one must find
conditional equations for the parameters of the reference system, by
demanding stationarity of $\Omega$ when varying the self-energy
through variation of the parameters.

The NE-SFT should furthermore recover the nonequilibrium DMFT if a
single-impurity Anderson model, with a continuum of bath degrees of
freedom, was chosen as a reference.  Apart from nonequilibrium
dynamical impurity approximations (DIA) resulting from Anderson models
with a finite number of bath sites, the NE-SFT should also allow for
the construction of cluster approximations, such as a nonequilibrium
generalization of the variational cluster approach (VCA).  Adding
baths one should, in the limit of a continuum of bath degrees of
freedom, also recover nonequilibrium analogues of the cellular DMFT
\cite{kotliar2001} and the dynamical cluster
approximation. \cite{hettler1998} Finally, it will be interesting to
see how the standard SFT is recovered within the general NE-SFT setup
in case of an equilibrium situation.

The most important question in the context of any method addressing
real-time dynamics, however, concerns macroscopic conservation laws.
Do approximations derived within the NE-SFT framework respect the
conservation of the total particle number, the total spin and the
total energy for a U(1) and SU(2) symmetric and time-independent
Hamiltonian?  This ``conserving'' nature of approximations is not
easily obtained.  The seminal work of Baym and Kadanoff
\cite{baym1961,baym1962} answers this question for approximations that
are ``$\Phi$ derivable'', including DMFT and self-consistent
perturbation theory, such as the second-order Born
approximation. While the construction of the NE-SFT makes use of the
Luttinger-Ward functional $\Phi$, the question whether it is
conserving must be addressed carefully since generic approximations
within the NE-SFT cannot be obtained by re-summations of diagram
classes.

The paper is organized as follows: After summarizing some concepts of
nonequilibrium Green's functions that are needed to set up the theory
in Sec.\ \ref{sec:non-equil-greens}, we discuss the essential
properties of the Luttinger-Ward functional for the nonequilibrium
case in Sec.\ \ref{sec:lutt-ward-funct} which is necessary to
construct the dynamical variational principle of nonequilibrium SFT in
Sec.\ \ref{sec:dynam-vari-princ}.  This is followed by a discussion of
how to construct approximations within the NE-SFT in Sec.\
\ref{sec:constr-appr}.  Sec.\ \ref{sec:dynamical-mean-field} then
shows the relation to nonequilibrium DMFT, in particular.

Some of the above steps are preparatory and will be presented in
analogy to the equilibrium SFT as far as possible.  The reader may
compare the central Eqs.\ (\ref{eq:NESEfcl}) and (\ref{eq:EulerEqn})
with their equilibrium counterparts (cf.\ Ref.\
\onlinecite{potthoff2012}, for example).  They do not, however, give
sufficient consideration to the intrinsic formal structure of the full
nonequilibrium SFT.  The essential following part of the paper is
therefore concerned with questions related to the causal structure of
the theory, with the concept of variations in ``unphysical''
directions as well as with the need to carry out the (functional)
derivatives with respect to the variational parameters {\em
  analytically} (see the discussion following Eq.\
(\ref{eq:nontrivVar})).  This paves the way for an efficient numerical
evaluation of different impurity or cluster approximations, which will
be published independently.  Finally, the analytical proof of the
conserving nature of {\em any} approach that is constructed within the
framework of the NE-SFT represents an important result.

The concept of physical and transverse variations is introduced in
Sec.\ \ref{sec:triv-nontriv-var}.  The Euler equation of the NE-SFT is
worked out in Sec.\ \ref{sec:eval-euler-equat} and used to understand
the relation of the NE-SFT to the conventional equilibrium SFT in
Sec.\ \ref{sec:therm-equil-init} and for setting up a concept for the
numerical evaluation of the theory in Sec.\ \ref{sec:numerics}.  Its
internal consistency is addressed in Sec.\ \ref{sec:therm-cons}.
Finally, the question of macroscopic conservation laws is discussed in
detail in Sec.\ \ref{sec:conservation-laws}.  Conclusions are given in
Sec.\ \ref{sec:conclusions-outlook}.

% ----------------------------------------------------
\section{Nonequilibrium Green's function}
\label{sec:non-equil-greens}

The self-energy functional approach relies on functionals that are
formally defined by means of all-order perturbation theory.
Therefore, we first summarize the concept of (nonequilibrium) Green's
functions \cite{kubo1957,matsubara1955,schwinger1961,keldysh1965} as
far as necessary for our purposes.  Out of the various available
formulations,
\cite{danielewicz1984,wagner1991,leeuwen2006c,rammer2007,kamenev2011}
we will basically follow the formal setup by Wagner. \cite{wagner1991}

We assume that the system at initial time $t_0$ is prepared in a
thermal state with inverse temperature $\beta$ and chemical potential
$\mu$, as given by a density operator
\begin{equation}
  \label{eq:iniState}
  \rho = \frac{\exp(-\beta \gop H_{\rm ini})}{\tr \exp(-\beta \gop H_{\rm ini})} \,,
\end{equation}
with $\gop H_{\rm ini} = \op H_{\rm ini} - \mu \op N$, where
\begin{equation}
  \label{eq:Hini-2ndQuant}
  \op H_{\rm ini}  =
  \sum_{\alpha\beta}T^{\rm(ini)}_{\alpha\beta}\cc{\alpha}\ac{\beta} +
  \frac{1}{2}\sum_{\alpha\beta\gamma\delta}
  U^{\rm(ini)}_{\alpha\beta\delta\gamma}\cc{\alpha}\cc{\beta}\ac{\gamma}\ac{\delta} \, 
\end{equation}
is the initial Hamiltonian and $\op N$ the total particle-number
operator. Greek indices refer to one-particle basis states which
typically are characterized by a lattice site, an orbital index and a
spin-projection quantum number.  For times $t>t_0$ the system's time
evolution shall be governed by the possibly time-dependent Hamiltonian
\begin{equation}
  \label{eq:Hfin-2ndQuant}
  \op H_{\rm fin} (t) =
  \sum_{\alpha\beta}T^{\rm(fin)}_{\alpha\beta}(t)\cc{\alpha}\ac{\beta} +
  \frac{1}{2}\sum_{\alpha\beta\gamma\delta}
  U^{\rm(fin)}_{\alpha\beta\delta\gamma}(t)\cc{\alpha}\cc{\beta}\ac{\gamma}\ac{\delta} \,.
\end{equation}
For the sets of time-dependent hopping and interaction parameters we
write $\mat T$ and $\mat U$ for short, and, whenever necessary or
convenient, we indicate the dependence of the Hamiltonian on those
parameters as $\op H_{\mat T,\mat U}$.

\begin{figure}[t]
  \centering
  \includegraphics[width=0.7\columnwidth]{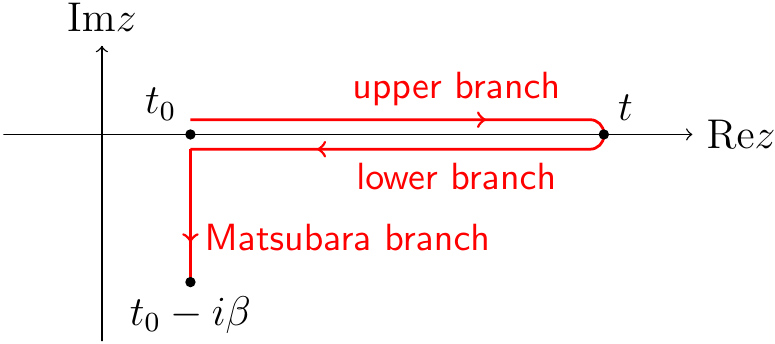}
  \caption{Three-branch contour $\mc C$ in the complex time plane, see
    text for discussion.}
  \label{fig:contour}
\end{figure}

In the Heisenberg picture with respect to $\gop H(t) \equiv \op H_{\rm
  fin}(t) - \mu \op N$, an arbitrary, possibly time-dependent
observable $\op A(t)$ is given by
\begin{equation}
  \label{eq:ObservHeisenbergPic}
  \op A_{\gop H}(t) = \op U(t_0,t) \op A(t) \op U(t,t_0) \,.
\end{equation}
Here, $\op U(t,t') = \ord \exp \left( -i \int_{t'}^t d z\, \gop
  H(z)\right)$ is the time-evolution operator for real times $t > t'$
and $\op U(t,t') = \widetilde\ord \exp \left( -i \int_{t'}^t d z\,
  \gop H(z) \right)$ for $t < t'$, where $\ord$ $(\widetilde\ord)$ is
the chronological (anti-chronological) time-ordering operator.  
For a complex ``time'' $t_{0}-i\tau$ with $0\le \tau \le \beta$, we define 
$\op U(t_{0}-i\tau,t_{0}) = \exp\left(-\gop H_{\rm ini} \tau \right)$.
Noting that $\exp\left(-\beta\gop H_{\rm ini}\right) = \op
U(t_{0}-i\beta,t_{0})$, the time-dependent expectation value of the
observable $\op A(t)$, namely $\langle \op A \rangle(t) = \tr(\rho \op
A_{\gop H}(t))$, can be written as:
\begin{equation}
  \label{eq:timedepexp}
  \ev<\op A>_{\mat T,\mat U}(t) = \frac{\tr\left( \ord_{\mc C}
      \exp\left(-i\int_{\mc C} d z' \, \gop H_{\mat T,\mat U}(z')\right)
      \op A(t) \right)}{\tr\left( \ord_{\mc C}
      \exp\left(-i\int_{\mc C} d z' \, \gop H_{\mat T,\mat
          U}(z')\right) \right)} \,.
\end{equation}
Here, the time integration is carried out along the contour $\mc C$ in
the complex time plane, see Fig.~\ref{fig:contour}, which extends
from ${z'}=t_0$ to ${z'}=\infty$ along the real axis (upper branch)
and back to ${z'}=t_0$ (lower branch) and finally from ${z'}=t_0$ to
${z'}=t_0-i \beta$ along the imaginary axis (Matsubara branch).  We
also refer to the upper and the lower branch as the Keldysh contour.
For a concise notation, we define $\op H(z)$ for contour times $z$ as
$\op H(z) = \op H_{\rm fin}(t)$ if $z = t > t_0$ and as $\op H(z) =
\op H_{\rm ini}$ if $z =t_{0} -i \tau$ with $0\leq\tau\leq\beta$.  In
the same way, we define $T_{\alpha\beta}(z)$ and
$U_{\alpha\beta\delta\gamma}(z)$.  $\ord_{\mc C}$ denotes the ordering
operator along the contour and, after expanding the exponential,
places an operator $\gop H(z_1)$ to the left of $\gop H(z_2)$ if $z_1$
is ``later'' than $z_2$, where $t_0 - i\beta$ is the ``latest'' time.
Obviously, $\ord_{\mc C}$ replaces $\ord$ on the upper and
$\widetilde\ord$ on the lower branch.

% -----------------------------------------------------------------------------------------------
% -----------------------------------------------------------------------------------------------

When the contour ordering operator $\ord_{\mc C}$ acts on $A(t)$ in
the numerator of Eq.~(\ref{eq:timedepexp}), it places $A(t)$ at the
position $z=t$ on $\mc C$ where the expectation value is
evaluated. Because the integrations along the upper and the lower
branches cancel each other in the interval $t<z'<\infty$, the
integration along the Keldysh branch is limited to ${z'} < t$ (see
Fig.~\ref{fig:contour}), and it does not matter whether $A(t)$ is
placed at $z=t$ on the upper or the lower branch of the contour.  For
the denominator, only the Matsubara branch contributes and results in
$\tr \exp(-\beta \gop H_{\rm ini})$.

For a system specified by the parameters $\mat T$ and $\mat U$, we
define the elements of the contour-ordered Green's function $\mat
G_{\mat T,\mat U}$ as
\begin{equation}
  \label{eq:DefNEGF}
  i G_{\mat T,\mat U;\alpha\alpha'}(z,z') = \ev< \ord_{\mc C} \ac{\alpha,\gop
    H}(z)\cc{\alpha',\gop H}(z') > \, .
\end{equation}
Here $\ev<\cdots> = \tr(\rho\,\cdots)$ denotes the expectation value
in the initial state.  Furthermore, the annihilation and creation
operators are given in their Heisenberg picture with respect to $\gop
H(t)$, $z,z'$ denote arbitrary points on the contour, and $\ord_{\mc
  C}$ is the time ordering of annihilation and creation operators on
$\mc C$ which yields an additional (fermionic) sign for each
transposition.  Note that the Green's function also depends on $\beta$
and $\mu$ via the initial thermal state.  These dependencies are
implicit in the notations.

The ``free'' Green's function $\mat G_{\mat T,0}$ is obtained by
setting $\mat U=0$ in Eq.~(\ref{eq:DefNEGF}).  Using the Heisenberg
equation of motion for the annihilation operator, we find
\begin{multline}
  \label{eq:inverseFreeNEGF}
  G^{-1}_{\mat T,0;\alpha\alpha'}(z,z') =
  \delta_{\alpha\alpha'}\delta_{\mc C}(z,z') i \partial_{z'} \\ -
  \delta_{\mc C}(z,z') \left(T_{\alpha\alpha'}(z') - \mu
    \delta_{\alpha\alpha'}\right) \,,
\end{multline}
where $\delta_{\mc C}$ is the contour delta-function, and the matrix
inverse refers to both one-particle basis indices and time variables.  With the
help of the free and the interacting Green's functions we can also
introduce the self-energy via the Dyson equation
\begin{equation}
  \label{eq:DysonEqnShort}
  \mat G_{\mat T,\mat U} = \mat G_{\mat T,0} + \mat G_{\mat
    T,0} \circ \mat \Sigma_{\mat T,\mat U} \circ \mat G_{\mat T,\mat
    U} \,,
\end{equation}
which is short for
\begin{multline}
  \label{eq:DysonEqnFull}
  G_{\mat T,\mat U;\alpha\alpha'}(z,z') = G_{\mat
    T,0;\alpha\alpha'}(z,z') + \sum_{\beta\beta'}\int_{\mc C} d\bar z
  d\bar{\bar z} \\ G_{\mat T,0;\alpha\beta}(z,\bar z) \Sigma_{\mat
    T,\mat U;\beta\beta'}(\bar z,\bar{\bar z}) G_{\mat T,\mat
    U;\beta'\alpha'}(\bar{\bar z},z') \,,
\end{multline}
i.e., the circle $\circ$ stands for the convolution along $\mc C$.

By switching to the interaction picture, the interacting Green's
function can be cast into the form:
\begin{equation}
  \label{eq:IntPicNEGF}
  i  G_{\mat T,\mat U;\alpha\alpha'}(z,z') 
  = 
  \frac{\ev< \ord_{\mc C} e^{-i\int_{\mc C} d z'' \, \gop H_{0,\mat U}(z'')}
    \ac{\alpha}(z)\cc{\alpha'}(z') >_{\mat T,0}}{\ev< \ord_{\mc C} e^{-i\int_{\mc
        C} d z'' \, \gop H_{0,\mat U}(z'')} >_{\mat T,0}} \, .
\end{equation}
Here the time dependence of all operators is due to $\gop H_{\mat
  T,0}$ only.  Likewise, the expectation value $\ev<\cdots>_{\mat
  T,0}$ is defined with the ``free'' density operator
$\exp\left(-\beta\gop H_{\mat T,0}\right)/\tr \exp\left(-\beta\gop
  H_{\mat T,0}\right)$.  Hence, Wick's theorem applies and therewith
the standard techniques of perturbation theory. \cite{wagner1991}

% ----------------------------------------------------
\section{Luttinger-Ward functional}
\label{sec:lutt-ward-funct}

% -----------------------------------------------------------------------------------------------
\begin{figure}[t]
  \centerline{\includegraphics[width=0.8\columnwidth]{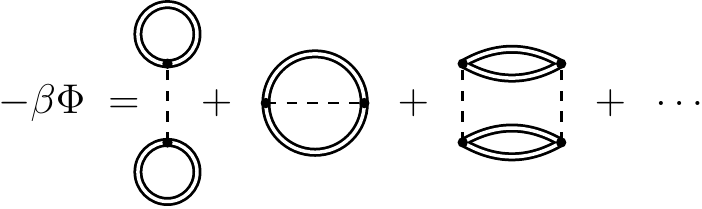}}
  \caption{ Diagrammatic definition of the Luttinger-Ward functional
    $\fcl\Phi_{\mat U}[\mat G]$.  Double lines: fully interacting
    propagator $\mat G$.  Dashed lines: interaction $\mat U$.  See
    text for discussion.}
  \label{fig:LW}
\end{figure}
% -----------------------------------------------------------------------------------------------

The nonequilibrium Luttinger-Ward functional $\fcl\Phi_{\mat U}[\mat
G]$ can be defined by means of all-order perturbation theory in close
analogy to the equilibrium case. \cite{luttinger1960} It is obtained
as the limit of the infinite series of closed renormalized skeleton
diagrams (see Fig.~\ref{fig:LW}), and is thus given as a functional of
the contour-ordered Green's function.  Note that functionals are
indicated by a hat. Usually the skeleton-diagram expansion
  cannot be summed up to get a closed form for $\fcl\Phi_{\mat U}[\mat
  G]$, and the explicit functional dependence is unknown even for the most
simple types of interactions like the Hubbard interaction.  As an
alternative to the diagrammatic definition of the Luttinger-Ward
functional, a nonequilibrium path-integral formalism may be used for
an entirely non-perturbative construction.  Again, this can be done
analogously to the equilibrium case. \cite{potthoff2006b} Both
variants allow to derive the following four properties that will be
used extensively for constructing the nonequilibrium SFT:

(i) The Luttinger-Ward functional vanishes in the non-interacting
limit:
\begin{equation}
  \label{eq:NELWfnclFreeCase}
  \fcl\Phi_{\mat U}[\mat G] \equiv 0 \quad \text{for} \quad \mat U = 0 \, , 
\end{equation}
since there is no zeroth-order diagram.

(ii) The functional derivative of the Luttinger-Ward functional with
respect to its argument is:
\begin{equation}
  \label{eq:FclDerivativeNELWfcl}
  \frac{\delta \fcl\Phi_{\mat U}[\mat G]}{\delta G(1,2)} =
  \frac{1}{\beta} \fcl\Sigma_{\mat U}[\mat G](2,1) \, ,
\end{equation}
with the short-hand notation $i \equiv (\alpha_i,z_i)$.
Diagrammatically, the functional derivative corresponds to the removal
of a propagator from each of the $\Phi$ diagrams.  Taking care of
topological factors, \cite{luttinger1960} one ends up with the
skeleton-diagram expansion of the self-energy which, independently
from the definition, Eq.~(\ref{eq:DysonEqnShort}), gives the
self-energy as a functional of the Green's function $\fcl{\mat
  \Sigma}_{\mat U}[\mat G]$.  Evaluating the functional $\fcl{\mat
  \Sigma}$ at the exact (``physical'') Green's function $\mat G_{\mat
  T, \mat U}$ yields the physical self-energy:
\begin{equation}
  \label{eq:SEfclEvaluatedPhysGreenFnc}
  \fcl{\mat \Sigma}_{\mat U}[\mat G_{\mat T,\mat U}] =
  \mat\Sigma_{\mat T,\mat U} \, .
\end{equation}

(iii) Since any diagram in the series depends on $\mat U$ and on $\mat
G$ \emph{only}, the Luttinger-Ward functional is ``universal'', i.e.,
it is independent of $\mat T$.  Two systems with the same interaction
$\mat U$ but different one-particle parameters $\mat T$ are described
by the same Luttinger-Ward functional.  This implies that the
functional $\fcl{\mat \Sigma}_{\mat U}[\mat G]$ is universal, too.

(iv) If evaluated at the physical Green's function $\mat G_{\mat T,
  \mat U}$ of the system with Hamiltonian $\op H_{\mat T, \mat U}$,
the Luttinger-Ward functional provides a quantity
\begin{equation}
  \fcl\Phi_{\mat U} [\mat G_{\mat T,\mat U}] = \Phi_{\mat T, \mat U} \, .
\end{equation}
Note that $ \Phi_{\mat T, \mat U}$ depends on the initial equilibrium
state of the system only, as contributions from the Keldysh branch
cancel each other (for details, see Sec.~\ref{sec:triv-nontriv-var}).
$\Phi_{\mat T, \mat U}$ is related to the grand potential of the
system via the expression
\begin{multline}
  \label{eq:GrandPotLWfcl}
  \Omega_{\mat T,\mat U} = \Phi_{\mat T,\mat U} +
  \frac{1}{\beta}\Tr\ln \left( \mat G_{\varepsilon_0,0}^{-1}\circ \mat
    G_{\mat T,\mat U} \right) \\ - \frac{1}{\beta}\Tr
  (\mat\Sigma_{\mat T,\mat U}\circ\mat G_{\mat T,\mat U}) \,.
\end{multline}
Here, we defined the trace as
\begin{equation}
  \label{eq:ContourTrace}
  \Tr \mat A = \sum_\alpha \int_{\mc C} d z\, A_{\alpha\alpha}(z,z^+),
\end{equation}
where $z^+$ is infinitesimally later than $z$ on $\mc C$.  The factor
$\mat G^{-1}_{\varepsilon_0,0}$ with $\varepsilon_{0}\to\infty$ has to
be introduced to regularize the $\Tr\ln$ term as discussed in
appendix~\ref{sec:analyt-funct-cont}.  It will be omitted in the
following as it does not affect the results.  Equation
(\ref{eq:GrandPotLWfcl}) can be derived using a coupling-constant
integration \cite{luttinger1960} or by integrating over the chemical
potential $\mu$. \cite{potthoff2006b} The proof is completely
analogous to the equilibrium case.

% ----------------------------------------------------
\section{Dynamical variational principle}
\label{sec:dynam-vari-princ}

We assume the functional $\fcl {\mat \Sigma}_{\mat U}[\mat G]$ is
invertible \emph{locally} to construct the Legendre transform of the
Luttinger-Ward functional:
\begin{equation}
  \label{eq:NELWfclLegendre}
  \fcl F_{\mat U} [\mat \Sigma] = \fcl\Phi_{\mat U}[\fcl{\mat G}_{\mat U}[\mat
  \Sigma]] - \frac{1}{\beta}\Tr (\mat\Sigma\circ\fcl{\mat G}_{\mat U}[\mat
  \Sigma]) \,.
\end{equation}
Here, $\fcl{\mat G}_{\mat U}[\fcl{\mat \Sigma}_{\mat U}[\mat G]] =
\mat G$.  With Eq.~(\ref{eq:FclDerivativeNELWfcl}) one has:
\begin{equation}
  \label{eq:LegendreLWfclDerivative}
  \frac{\delta \fcl F_{\mat U}[\mat \Sigma]}{\delta \Sigma(1,2)} =
  - \frac{1}{\beta} \fcl G_{\mat U}[\mat \Sigma](2,1) \,.
\end{equation}
We now define the self-energy functional as:
\begin{multline}
  \label{eq:NESEfcl}
  \fcl\Omega_{\mat T,\mat U}[\mat\Sigma] = \frac{1}{\beta} \Tr\ln
  \left( \mat G_{\mat T,0}^{-1} - \mat\Sigma \right)^{-1} + \fcl
  F_{\mat U} [\mat \Sigma] \, ,
\end{multline}
Its functional derivative is (use Eq.~\ref{eq:Tr(f(X))_contour}):
\begin{equation}
  \label{eq:NESEfclDerivativeSE}
  \frac{\delta \fcl\Omega_{\mat T,\mat U}[\mat
    \Sigma]}{\delta \mat \Sigma} = \frac{1}{\beta} \left( \mat G_{\mat T,0}^{-1} - \mat\Sigma
  \right)^{-1} - \frac{1}{\beta} \fcl {\mat G}_{\mat U}[\mat \Sigma]\,.
\end{equation}
The equation
\begin{equation}
  \fcl{\mat G}_{\mat U}[\mat \Sigma] = \left(\mat G_{\mat T,0}^{-1} - \mat \Sigma\right)^{-1}
  \label{eq:sig}
\end{equation}
is a (highly non-linear) conditional equation for the self-energy of
the system $\op H_{\mat T,\mat U}$.
Equations~(\ref{eq:DysonEqnShort}) and
(\ref{eq:SEfclEvaluatedPhysGreenFnc}) show that it is satisfied by the
physical self-energy $\mat \Sigma = \mat \Sigma_{\mat T, \mat U}$.
Note that the left-hand side of Eq.~(\ref{eq:sig}) is independent of
$\mat T$ but depends on $\mat U$ (due to the universality of
$\fcl{\mat G}_{\mat U}[\mat \Sigma]$), while the right-hand side is
independent of $\mat U$ but depends on $\mat T$ via $\mat G_{\mat
  T,0}^{-1}$.

The obvious problem of finding a solution of Eq.~(\ref{eq:sig}) is
that there is no closed form for the functional $\fcl{\mat G}_{\mat
  U}[\mat \Sigma]$.  Solving Eq.~(\ref{eq:sig}) is equivalent,
however, to a search for the stationary point of the grand potential
as a functional of the self-energy:
\begin{equation}
  \frac{\delta \fcl{\Omega}_{\mat T, \mat U}[\mat \Sigma]}
  {\delta \mat \Sigma} = 0 \;.
  \label{eq:StatPointNESEfcl}
\end{equation}
This equation is the starting point for nonequilibrium
self-energy functional theory.

Note that, while there are various symmetry relations between the
elements $\Sigma_{\alpha\alpha'}(z,z')$ of the self-energy at
different times $z$ and $z'$, the elements of $\mat \Sigma$ have to be
treated as \emph{independent of each other} for the functional
differentiation to ensure the equivalence of the variational principle
Eq.~(\ref{eq:StatPointNESEfcl}) with the fundamental Dyson equation
Eq.~(\ref{eq:sig}).  As will become clear below, the stationarity with
respect to some of the variational directions just ensures the correct
symmetry relations between the elements of
$\Sigma_{\alpha\alpha'}(z,z')$, while the other variational directions
fix the actual value of $\Sigma_{\alpha\alpha'}(z,z')$.

% ----------------------------------------------------
\section{Constructing approximations}
\label{sec:constr-appr}

Even though the Luttinger-Ward functional and its Legendre transform
$\fcl F_{\mat U}[\mat\Sigma]$ are generally unknown, it is possible to
evaluate the self-energy functional Eq.~(\ref{eq:NESEfcl}) exactly on
a certain subspace of self-energies: To this end we compare the
self-energy functional of the original system with the self-energy
functional of a reference system, given by a Hamiltonian $\op H'
\equiv \op H_{\mat \lambda',\mat U}$, which differs from the original
Hamiltonian $\op H_{\mat T,\mat U}$ only in its one-particle
parameters $\mat \lambda'$, but shares its interaction part. In the
following, primed quantities refer to the reference system. The
respective self-energy functional is
\begin{multline}
  \label{eq:NESEfclRef}
  \fcl\Omega_{\mat \lambda',\mat U}[\mat\Sigma] = \frac{1}{\beta}
  \Tr\ln \left( \mat G_{\mat \lambda',0}^{-1} - \mat\Sigma
  \right)^{-1} + \fcl F_{\mat U} [\mat \Sigma] \,.
\end{multline}
Since $\fcl F_{\mat U}[\mat\Sigma]$ is universal, we can eliminate
$\fcl F_{\mat U}[\mat\Sigma]$ and write
\begin{multline}
  \label{eq:NESEfclDiff}
  \fcl\Omega_{\mat T,\mat U}[\mat\Sigma] = \fcl\Omega_{\mat
    \lambda',\mat U}[\mat\Sigma] + \frac{1}{\beta} \Tr\ln
  \left( \mat G_{\mat T,0}^{-1} - \mat\Sigma \right)^{-1}  \\
  - \frac{1}{\beta} \Tr\ln \left( \mat G_{\mat \lambda',0}^{-1} -
    \mat\Sigma \right)^{-1} \,.
\end{multline}

The previous expression is still exact, but the self-energy functional
for the reference system is not available in a closed form, even for
very simple cases, as e.g.\ the atomic limit of the Hubbard model.
However, we can nevertheless make use of Eq.~(\ref{eq:NESEfclDiff}),
if both the exact self-energy $\mat \Sigma_{\mat \lambda', \mat U}$
and the self-energy functional of the reference system, evaluated at
the exact self-energy, i.e., $\fcl\Omega_{\mat \lambda', \mat U}[\mat
\Sigma_{\mat \lambda',\mat U}]= \Omega_{\mat \lambda',\mat U}$, are
accessible. Using Dyson's equation (Eq.~\ref{eq:DysonEqnShort}) for
the reference system, we find for the self-energy functional of the
original system if evaluated at a trial self-energy taken from the
reference system and parametrized by the set of variational parameters
$\mat \lambda'$:
\begin{multline}
  \label{eq:SFTfcl}
  \fcl\Omega_{\mat T,\mat U}[\mat\Sigma_{\mat \lambda',\mat U}] =
  \Omega_{\mat \lambda',\mat U} + \frac{1}{\beta} \Tr\ln \left( \mat
    G_{\mat T,0}^{-1} - \mat\Sigma_{\mat \lambda',\mat U} \right)^{-1}
  \\ - \frac{1}{\beta} \Tr\ln \left( \mat G_{\mat \lambda',\mat U}
  \right) \,.
\end{multline}
This shows that an exact evaluation of the general nonequilibrium
self-energy functional is possible on the restricted space of trial
self-energies spanned by any reference system with the same
interaction part, provided that the contour-ordered self-energy and
Green's function as well as the initial-state grand potential of the
reference system can be computed exactly.

The time-dependent optimal variational parameters $\mat \lambda'_{\rm
  opt}(z)$ have to be determined via the Euler equation:
\begin{equation}
  \label{eq:EulerEqn}
  \left. \frac{\delta \fcl\Omega_{\mat T,\mat U}[\mat\Sigma_{\mat \lambda',\mat U}]}{\delta \mat \lambda'(z)} \right|_{\mat\lambda'(z) =
    \mat\lambda'_{\rm opt}(z)} = 0 \, .
\end{equation}
We thus have (approximate) access to the initial-state grand potential
$\fcl\Omega_{\mat T, \mat U}[\mat \Sigma_{\mat \lambda'_{\rm opt},\mat
  U}]$ as well as to the final-state dynamics via the one-particle
Green's function
\begin{equation}
  \label{eq:gsft}
  \mat G^{\rm SFT}
  \equiv
  (\mat G_{\mat T,0}^{-1} - \mat
  \Sigma_{\mat \lambda'_{\rm opt},\mat U})^{-1} 
\end{equation}
on the Keldysh branch.  The choice of the reference system specifies
the type of approximation. Approximations generated in this way are
non-perturbative by construction.

\begin{figure}[t]\centering
  \includegraphics[width=0.8\columnwidth ]{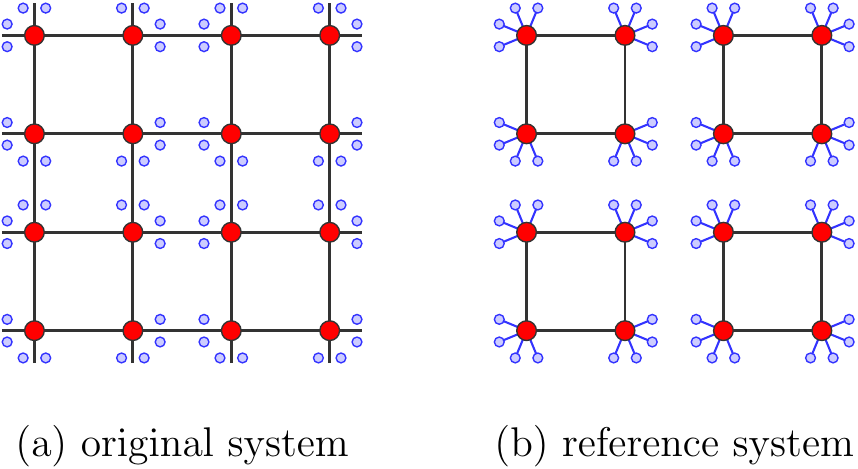}
  \caption{Schematic representation of the original system (a) and of
    a generic reference system (b). Large red circles: correlated
    sites with Hubbard-like local interaction $U$. Small blue circles:
    uncorrelated ``bath'' sites, i.e., $U = 0$. Bold black lines:
    intra-cluster hopping. Thin blue lines: hybridization, i.e.,
    hopping between correlated and bath sites in the reference system.
    Note that in the original system (a) bath sites are decoupled from
    the correlated ones. Their presence is helpful for formal reasons
    to ensure equal Hilbert space dimensions in (a) and (b).}
  \label{fig:orgrefsys}
\end{figure}

The Hamiltonian $\op H_{\mat \lambda',\mat U}$ of the reference system
must have the same interaction part as the one of the original system
and, for any practical application, must allow for an exact
calculation of the trial self-energy $\mat \Sigma_{\mat \lambda', \mat
  U}$ and of the Green's function $\mat G_{\mat \lambda',\mat U}$ by
analytical or numerical means.  Typically, this is achieved by cutting
the original lattice into disconnected clusters with a small number of
sites $L_c$ (Fig.~\ref{fig:orgrefsys}).  To enlarge the number of
variational degrees of freedom locally without changing the
interaction part, a number $L_{b}$ of uncorrelated ``bath sites'' may
be added to each of the reference system's correlates sites and
coupled to the correlated sites via a finite hybridization.  It is
convenient to have equal Hilbert spaces and thus to formally include
the bath sites in the original system as well but without a coupling
to the physical sites (Fig.~\ref{fig:orgrefsys}).  In the case of a
local (Hubbard-type) interaction and for sufficiently small $L_{c}$
and $L_{b}$, the reference system can be treated by
exact-diagonalization techniques.

% ----------------------------------------------------
\section{Dynamical mean-field theory}
\label{sec:dynamical-mean-field}

Nonequilibrium dynamical mean-field theory is recovered within the SFT
framework when we choose the reference system as a set of completely
decoupled correlated sites ($L_{c}=1$) with an infinite number of bath
sites ($L_{b}=\infty$), i.e., as a set of decoupled single-impurity
Anderson models.  For $L_{c}=1$ the trial self-energies are local,
i.e., diagonal with respect to the spatial indices, and the Euler
equation (\ref{eq:EulerEqn}) thus explicitly reads as:
\begin{widetext}
  \begin{equation}
    \label{eq:DMFTEulerEqn}
    0 
    =
    \frac{\delta \fcl\Omega_{\mat T, \mat
        U}[\mat\Sigma_{\mat \lambda',\mat U}]}{\delta\mat\lambda'(z)} = \frac{1}{\beta}
    \sum_{i,\sigma_{1}\sigma_{2}}\int_{\mc C} dz_1dz_2\, \left( \left(\mat G_{\mat
          T,0}^{-1} - \mat\Sigma_{\mat\lambda',\mat U} \right)^{-1} -
      \mat G_{\mat\lambda',\mat U}\right)_{ii,\sigma_{1}\sigma_{2}}
    (z_1,z_2)
    \frac{\delta\Sigma_{\mat\lambda',\mat
        U;ii,\sigma_{2}\sigma_{1}}(z_2,z_1^+)}{\delta\mat\lambda'(z)} \, .
  \end{equation}
\end{widetext}
Here, $i$ is a site index and $\sigma_{i}$ refers to the local orbital
and spin degrees of freedom.

Equation~(\ref{eq:DMFTEulerEqn}) would be trivially satisfied if the
bracket in the integrand vanished. Because the vanishing of the
bracket is nothing but the standard self-consistency equation of DMFT,
\cite{georges1996,freericks2006,schmidt2002} we see that
nonequilibrium SFT yields (nonequilibrium) DMFT as a stationary point
-- provided that the DMFT self-energy can be represented as the
self-energy $\mat\Sigma_{\mat\lambda',\mat U}$ of a single-impurity
Anderson Hamiltonian with single-particle (bath) parameters
$\mat\lambda'$. The representability of the DMFT action by an actual
impurity Hamiltonian with $L_b=\infty$ is not straightforward to see
for nonequilibrium Green's functions but can be shown under rather
general conditions.\cite{gramsch2013}

When one considers finite single-impurity models with a small number
of bath orbitals, the bracket in Eq.~(\ref{eq:DMFTEulerEqn}) will in
general not vanish because the discrete pole structure of the impurity
Green's function cannot be reconciled with the branch cuts of the
Green's function for the original model.  Due to the presence of the
projector $\delta\Sigma_{\mat\lambda'}/\delta\mat\lambda'$, however,
stationarity of the self-energy functional is nevertheless possible.
This allows to generate non-perturbative and consistent approximations
to DMFT by solving reference systems with a few degrees of freedom
only.  In the equilibrium case, this has been shown to be a highly
efficient strategy (see, e.g.,
Refs.~\onlinecite{pozgajcic2004,eckstein2007}).

% ----------------------------------------------------
\section{Physical and transverse variations}
\label{sec:triv-nontriv-var}

The variational problem, Eq.~(\ref{eq:EulerEqn}), is posed on the
whole contour $\mc C$, i.e., the self-energy functional must be
stationary with respect to variations of the parameters
$\mat\lambda'(z)$ \emph{separately} on the Matsubara branch and on
both branches of the Keldysh contour. This generates one
imaginary-time and two independent real-time Euler equations which are
obtained by writing $\fcl\Omega_{\mat T,\mat U}[\mat\Sigma_{\mat
  \lambda',\mat U}] \equiv \fcl\Omega_{\mat T,\mat U}[\mat\Sigma_{\mat
  \lambda'_{+},\mat \lambda'_{-},\mat \lambda'_{\rm M},\mat U}]$ as a
functional of the single particle parameters $\mat\lambda_{\pm}(t)$ on
the upper/lower branch of the contour (for real $t$), as well as of
the parameters $\mat \lambda'_{\rm M}(t_0-i\tau)$ on the Matsubara
branch.  Using a simple transformation of variables,
\begin{align}
  \mat\lambda'_{\rm phys}(t) &= \frac{1}{2} (\mat\lambda'_{+}(t) +
  \mat\lambda'_{-}(t)) \,,
  \nonumber \\
  \mat\lambda'_{\rm trans}(t) &= \frac{1}{2} (\mat\lambda'_{+}(t) -
  \mat\lambda'_{-}(t)) \,, \label{eq:lampm}
\end{align}
the real-time equations become equivalent to $\delta \fcl\Omega_{\mat
  T,\mat U}[\mat\Sigma_{\mat \lambda'_{\rm phys},\mat \lambda'_{\rm
    trans},\mat \lambda'_{\rm M},\mat U}] / \delta \mat \lambda'_{\rm
  phys / trans}(t) = 0$.
  
The separation into variations with respect to $\mat \lambda'_{\rm
  phys}$ (``physical variations'') and $\mat \lambda'_{\rm trans}$
(``transverse variations'') has a simple motivation: In the end, we
are only interested in solutions of the Euler equation by a physical
parameter set $\mat\lambda'(z)$, i.e., one that corresponds to an
actual \emph{Hamiltonian}. These parameters must thus satisfy
$\mat\lambda'_{+}(t)=\mat\lambda'_{-}(t)$, i.e., $\mat\lambda'_{\rm
  trans}(t)=0$. In addition, $\mat\lambda_M'(t_0-i\tau)$ must not depend
on imaginary time (this is discussed in Sec.\
\ref{sec:therm-equil-init}).  Transverse variations $\delta
\mat\lambda'_{\rm trans}(t) \ne 0$ shift the parameters away from the
physical manifold, while physical variations remain therein.

Let us first consider variations of $\mat\lambda'_{\rm phys}(t)$.
Interestingly, one can show that the self-energy functional is always
stationary with respect to physical variations  when evaluated at
  a physical parameter set, which satisfies $\mat\lambda'_{\rm
    trans}(t)=0$, i.e., 
\begin{equation}
  \left. \frac{\delta \fcl\Omega_{\mat T,\mat U}[\mat\Sigma_{\mat \lambda'_{\rm phys},\mat \lambda'_{\rm trans},\mat \lambda'_{\rm M}, \mat U}]}{\delta \mat \lambda_{\rm phys}'(t)}  \right|_{\mat\lambda'_{\rm trans}(t) = 0}
  = 0 \:.
  \label{eq:delphys}
\end{equation}
To prove Eq.~(\ref{eq:delphys}), we first note that any Green's
function defined by Eq.~(\ref{eq:DefNEGF}) is symmetric with respect
to a shift of the largest time-argument on the Keldysh contour from
the upper to the lower branch, i.e.,
\begin{equation}
  \label{causal symmetry}
  \begin{split}
    \mat X(t_0-i\tau,t^+) &= \mat X(t_0-i\tau,t^-),
    \\
    \mat X(t',t^+) &= \mat X(t',t^-) \text{~for~} t > t' ,
  \end{split}
\end{equation}
and similar for the first time-argument ($t^\pm$ denotes a time
argument on the upper/lower branch at $t$). This symmetry relation,
which is often formulated as fundamental relation between between
retarded, advanced, and time-ordered components of the Green's
functions, \cite{leeuwen2006c} immediately follows from the fact that
the forward and backward time-evolution cancel each other after the
right-most operator on the Keldysh contour (see also the discussion of
Fig.~\ref{fig:contour}). The same property holds for the convolution
$\mat A \circ \mat B$ of any two contour functions $\mat A$ and $\mat
B$ if it holds for $\mat A$ and $\mat B$ individually, and thus for
any function of $\mat X$
(cf. Eq.~(\ref{eq:AnalytFctContFct})). Furthermore, it is easy to see
that in the expression Eq.~(\ref{eq:ContourTrace}) for the trace all
integrations over the Keldysh branch cancel for any function with the
symmetry (\ref{causal symmetry}). Thus the self-energy functional
(\ref{eq:NESEfcl}), when evaluated at physical parameters, depends on
the Matsubara part of $\mat \Sigma_{\mat\lambda',\mat U}$ only.  This
immediately implies the stationarity condition (\ref{eq:delphys}).

Stationarity with respect to physical variations locally restricts the
solution to the physical manifold.  Thus, a second equation is needed
to fix the solution within the physical manifold.  This ``second''
equation is given by the condition that the self-energy functional be
stationary with respect to the transverse variations, if evaluated at
a physical parameter set:
\begin{equation}
  \label{eq:nontrivVar}
  \left. \frac{\delta \fcl\Omega_{\mat T,\mat U}[\mat\Sigma_{\mat \lambda'_{\rm phys},\mat \lambda'_{\rm trans},\mat \lambda'_{\rm M}, \mat U}]}{\delta \mat \lambda_{\rm trans}'(t)}  \right|_{\mat\lambda'_{\rm trans}(t) = 0}
  = 0 \: .
\end{equation}
Equation~(\ref{eq:nontrivVar}) is the central equation of the
nonequilibrium SFT.

Let us stress once more that the functional derivative with respect to
$\mat\lambda'_{\rm trans}(t)$ is a derivative into a ``non-physical''
direction in parameter space.  This has important conceptual
consequences for the numerical evaluation of the theory.  In the vast
majority of previous \emph{equilibrium} SFT studies, the grand
potential $\fcl\Omega_{\mat T,\mat U}[\mat\Sigma_{\mat \lambda',\mat
  U}]$ has been computed for different (static) parameter sets $\mat
\lambda'$, and algorithms to find a stationary point of a
multi-dimensional scalar function $\mat \lambda' \mapsto
\fcl\Omega_{\mat T,\mat U}[\mat\Sigma_{\mat \lambda',\mat U}]$ have
been employed (see Ref.\ \onlinecite{balzer2010}, for example).  In
the nonequilibrium case, a similar strategy would require to work
explicitly with Green's functions that are defined with a different
Hamiltonian for the forward and backward time-evolution. A more
convenient strategy, which is worked out in the following, is to carry
out the functional derivative analytically and to solve the resulting
Euler equation by numerical means.  The analytical expressions for the
functional derivatives are then given by higher order correlation
functions evaluated at the physical parameters.

% ----------------------------------------------------
\section{Evaluation of the Euler equation}
\label{sec:eval-euler-equat}

We focus on Eq.~(\ref{eq:SFTfcl}) again and perform the functional
derivative in Eq.~(\ref{eq:EulerEqn}) analytically.  This is most
conveniently done by considering the variational parameters as
functions of the contour variable, i.e., $\mat \lambda'(z)$ with $z
\in \mc C$, instead of treating $\mat\lambda_{\pm}(t)$ and
$\mat\lambda_{\rm M}(t_0-i\tau)$ separately.

Using the chain rule, we find:
\begin{equation}
  \label{eq:EulerEqnChainRule}
  \frac{\delta \fcl\Omega_{\mat T,\mat U}[\mat\Sigma_{\mat\lambda',\mat U}]}{\delta \lambda'_{\alpha_{1}\alpha_{2}}(z)} = 
  \Tr \left( \frac{\delta \fcl\Omega_{\mat T,\mat U}[\mat\Sigma_{\mat\lambda',\mat U}]}{\delta\mat \Sigma_{\mat
        \lambda',\mat U}} \circ \frac{\delta \mat \Sigma_{\mat
        \lambda',\mat U}}{\delta \lambda'_{\alpha_{1}\alpha_{2}}(z)} \right) \,.
\end{equation}
The first factor is given by Eq.~(\ref{eq:NESEfclDerivativeSE}) but
can be rewritten in a more convenient way.  We define the difference
between the one-particle parameters of the original and of the
reference system as
\begin{equation}
  \mat V(z) = \mat T(z) - \mat \lambda'(z) \: .
\end{equation} 
With this we immediately have (see Eq.\ (\ref{eq:inverseFreeNEGF})):
\begin{equation}
  \label{eq:freeGreensFncsOrgRef}
  G^{-1}_{\mat \lambda',0}(1,2) = G^{-1}_{\mat T,0}(1,2) +
  \delta_{\mc C}(z_1,z_2) V_{\alpha_1\alpha_2}(z_{2}) \,.
\end{equation}
Here, we use the standard notation $1 \equiv (\alpha_{1},z_{1})$ etc.
With the definition of the SFT Green's function, Eq.\ (\ref{eq:gsft}),
and with Dyson's equation for the reference system we get
\begin{equation}
  \label{eq:CPTequation}
  \mat G^{\rm SFT} = \mat G_{\mat\lambda',\mat U} + \mat G_{\mat\lambda',\mat U} \mat V \circ \mat G^{\rm SFT} \, .
\end{equation}
This equation constitutes the nonequilibrium cluster-perturbation
theory. \cite{potthoff2011c} 
One may formally consider perturbation theory with respect to $\mat V$
and define the corresponding $T$-matrix as
\begin{eqnarray}
  \mat Y_{\mat\lambda',\mat
    T,\mat U}(z_{1},z_{2}) 
  &=& 
  \mat V(z_{1}) \delta_{\mc C}(z_{1},z_{2}) 
  \nonumber \\
  &+&
  \mat V(z_{1}) \mat G^{\rm SFT}(z_{1},z_{2}) \mat V(z_{2}) 
  \: .  
\end{eqnarray}
The related Lippmann-Schwinger equation is:
\begin{equation}
  \label{eq:LippmannSchwinger}
  \mat G^{\rm SFT} = \mat
  G_{\mat\lambda',\mat U} +  \mat G_{\mat\lambda',\mat U} \circ \mat
  Y_{\mat\lambda',\mat T,\mat U} \circ \mat G_{\mat\lambda',\mat U} \, .
\end{equation}
This eventually yields
\begin{equation}
  \label{eq:first}
  \frac{\delta \fcl \Omega_{\mat T,\mat U}[\mat\Sigma_{\mat \lambda',\mat U}]}{\delta\mat \Sigma_{\mat \lambda',\mat U}} = \frac{1}{\beta} \mat G_{\mat\lambda',\mat U} \circ \mat
  Y_{\mat\lambda',\mat T,\mat U} \circ \mat G_{\mat\lambda',\mat U} \:
\end{equation}
for the first factor in Eq.~(\ref{eq:EulerEqnChainRule}).

To evaluate the second factor, the Dyson equation for the reference
system is used once more to get $\mat\Sigma_{\mat\lambda',\mat U} =
\mat G_{\mat\lambda',0}^{-1} - \mat G_{\mat\lambda',\mat U}^{-1}$.
The $\mat \lambda'$-dependence of the inverse free Green's function is
simple, $G_{\mat\lambda',0}^{-1}(1,2) = \delta_{\mc C}(z_{1},z_{2})
\delta_{\alpha_{1},\alpha_{2}} i \partial_{z_2} - \delta_{\mc
  C}(z_1,z_2) ( \lambda'_{\alpha_1\alpha_2}(z_2) -
\delta_{\alpha_1\alpha_2} \mu)$.  We thus get:
\begin{multline}
  \label{eq:NEdSEdLambda}
  \frac{\delta \Sigma_{\mat\lambda',\mat
      U}(3,4)}{\delta\lambda'_{\alpha_1\alpha_2}(z_1)} =
  - \delta_{\mc C}(z_3,z_4) \delta_{\alpha_3\alpha_1} \delta_{\mc C}(z_{4},z_{1}) \delta_{\alpha_{4}\alpha_{2}} \\
  + \iint d5d6\, G_{\mat\lambda',\mat U}^{-1}(3,5) \frac{\delta
    G_{\mat\lambda',\mat U}(5,6)}{\delta
    \lambda'_{\alpha_1\alpha_2}(z_1)} G_{\mat\lambda',\mat
    U}^{-1}(6,4)\,.
\end{multline}
The functional derivative of the Green's function is computed in the
appendix \ref{sec:phys} and given by Eq.~(\ref{eq:dGdlambda}).

Combining this with Eq.~(\ref{eq:first}), we finally get the
derivative of the self-energy functional with respect to $\mat
\lambda'(z_{1})$ in the form:
\begin{multline}
  \label{eq:EulerEqnReform}
  \frac{\delta \fcl \Omega_{\mat T,\mat U}[\mat\Sigma_{\mat
      \lambda',\mat U}]}{\delta \lambda'_{\alpha_1\alpha_2}(z_1)} =
  -\frac{1}{\beta} \iint d3d4 \, \\ Y_{\mat \lambda',\mat T,\mat
    U}(4,3) \left. L_{\mat\lambda',\mat U} (3,2,1^+,4)
  \right|_{z_2=z_1} \,,
\end{multline}
where
\begin{multline}
  \label{eq:4PointVertexFnc}
  L_{\mat\lambda',\mat U}(1,2,3,4) = G_{\mat\lambda',\mat
    U}(2,4)G_{\mat\lambda',\mat U}(1,3) \\ - G_{\mat\lambda',\mat
    U}(1,4)G_{\mat\lambda',\mat U}(2,3) + G_{\mat\lambda',\mat
    U}^{(2)}(1,2,3,4)
\end{multline}
is the two-particle (four-point) vertex function with external legs
and $\mat G_{\mat\lambda',\mat U}^{(2)}$ is the two-particle Green’s
function of the reference system, see Eq.~(\ref{eq:TwoPartGreensFnc}).

Therewith, we have the Euler equation of the nonequilibrium SFT:
\begin{equation}
  \label{eq:sfteuler}
  \iint d3d4 \, Y_{\mat \lambda'_{\rm opt},\mat T,\mat
    U}(4,3) \left. L_{\mat\lambda'_{\rm opt},\mat U} (3,2,1^+,4)
  \right|_{z_2=z_1} = 0 \: .
\end{equation}
This result will be needed both for the numerical determination of the
stationary point and for working out the relation between
nonequilibrium and conventional equilibrium SFT.

% ----------------------------------------------------
\section{Thermal equilibrium and initial state}
\label{sec:therm-equil-init}

Nonequilibrium SFT reduces to the conventional equilibrium formalism
for a system where $\mat T(z)$ and $\mat U(z)$ are constant on the
entire contour $\mc C$, i.e., for the case $\op H_{\rm fin}(t) = {\rm
  const.} = \op H_{\rm ini}$.  To prove this fact explicitly, we have
to show that a stationary point of the equilibrium SFT functional,
which determines \emph{time-independent} optimal parameters $\mat
\lambda'_{\rm opt}$, is also a stationary point of the more general
nonequilibrium Euler equation (\ref{eq:EulerEqn}), i.e., of Eq.\
(\ref{eq:sfteuler}), when $\mat T(z)$ and $\mat U(z)$ are constant.

Equilibrium SFT is obtained from the more general nonequilibrium formalism by
restricting the functional (\ref{eq:SFTfcl}) to the Matsubara branch
of the contour, and furthermore, by considering time-independent and
physical variations only, i.e., the trial self-energy $\mat
\Sigma_{\mat \lambda',\mat U}$ is obtained as the Matsubara
self-energy of a Hamiltonian with constant parameters $\mat \lambda'$,
and the parameters are varied to make $\fcl\Omega_{\mat T,\mat
  U}[\mat\Sigma_{\mat \lambda',\mat U}]$ stationary.  In the language
of the more general nonequilibrium SFT formalism, those variations
correspond to a variation $\delta \mat\lambda(z)$ which is constant
along the whole contour, i.e.,
\begin{equation}
  \label{eq:dOmegaInt}
  \frac{\partial\fcl\Omega_{\mat T,\mat
      U}[\mat\Sigma_{\mat \lambda',\mat U}]}{\partial\mat\lambda'}
  = \int_{\mc C}dz\, 
  \frac{\delta \fcl\Omega_{\mat T, \mat
      U}[\mat\Sigma_{\mat \lambda',\mat U}]}{\delta\mat\lambda'(z)}
  \,.
\end{equation}
Note that the integrations over the upper and lower branch of the
Keldysh contour cancel, as discussed in connection with Eq.\
(\ref{eq:delphys}).  We now suppose that the original Hamiltonian is
time-independent, and that $\mat \lambda'_{\rm opt}$ is a solution of
the equilibrium SFT formalism, i.e., the single variational equation
$\partial\fcl\Omega_{\mat T,\mat U}[\mat\Sigma_{\mat \lambda',\mat U}]
/\partial\mat\lambda' |_{\mat\lambda'_{\rm opt}} =0 $ is satisfied.

To see that the parameters $\mat \lambda'_{\rm opt}$ also represent a
solution of the nonequilibrium SFT, we must show that all other
variations, including physical, transverse, and Matsubara ones, vanish
as well. For this it is sufficient to show that the general
variational equation becomes time-translationally invariant, i.e.,
that the expression
\begin{equation}
  \label{eq:timeindepEuler}
  \left.\frac{\delta\fcl\Omega_{\mat T,\mat
        U}[\mat\Sigma_{\mat \lambda',\mat U}]}{\delta\mat\lambda'(z)}\right|_{
    \mat\lambda'(z)=\mat\lambda'_{\rm
      opt}} 
\end{equation}
does not depend on $z$.

Consider a $z$ on the Matsubara branch first.  Invariance under
translations of imaginary time is most easily seen from the explicit
expression (\ref{eq:EulerEqnReform}) for the variational derivative:
For $z_1=t_0-i\tau_1$, the integrals in Eq.\ (\ref{eq:EulerEqnReform})
reduce to the Matsubara branch.  Furthermore, the two functions $\mat
L$ and $\mat Y$ in the integrand are translationally invariant in
imaginary time as they are evaluated at $\tau$-independent parameters
$\mat \lambda'$.  More precisely, we can write $\mat
L(\tau_3,\tau_1,\tau_1^+,\tau_4) \equiv \tilde {\mat
  L}(\tau_3-\tau_1,\tau_4-\tau_1)$ and $\mat Y(\tau_3,\tau_4) \equiv
\tilde{\mat Y}(\tau_3-\tau_4)$ with functions $\tilde{\mat L}$ and
$\tilde{\mat Y}$ that are anti-periodic under $\tau \to \tau+\beta$.
After a shift of variables it is easily seen that the integral
in Eq.~(\ref{eq:EulerEqnReform}) does not depend on $z_1$.

For $z_1$ on the Keldysh branch, on the other hand, time-translational
invariance of Eq.~(\ref{eq:timeindepEuler}) can be seen from a Lehmann
representation (or spectral representation) of the functions $\mat L$
and $\mat Y$. The explicit calculation is more tedious and presented
in appendix \ref{sec:time-indep-euler}.

For a general nonequilibrium situation with $\op H_{\rm fin}(t) \ne
{\rm const.}$ the above argument can be used to show that the
causality principle is satisfied by the nonequilibrium SFT:
Satisfying the general variational equation (\ref{eq:EulerEqn}) for
all variations of $\mat\lambda'(t_{0}-i\tau)$ on the Matsubara branch
requires that the optimal parameters on the Matsubara branch are
$\tau$-independent and must be given by a solution of the equilibrium
SFT.  This shows that the description of the initial state is
independent from the final-state dynamics.

We also note that, as in the equilibrium case, the self-energy
functional evaluated at the stationary point, $\fcl\Omega_{\mat T,
  \mat U}[\mat\Sigma _{\mat\lambda'_{\rm opt},\mat U}]$, has a clear
physical meaning: It represents the (approximate) grand potential of
the initial thermal state.  Provided that there are several stationary
points for a given set of (time-dependent) parameters of the original
system, the one with the lowest grand potential in the initial state
describes the thermodynamically stable initial state and the emerging
final-state dynamics.  Furthermore, provided that the same type of
reference system is considered, the (approximate) description of the
initial state is on equal footing with the one for the final state.
Concluding, the nonequilibrium SFT is a true extension of the
equilibrium SFT.

% ----------------------------------------------------
\section{Propagation scheme}
\label{sec:numerics}

A numerical evaluation of the Euler equation (\ref{eq:EulerEqn}) seems
like a formidable task because already the time dependence of a single
one-particle parameter of the reference system provides an infinite
variational space.  However, the variational principle of the
nonequilibrium SFT has an inherent causal structure which allows it
to determine the optimal parameters at successively increasing
(physical) times, without modifying the result at earlier times.  This
causal structure is most easily visible from Eq.~(\ref{eq:sfteuler}):
The integrals over $z_{3}$ and $z_{4}$ extend over the entire contour
$\mc C$ but can be cut at $z_{1}=t_{1}^\pm$ such that $t_{1}$ is the
(physically) latest time (see discussion in
Sec.~\ref{sec:triv-nontriv-var}).  As all $\mat \lambda'(z)$-dependent
quantities in the integrand are exact correlation functions of the
reference system, Eq.~(\ref{eq:sfteuler}) involves $\mat \lambda'(z)$
at earlier times $t<t_{1}$ only.  Hence, the conditional equation for
$\mat \lambda'_{\rm opt}(z_{1})$ and thus $\mat \lambda'_{\rm opt}(z_{1})$ itself depends
on $\mat \lambda'_{\rm opt}(z)$ with $t<t_{1}$ only.

For a numerical evaluation of the theory, one has to start from the
Euler equation on the Matsubara branch only and perform a conventional
equilibrium SFT calculation (cf.~Sec.~\ref{sec:therm-equil-init}).
This sets the initial conditions for determination of the
time-dependent optimal variational parameters $\mat \lambda'_{\rm
  opt}(t)$.  Provided that the parameters have already been determined
at times earlier than a given physical time $t$, one has to fix $\mat
\lambda'_{{\rm opt}}(t)$ by solving Eq.~(\ref{eq:sfteuler}) with
$z_1=t_1^\pm$.  This is somewhat inconvenient as the integrand in
Eq.~(\ref{eq:sfteuler}) only implicitly depends on $\mat
\lambda'_{{\rm opt}}(t)$.  The dependence can be made explicit,
however, by means of a simple trick: Since Eq.~(\ref{eq:sfteuler})
must hold for all $z_1$, and since it holds at the initial time $t_0$
(the starting point is a stationary point of the equilibrium SFT), it
suffices to require the \emph{time derivative} of $\delta
\fcl\Omega_{\mat T,\mat U}[\mat \Sigma_{\mat\lambda',\mat U}] / \delta
\mat \lambda'(z)$, as given by Eq.~(\ref{eq:EulerEqnReform}), to
vanish instead of the function itself.  This will lead to an
expression which involves $\mat \lambda'_{{\rm opt}}(t)$ explicitly.

According to Eqs.~(\ref{eq:EulerEqnReform}) and
(\ref{eq:4PointVertexFnc}), the time derivative $(d/dt) \delta
\fcl\Omega_{\mat T,\mat U}[\mat \Sigma_{\mat\lambda',\mat U}]/\delta
\mat \lambda'(t^\pm)$ can be obtained from the corresponding equations
of motion for the four-point vertex function $\mat L_{\mat
  \lambda',\mat U}$.  Commuting the respective annihilation and
creation operators with the one-particle part of the Hamiltonian
results in matrix products with $\mat \lambda'$. Commuting with the
interacting part, however, gives rise to higher-order products of
annihilation and creation operators which we denote by $\hat\psi$ or
$\hat\psi^{\dagger}$, respectively: $\com[\ac{}(1),\op H'_1(1)] \equiv
\hat\psi(1)$ and $\com[\op H'_1(1),\cc{}(1)] \equiv
\hat\psi^\dagger(1)$.  After differentiating with respect to time, the
Euler equation on the Keldysh branch acquires the form:
\begin{equation}
  0 = i \beta \partial_{z} \frac{\delta
    \fcl\Omega_{\mat T,\mat U}[\mat \Sigma_{\mat\lambda',\mat
      U}]}{\delta
    \mat \lambda'(z)} 
  = \com[\mat K^{(0)}_{\mat\lambda'},\mat \lambda'](z) + \mat
  K^{(1)}_{\mat\lambda'}(z) \,, \label{eq:DiffEulerEqn}
\end{equation}
where we have defined
\begin{multline}
  K^{(0)}_{\mat\lambda';\alpha_2\alpha_1}(z_1) = \\
  \iint d3d4\, Y_{\mat \lambda',\mat T,\mat U}(4,3)
  \left. L_{\mat\lambda',\mat U} (3,2,1^+,4) \right|_{z_2=z_1} \,,
\end{multline}
and
\begin{multline}
  K^{(1)}_{\mat\lambda';\alpha_2\alpha_1}(z_1) = \iint d3d4\,
  Y_{\mat \lambda',\mat T,\mat U}(4,3) \times \\ \times \left[
    L_{\mat\lambda',\mat U} (3,2,1^+_{\psi},4) - L_{\mat\lambda',\mat
      U} (3,2_\psi,1^+,4) \right]_{z_2=z_1} \,.
\end{multline}
Here, indexing orbital and time arguments with $\psi$ means that the
associated operators in the respective correlation functions are
replaced by $\hat\psi$ or $\hat\psi^\dagger$, respectively.  For
example,
\begin{equation}
  i G_{\mat\lambda',\mat U}(1_\psi,2) = \ev< \ord_{\mc C} \hat\psi(1)\cc{}(2) > \,,
\end{equation}
and
\begin{equation}
  i G_{\mat\lambda',\mat U}(1,2_\psi) = \ev<  \ord_{\mc C} \ac{}(2)\hat\psi^\dagger(2) > \,.
\end{equation}

The contour integrations in $\mat K^{(0)}_{\mat\lambda'}$ and $\mat
K^{(1)}_{\mat\lambda'}$ are confined to times (physically) earlier
than $z_1$.  Hence Eq.~(\ref{eq:DiffEulerEqn}) provides an explicit
equation for the optimal parameters $\mat\lambda'_{\rm opt}(t)$ at a
given time $t$ in terms of the parameters at earlier times which can
be used to obtain the optimal solution by successively increasing $t$,
starting from the equilibrium SFT solution for the initial state.

% ----------------------------------------------------
\section{Internal consistency}
\label{sec:therm-cons}

The SFT provides access to time-dependent expectation values of
arbitrary one-particle observables as well as to the grand potential
of the initial thermal state.  An \emph{exact} relation between both
quantities can be derived by formally extending the grand canonical
density operator to the whole Keldysh-Matsubara contour, such that the
partition function reads as $Z_{\mat T,\mat U} = \tr (\ord_{\mc C}
e^{-i\int_{\mc C}dz\,\gop H_{\mat T,\mat U}(z)})$ (see also discussion
in Sec.~\ref{sec:non-equil-greens}).  The grand potential
$\Omega_{\mat T,\mat U} = -\beta^{-1} \ln Z_{\mat T,\mat U}$ then
becomes a functional of the (contour)-time dependent single-particle
parameters of the model.  We now consider an arbitrary one-particle
observable of the form $\op A(z) = \sum_{\alpha\beta}
a_{\alpha\beta}(z) \cc{\alpha}\ac{\beta}$ which couples linearly to
the Hamiltonian $\op H_{{\mat T},\mat U}(z) = \op
H^{{(0)}}_{\widetilde{\mat T},\mat U}(z) + \lambda_A(z) \op A(z)$ via
a time-dependent parameter $\lambda_A(z)$.  The set of one-particle
parameters $\mat T(z)$ comprises $\lambda_{A}(z)$ as well as the
remaining parameters $\widetilde{\mat T}(z)$.  Then, the expectation
value of $\op A(z)$ can be obtained via the linear-response relation
\begin{equation}
  \label{eq:thermodynEV}
  \ev< \op A (z)>_{\mat T,\mat U} = -i \beta 
  \left.
    \frac{\delta \Omega_{\mat T,\mat U}}{\delta
      \lambda_A(z)} 
  \right|_{\lambda_A(z)=0}\, ,
\end{equation}
where only the variational derivative in the ``transverse'' but not in
the ``physical'' contributes, as discussed in Sec.\
\ref{sec:triv-nontriv-var}.

On the other hand the expectation value may be computed from the
one-particle Green's function as:
\begin{equation}
  \label{eq:ExpValLesserSFTGreensFnc}
  \ev< \op A(z) >_{\mat T,\mat U} = -i \tr\left(\mat a(z) \mat
    G_{\mat T,\mat U}(z,z^+)\right) \, .
\end{equation}

The SFT provides approximate expressions for the grand potential as
well as for the expectation value.  However, one can show that these
approximations are consistent, i.e.:
\begin{equation}
  \label{eq:consistent}
  \frac{\delta \fcl \Omega_{\mat T,\mat U}[\mat\Sigma_{\mat \lambda'_{\rm opt},\mat U}]}{\delta \lambda_A(z)} 
  \Bigg|_{\lambda_A(z)=0}
  =
  \frac{1}{\beta} \tr\left( \mat a(t)\mat G^{\rm SFT}(z,z^+) \right) \, ,
\end{equation}
where $\mat G^{\rm SFT}$ is the SFT Green's function, Eq.\
(\ref{eq:gsft}).  Here $\fcl \Omega_{\mat T,\mat U}[\mat\Sigma_{\mat
  \lambda'_{\rm opt},\mat U}]$ is the grand potential at the optimal
parameters of the reference system which still can be considered as a
functional of the time-dependent parameters of the original system and
of $\lambda_A(z)$ in particular.  Eq.~(\ref{eq:consistent}) represents
a generalization of the ``thermodynamical consistency'' that has been
shown in the context of the equilibrium formalism
already. \cite{aichhorn2006b}

To prove Eq.~(\ref{eq:consistent}), we note that its left-hand side
has a twofold dependence on $\lambda_A(z)$: (i) via the free Green's
function of the original model, $\mat G_{{\mat T},0}^{-1}$, which
enters the second term in Eq.~(\ref{eq:SFTfcl}), and (ii) via the
optimized parameters $\mat\lambda'_{\rm opt}(z)$ which depend on the
time-dependent parameters in the final state of the original system.
Consequently, there are two terms resulting from the derivative:
\begin{eqnarray}
  \frac{\delta \fcl \Omega_{\mat T,\mat U}[\mat\Sigma_{\mat \lambda'_{\rm opt},\mat U}]}{\delta \lambda_A(z)} 
  &=&
  \frac{\delta \fcl \Omega_{\mat T,\mat U}[\mat\Sigma_{\mat \lambda'_{\rm opt},\mat U}]}{\delta \mat\lambda'_{\rm opt}} 
  \circ 
  \frac{\delta \mat\lambda'_{\rm opt}}{\delta \lambda_A(z)}
  \nonumber \\
  &+& 
  \frac{\delta \fcl \Omega_{\mat T,\mat U}[\mat\Sigma_{\mat \lambda'_{\rm opt},\mat U}]}{\delta \mat T} 
  \circ 
  \frac{\delta \mat T}{\delta \lambda_A(z)}
  \: .
\end{eqnarray}
Internal consistency is achieved because of the stationarity of the
self-energy functional at $\mat \lambda'_{\rm opt}(z)$, which implies
that the first term must vanish.  Using Eqs.~(\ref{eq:NESEfcl}) and
(\ref{eq:Tr(f(X))_contour}), the functional derivative with respect to
$\mat T(z)$ in the second term is found to be:
\begin{equation}
  \frac{\delta \fcl \Omega_{\mat T,\mat U}[\mat\Sigma_{\mat \lambda'_{\rm opt},\mat U}]}{\delta \mat T(z)} 
  = \frac{1}{\beta} \mat G^{\rm SFT}(z,z^+) \, . 
\end{equation}
The second factor yields $a_{\alpha\beta}(t)$, which proves
Eq.~(\ref{eq:consistent}).

% ----------------------------------------------------
\section{Conservation laws}
\label{sec:conservation-laws}

Approximations cannot be expected \emph{a priori} to respect
fundamental conservation laws that result from the invariance of the
Hamiltonian under certain continuous groups of unitary
transformations.  In fact, conservation of the total particle number,
the total spin or the total energy are certainly violated within
simple non-self-consistent or non-variational schemes such as the
nonequilibrium cluster-perturbation theory -- apart from certain
highly symmetric situations such as given by the Hubbard model on a
bipartite lattice at
half-filling. \cite{potthoff2011c,balzer2012,jurgenowski2013} A
general theory for real-time dynamics must therefore address the
question under which conditions an approximation is conserving.

With respect to self-consistent perturbative approximations, this
question has been answered by Baym and Kadanoff:
\cite{baym1961,baym1962} A diagrammatic approximation is defined by a
certain truncation of the skeleton-diagram expansion of the
self-energy, which yields the self-energy as a functional of the
Green's function.  Combined with Dyson's equation, which provides an
independent relation between self-energy and Green's function, the
problem can be solved using an iterative and self-consistent approach.
A perturbative approximation is found to be conserving if the
(truncated) skeleton-diagram expansion of the self-energy is obtained
as the functional derivative of an approximate Luttinger-Ward
functional that itself is constructed by truncations and re-summations
within diagrammatic weak-coupling perturbation theory, i.e., the
self-energy must be $\Phi$-derivable.  $\Phi$-derivable approximations
are conserving.

Contrary, approximations generated within the framework of the SFT are
non-perturbative and do not rely on diagrammatic re-summations.  While
the Luttinger-Ward functional is essential for the construction of the
SFT, and while the SFT self-energy is obtained as its functional
derivative, approximations are generated in a very different way as
compared to perturbation theory.  Namely, instead of truncating the
Luttinger-Ward functional diagrammatically, it is restricted to a
sub-manifold of self-energies generated by some (simpler) reference
system.  The SFT self-energy is derived from this restricted $\Phi$
functional.  Hence, approximations constructed within the SFT are
``$\Phi$-derivable'' but in a different sense as compared to
weak-coupling theory.

Note that the DMFT, as the most prominent approximation in this
context, represents an exception.  DMFT can be understood as an
approximation generated within the SFT framework (see
Sec.~\ref{sec:dynamical-mean-field}).  At the same time, DMFT is a
$\Phi$-derivable approximation in the spirit of Baym and Kadanoff as
it can be constructed diagrammatically from a truncated Luttinger-Ward
functional involving local propagators only.

In the following we will modify and adapt the essential ideas of Baym
and Kadanoff to analyze under which circumstances an \emph{arbitrary}
approximation constructed within the SFT framework is conserving.  The
important point observed by Baym and Kadanoff is that the fundamental
conservation laws, reformulated in terms of the self-energy and the
Green's function, result from invariances of the Luttinger-Ward
functional under appropriate gauge transformations of the Green's
function:
\begin{equation}
  \label{eq:LWgaugeinv}
  0 = \delta\fcl\Phi_{\mat U}[\mat G_{\mat T,\mat U}] =
  \frac{1}{\beta} \Tr\left(
    \mat\Sigma_{\mat T,\mat U} \circ \delta \mat G_{\mat T,\mat U} \right) \,.
\end{equation}

Within SFT, the self-energy functional is in fact constructed with the
help of the Luttinger-Ward functional, see
Eqs. (\ref{eq:NELWfclLegendre}) and (\ref{eq:NESEfcl}).  However, it
does not inherit its gauge invariance.  Nevertheless, the Euler
equation provides the analog of Eq.~(\ref{eq:LWgaugeinv}) at the
stationary point:
\begin{eqnarray}
  \label{eq:EulerEqnVar}
  0 &=& \left. \delta\fcl\Omega_{\mat\lambda',\mat U}[\mat
    \Sigma_{\mat\lambda',\mat U}] \right|_{\mat
    \lambda'=\mat\lambda'_{\rm opt}} 
  \nonumber \\ 
  &=& \frac{1}{\beta}
  \left. \Tr\left( \left(\mat G^{\rm SFT} - \mat
        G_{\mat\lambda',\mat U}\right) \circ \delta \mat
      \Sigma_{\mat\lambda',\mat U}\right) \right|_{\mat
    \lambda'=\mat\lambda'_{\rm opt}} \,,
\end{eqnarray}
i.e., by construction the variation of the grand potential
$\fcl\Omega_{\mat T, \mat U}[\mat \Sigma_{\mat\lambda',\mat U}]$ with
respect to an \emph{arbitrary} set of one-particle parameters of the
reference system $\mat\lambda'$ vanishes, if evaluated at the optimal
parameters.  Thus, the goal is to identify a certain class of
parameter variations which generates, via Eq.~(\ref{eq:EulerEqnVar}),
the necessary conditions on the SFT Green's function and the
self-energy from which the conservation laws derive.

% ----------------------------------------------------
\subsection{Particle number and spin}
\label{sec:particle-number-spin}

Particle-number and spin conservation can be treated simultaneously.
The integral quantities $N_{\rm tot}$ and $\mat S_{\rm tot}$ can be
expressed as
\begin{equation}
  A = \sum_{i} A_{i}
\end{equation}
in terms of local quantities $A_{i}$, the local occupation number and
the local spin, $\op n_i$ and $\op S_i^{(\eta)}$ ($\eta\in\{x,y,z\}$),
\begin{equation}
  \op A_i = \sum_{\sigma\sigma'} a_{\sigma\sigma'} \cc{i\sigma}\ac{i\sigma'} \,,
\end{equation}
using the notation
\begin{equation}
  \label{eq:casen}
  a_{\sigma\sigma'} = \delta_{\sigma\sigma'} 
\end{equation}
in the case $\op A_i = \op n_i$ and
\begin{equation}
  \label{eq:cases}
  a_{\sigma\sigma'} = \frac{1}{2} \sigma^{(\eta)}_{\sigma\sigma'} \,
\end{equation}
in the case $\op A_i = \op S_i^{(\eta)}$.  Here, $i$ refers to the
sites of the lattice model, $\sigma,\sigma' = \uparrow,\downarrow$ to
the spin degrees of freedom, and $\mat\sigma^{(\eta)}$ stands for the
three Pauli matrices.

Consider a system with Hamiltonian $\op H_{\mat T,\mat U}$ and the
time-dependent expectation value of the local quantity $\op A_{i}$ as
given by Eq.~(\ref{eq:timedepexp}).  With the help of the one-particle
Green's function, the expectation value can be written as:
\begin{equation}
  \label{eq:ExpValLesserGreensFnc}
  \ev< \op A_{i} >_{\mat T,\mat U}(t) = -i \sum_{\sigma\sigma'}  a_{\sigma\sigma'} G_{\mat T,\mat U, ii,\sigma'\sigma}(t,t^+) \,.
\end{equation}
Its equation of motion is readily obtained from the equation of motion
for the Green's function (i.e., from Eqs.~(\ref{eq:inverseFreeNEGF})
and (\ref{eq:DysonEqnShort}), or see Ref.~\onlinecite{rammer2007}).
We find:
\begin{eqnarray}
  \label{eq:EOMexpVal}
  \partial_t \ev< \op A_{i} >_{\mat T,\mat U}(t) 
  &=& 
  \sum_{\sigma\sigma'}
  a_{\sigma\sigma'} \com[\mat G_{\mat T,\mat U},\mat T]_{ii,\sigma'\sigma}(t,t^+) 
  \nonumber \\ 
  &+& 
  \sum_{\sigma\sigma'} a_{\sigma\sigma'} \contcom[\mat G_{\mat T,\mat U},\mat
  \Sigma]_{ii,\sigma'\sigma}(t,t^+) \,,
\end{eqnarray}
where $[\cdot,\cdot]$ is the commutator and $\contcom[\cdot,\cdot]$
indicates that besides the commutator a contour integration is
implied.

For a Hubbard-type model with local interaction, the second commutator
vanishes identically.  Eq.~(\ref{eq:EOMexpVal}) thus attains the form
of a continuity equation where the first commutator represents the
divergence of the charge current or spin current.  It vanishes if
summed over all sites $i$ due the cyclic property of the trace, and we
are left with $\partial_t \ev< \op A >_{\mat T,\mat U}(t) = 0$, i.e.,
conservation of the total particle number or spin.
  
Within SFT the real-time dynamics of one-particle observables is
determined by the approximate Green's function $\mat G^{\rm SFT}$, as
given by Eq.~(\ref{eq:gsft}).  The SFT self-energy is the self-energy
of a reference system with one-particle parameters $\mat \lambda'$.
Both are taken at optimal parameter values $\mat \lambda'_{\rm opt}$
satisfying the SFT Euler equation, Eq.~(\ref{eq:EulerEqn}).  Thus, our
goal is to show that
\begin{equation}
  \label{eq:CondSFTconsApprox}
  \sum_{\sigma\sigma'} a_{\sigma\sigma'} \contcom[\mat G^{\rm SFT},\mat\Sigma_{\mat\lambda'_{\rm opt},\mat U}]_{ii,\sigma'\sigma}(t,t^+)
  = 0 \,  . 
\end{equation}
This would be sufficient to ensure that an approximation constructed
within the SFT framework respects the conservation of particle number
and spin even locally.

To this end we consider the following gauge transformations of the
one-particle parameters of the reference system $\mat \lambda' \mapsto
\mat{\bar \lambda}'$,
\begin{eqnarray}
  \mat \varepsilon'(z) &\mapsto& \mat{\bar \varepsilon}'(z) = \mat
  \varepsilon'(z) - \partial_z\mat
  \chi(z)\,, 
  \nonumber \\
  \mat T'(z) &\mapsto& \mat{\bar T}'(z) = e^{i\mat\chi(z)}\mat
  T'(z)e^{-i\mat\chi(z)} \, ,
  \label{eq:gaugetrans1partpar}
\end{eqnarray}
where $\mat\varepsilon'$ denotes the (spatially) diagonal part of
$\mat\lambda'$ and $\mat T'$ its off-diagonal part.  The gauge
transformation is generated by a spatially diagonal contour function
$\mat \chi$ of the form
\begin{equation}
  \chi_{ij,\sigma\sigma'}(z) = \delta_{ij} \chi_i(z)
  a_{\sigma\sigma'} \, . \label{eq:chi}
\end{equation}
To ensure a Hermitian reference system, $\mat \chi$ must be real but
can be chosen arbitrary in other respects.  Note that $\mat \chi$
commutes with $\mat\varepsilon'$, which will become important later.
This is trivially satisfied in the case $\op A_i = \op n_i$, see
Eq.~(\ref{eq:casen}), and also holds in the case $\op A_i = \op
S_i^{(\eta)}$, see Eq.~(\ref{eq:cases}), provided that
$\mat\varepsilon'$ is independent of spin indices.  The latter is a
necessary condition to ensure total spin conservation \emph{in the
  reference system}.

The next step is to show that the above gauge transformation of the
one-particle parameters $\mat \lambda'$ implies that the exact Green's
function $\mat G' \equiv \mat G_{\mat \lambda',\mat U}$ and the exact
self-energy $\mat \Sigma' \equiv \mat \Sigma_{\mat \lambda',\mat U}$
of the reference system transform as:
\begin{equation}
  \label{eq:transformedGreenFnc}
  \mat{G}'(z_1,z_2) \mapsto
  \mat{\bar G}'(z_1,z_2) = e^{i\mat\chi(z_1)} \mat G'(z_1,z_2)
  e^{-i\mat\chi(z_2)} \,,
\end{equation}
and
\begin{equation}
  \label{eq:transformedSE}
  \mat{\Sigma}'(z_1,z_2) \mapsto
  \mat{\bar \Sigma}'(z_1,z_2) = e^{i\mat\chi(z_1)} \mat \Sigma'(z_1,z_2)
  e^{-i\mat\chi(z_2)} \, .
\end{equation}
We first note that Eq.~(\ref{eq:transformedGreenFnc}) implies
Eq.~(\ref{eq:transformedSE}), which is verified by referring to the
(exact) skeleton-diagram expansion $\mat{\bar \Sigma}' = \fcl{\mat
  \Sigma}_{\mat U}[\mat{\bar G}']$: Inserting the transformed
$\mat{\bar G}'$, the phase factors of the incoming and the outgoing
propagators cancel at each internal vertex.  Only at the two links for
the external legs the phase factors do not find a counterpart.  This
leaves us with the two phase factors at the transformed self-energy in
Eq.~(\ref{eq:transformedSE}).  In order to verify
Eq.~(\ref{eq:transformedGreenFnc}), it is sufficient to show that the
transformed Green's function and the transformed self-energy satisfy
the equation of motion for the transformed parameters:
\begin{eqnarray}
  \label{eq:EOMRefTrans}
  i\partial_{z_1} \mat{\bar G}'(z_1,z_2) &=&
  \delta_{\mc C}(z_1,z_2) + \mat{\bar
    \lambda}'(z_1)\mat{\bar G}'(z_1,z_2)  
  \nonumber \\    
  &+& (\mat{\bar
    \Sigma}'\circ\mat{\bar G}')(z_1,z_2) \,.
\end{eqnarray}
This is a straightforward calculation which makes use of the fact that
$\mat \chi$ commutes with $\mat\varepsilon'$.  See
appendix~\ref{sec:auxcalc-cons} for details.

A first-order variation of the one-particle parameters of the
reference system, given by $\delta\mat \chi(z)$ leads to the following
first-order variation of the self-energy
(cf.~Eq.~(\ref{eq:transformedSE})):
\begin{equation}
  \label{eq:PartSpinVarSE}
  \delta \mat \Sigma'(z_1,z_2) = i  \delta\mat \chi(z_1) \mat \Sigma'(z_1,z_2) -
  i \mat \Sigma'(z_1,z_2) \delta\mat \chi(z_2)  \, .
\end{equation}
This leads to a first-order variation $\delta \fcl \Omega_{\mat T,\mat
  U} [\mat \Sigma_{\mat\lambda',\mat U}]$ which vanishes for optimal
values of the variational parameters $\mat\lambda'$, provided that the
variation $\delta \mat T' = i \delta \mat \chi \mat T' - i \mat T'\delta
\mat \chi $ and $\delta \mat \varepsilon' = - \partial_z \delta \mat
\chi$ of the reference parameters induced by
Eq.~(\ref{eq:gaugetrans1partpar}) are chosen to part of our
variational space.  We insert Eq.~(\ref{eq:PartSpinVarSE}) into the
SFT Euler equation, as given by Eq.~(\ref{eq:EulerEqnVar}), and use
Eq.~(\ref{eq:chi}) to get
\begin{eqnarray}
  0 
  &=&  
  \left.\beta\delta \fcl \Omega_{\mat T,\mat U}[\mat \Sigma_{\mat\lambda',\mat U}] \right|_{\mat \lambda'=\mat\lambda'_{\rm opt}} 
  \nonumber \\
  &=& 
  - i \sum_{i,\sigma\sigma'} \int_{\mc C} dz\, a_{\sigma\sigma'}
  \Big(\contcom[\mat G^{\rm SFT},\mat \Sigma_{\mat\lambda'_{\rm opt},\mat U}]
  \nonumber \\ 
  &+& 
  \contcom[\mat G_{\mat\lambda'_{\rm opt},\mat U},\mat
  \Sigma_{\mat\lambda'_{\rm opt},\mat U}] \Big)_{ii,\sigma'\sigma}(z,z^+) \delta
  \chi_i(z) 
  \Big|_{\mat \lambda'=\mat\lambda'_{\rm opt}} 
  \, .
  \nonumber \\
  \label{eq:PartNSpinVarEuler}
\end{eqnarray}
Since this holds for arbitrary first-order variations $\delta
\chi_i(z)$, the term $\sum_{\sigma\sigma'}a_{\sigma\sigma'}(\cdots)$ must vanish. Consider the second
term in the bracket: The condition $\sum_{\sigma\sigma'}
a_{\sigma\sigma'}[\mat G_{\mat\lambda'_{\rm opt},\mat U}
\stackrel{\circ}{,} \mat \Sigma_{\mat\lambda'_{\rm opt},\mat
  U}]_{ii,\sigma'\sigma}(z,z^+) =0$ is just equivalent with local
particle-number and spin conservation \emph{in the reference system}
(see the discussion after Eq.~(\ref{eq:EOMexpVal})). Therefore,
\emph{if} this is satisfied, the first term in the bracket must vanish
as well, i.e., Eq.~(\ref{eq:CondSFTconsApprox}) is inferred.  It is
quite intuitive that particle-number and spin conservation is
respected by an approximation within the SFT only if it is exactly
satisfied for the reference system that has been chosen to specify the
approximation.  We conclude that within the SFT particle-number and
spin conservation is proliferated from the reference system, where it
must hold exactly, to the original system, where it holds when
formulated with the approximate SFT Green's function and self-energy.

The conservation laws are ensured by stationarity of the SFT grand
potential with respect to the parameter variations defined by Eq.\
(\ref{eq:gaugetrans1partpar}).  Note that $\mat\varepsilon'$ and $\mat
T'$ are not varied independently, i.e., particle-number and spin
conservation requires stationarity with respect to variations along
certain \emph{directions} in the parameter space.  In particular,
complex hopping-parameter variations must be taken into account.
Stationarity with respect to other directions can, of course, be
imposed additionally.

Consider the Hubbard model and a variational cluster approximation
(VCA) as an example.  This results from the reference system shown in
Fig.~\ref{fig:orgrefsys} for a cluster consisting of $L_{c}$
correlated sites but no additional bath degrees of freedom.  The
conservation laws are respected if arbitrarily time-dependent and
mutually independent variations for each of the on-site energies are
considered as well as the resulting variations of the intra-cluster
hopping parameters as prescribed by Eq.\
(\ref{eq:gaugetrans1partpar}).  Essentially the same holds for
approximations where additional bath degrees of freedom are considered
to enlarge the parameter space.

The calculations above also show that conservation of the \emph{total}
particle-number and the \emph{total} spin are respected with
site-independent variations, i.e., with a site-independent $\chi_i(z)
= \chi(z)$ only.  For the case of the particle number, this is
equivalent with an arbitrarily time-dependent but \emph{spatially
  homogeneous} variation of the on-site energies only as the phase
factors in the transformation law for the off-diagonal parameters
cancel each other.  Analogously, the total spin is conserved within
SFT if an arbitrarily time-dependent but \emph{spatially homogeneous}
magnetic field coupling to the total spin of the reference system is
treated as a variational parameter.

For models with local interactions but several orbital degrees of
freedom $m$, i.e., in the case of more complicated Coulomb parameters
$U_{i; m_{1}m_{2}m_{3}m_{3}}$, the local variants of the conservation
laws refer to the total particle number at a site $N_{i} =
\sum_{m\sigma} c^\dagger_{im\sigma}c_{im\sigma}$ and the total spin at
a site $\mat S_{i} = \sum_{m} \mat S_{im}$ with $\mat S_{im} = (1/2)
\sum_{\sigma\sigma'} c^\dagger_{im\sigma} \mat \sigma_{\sigma\sigma'}
c_{im\sigma'}$ as well as to the corresponding charge and spin
currents.  Here, the relevant variational parameters are the
conjugated fields $\varepsilon_{i}$ and $\mat B_{i}$ coupling to
$N_{i}$ and $\mat S_{i}$, respectively.  Note that models with
off-site Coulomb-interaction terms are in principle beyond the scope
of the SFT (however, see Ref.~\onlinecite{aichhorn2004}) as the
presence of inter-site interactions prevents a simple decomposition of
the lattice problem into independent cluster problems.

% ----------------------------------------------------
\subsection{Energy}
\label{sec:energy}

The case of energy conservation is more elaborate.  This is related to
the fact that the SFT is a variational approach which focusses on
one-particle quantities, i.e., on the variational optimization of the
one-particle self-energy and thus of the one-particle Green's
function, while the interaction part of the total energy is a
two-particle quantity.  Fortunately, it can be expressed in terms of
the one-particle Green's function and self-energy using the equation
of motion.  We can therefore proceed analogously to particle-number
and spin conservation and again try to make use of the ideas of Baym
and Kadanoff. \cite{baym1961,baym1962} Complications are nevertheless
to be expected and found in fact.

The kinetic (and potential) energy $E_{\rm kin}(t) = \ev<\op H_{\mat
  T,0}(t)>$ and the interaction energy $E_{\rm int}(t) = \ev<\op
H_{0,\mat U}(t)>$ of the system can be written as (see Refs.\
\onlinecite{baym1961,baym1962}, for example):
\begin{eqnarray}
  E_{\rm kin}(t) &=& -i \tr\left( \mat T(t) \mat
    G(t,t^+) \right) \label{eq:Ekin} \,, \\
  E_{\rm int}(t) &=& -\frac{i}{4} \tr \left(
    (\mat\Sigma\circ\mat G + \mat G\circ\mat \Sigma)(t,t^+) \right)
  \,. \label{eq:Epot}
\end{eqnarray}
The former directly follows from the definition of the Green's
function.  For the latter, we made use of the equation of motion for
the Green's function and Dyson's equation.  Note that we have written
$\mat G \equiv \mat G_{\mat T,\mat U}$ and $\mat \Sigma \equiv \mat
\Sigma_{\mat T,\mat U}$ for short.  The total energy of the system is
$E_{\rm tot}(t) \equiv \ev<\op H(t)> = E_{\rm kin}(t) + E_{\rm
  int}(t)$.  In the following, we assume that the interaction
parameters $\mat U$ are time-independent (see also discussion in Sec.\
\ref{sec:discussion}).  Using $\partial_t E_{\rm tot}(t) =
\ev< \partial_t\op H(t)>$, this immediately implies the following
energy-balance relation:
\begin{equation}
  \label{eq:dtEtotUconst}
  \frac{\partial E_{\rm tot}(t)}{\partial t} 
  = 
  \sum_{\alpha\beta}\frac{\partial
    T_{\alpha\beta}(t)}{\partial t} \ev<\cc{\alpha}(t)\ac{\beta}(t)> \, .
\end{equation}

Next, we express both the left-hand side and the right-hand side of
Eq.~(\ref{eq:dtEtotUconst}) in terms of $\mat\Sigma$ and $\mat
G$. Using the equation of motion again, the time derivatives of
$E_{\rm kin}(t)$ and $E_{\rm int}(t)$ can be computed.  From
Eq.~(\ref{eq:Ekin}) we get:
\begin{widetext}
  \begin{eqnarray}
    \frac{\partial E_{\rm kin}(t)}{\partial t} 
    &=& \sum_{\alpha\beta} T_{\alpha\beta} \left( \com[\mat G,\mat
      T]_{\beta\alpha}(t,t^+) + \contcom[\mat G,\mat
      \Sigma]_{\beta\alpha}(t,t^+)  \right) -i \tr\left( \frac{\partial
        \mat T(t)}{\partial t} \mat G(t,t^+) \right)
    \nonumber \\
    &=& \sum_{\alpha\beta} T_{\alpha\beta} \contcom[\mat G,\mat
    \Sigma]_{\beta\alpha}(t,t^+) -i \tr\left( \frac{\partial
        \mat T(t)}{\partial t} \mat G(t,t^+) \right) \label{eq:dtEkin1} \, .
  \end{eqnarray}
  Here the first term in the first line vanishes due the cyclic
  property of the trace.  Exploiting once more the equation of motion
  and the complex conjugated equation, we find:
  \begin{equation}
    \label{eq:dtEkin2}
    \frac{\partial E_{\rm kin}(t_1)}{\partial t_1} = i
    \sum_{\alpha_1}\int d2\,\left( \left( \frac{\partial}{\partial t_1}
        G(1,2) \right) \Sigma(2,1^+)  
      + \Sigma(1,2) \frac{\partial}{\partial t_1} G(2,1^+) \right) -i \tr\left( \frac{\partial
        \mat T(t_1)}{\partial t_1} \mat G(t_1,t_1^+) \right) \,.
  \end{equation}
  Note, that the last summand just equals the right-hand side of
  Eq.~(\ref{eq:dtEtotUconst}).  This equation can easily be combined
  with the time derivative of the interaction energy
  (Eq.~\ref{eq:Epot}).  After applying the product rule, the
  energy-balance relation (Eq.~\ref{eq:dtEtotUconst}) is expressed
  as:\cite{baym1962}
  \begin{eqnarray}
    \label{eq:energycons_condition}
    - \frac{3}{4} \sum_{\alpha_1}\int d2\, \frac{\partial}{\partial
      t_1} ( \Sigma(1,2)G(2,1^+) + G(1,2)\Sigma(2,1^+) ) 
    + \sum_{\alpha_1}\int d2\,\left(  \frac{\partial \Sigma(1,2)}{\partial
        t_1} G(2,1^+) 
      + G(1,2)
      \frac{\partial \Sigma(2,1^+)}{\partial t_1}  \right)   
    = 0 \: . 
    \nonumber \\
  \end{eqnarray}

  An approximation constructed within the SFT will respect energy
  balance if Eq.~(\ref{eq:energycons_condition}) holds but with $\mat
  \Sigma$ replaced by $\mat \Sigma_{\mat\lambda'_{\rm opt},\mat U}$
  and with $\mat G$ replaced by $\mat G^{\rm SFT}$.  Thus, the goal is
  to find a class of transformations of the one-particle parameters
  such that their corresponding first-order variations around the
  stationary point generate the above equation as the SFT Euler
  equation. In principle, this can be achieved with
  \begin{equation}
    \mat\lambda'(z) \mapsto \mat{\bar\lambda}'(z) = i ( 1 -
    \dot\theta^{-1/2}) \partial_z + \frac{i}{4}
    \dot\theta^{-3/2}\ddot\theta + \dot\theta^{1/2}
    \mat\lambda'(\theta) \, , 
    \label{eq:timetrans1partpar}
  \end{equation}
  where $\theta(z)$ is an arbitrary real function on the contour with
  $\partial_z\theta(z) \neq 0$ which describes a transformation of the
  time scale.  Note that due to the term $\propto \partial_z$ the
  action of $\mat{\bar\lambda}'(z)$ is non-local in time.  This is a
  severe complication if $\mat{\bar\lambda}'(z)$ should represent
  parameters of an actual impurity Hamiltonian, as discussed in Sec.\
  \ref{sec:discussion} below. It is nevertheless illustrative to see
  how energy conservation can be derived if the self-energy functional
  is stationary under the variations defined by
  Eq.~(\ref{eq:timetrans1partpar}).
 
  The time-dependent transformation of the one-particle parameters
  induces a corresponding transformation of the exact Green's function
  $\mat G' \equiv \mat G_{\mat \lambda',\mat U}$ and of the exact
  self-energy $\mat \Sigma' \equiv \mat \Sigma_{\mat \lambda',\mat U}$
  of the reference system.  For $\mat G'$ we have:
  \begin{equation}
    \label{eq:TimeTransformedGreenFnc}
    \mat G'(z_1,z_2) \mapsto \mat{\bar G}'(z_1,z_2) =
    \dot\theta_1^{1/4} \mat
    G'(\theta_1,\theta_2) \dot\theta_2^{1/4} \,,
  \end{equation}
  where the short hand notation $\theta_1 = \theta(z_1)$ and
  $\dot\theta_1 = \partial_{z_1} \theta(z_1)$ etc.\ is used.  Via the
  skeleton-diagram expansion $\mat\Sigma' = \fcl{\mat\Sigma}_{\mat
    U}[\mat G']$, this induces the following transformation of the
  self-energy:
  \begin{align}
    \label{eq:TtransformedSE}
    \mat \Sigma'(z_1,z_2) \mapsto \mat{\bar \Sigma}'(z_1,z_2) =
    \dot\theta_1^{3/4} \mat \Sigma'(\theta_1,\theta_2)
    \dot\theta_2^{3/4} \,.
  \end{align}
  Namely, any internal vertex at time $z_{i}$ connects to four
  propagators and thereby collects a factor $\dot\theta_i$ by which
  the implicit $z_i$ integration can be transformed into a $\theta_i$
  integration.  The factors $\dot\theta_1^{3/4}$ and
  $\dot\theta_2^{3/4}$ in Eq.~(\ref{eq:TtransformedSE}) result from
  the three incoming and outgoing propagators at the two ``external''
  vertices.  Now, Eq.~(\ref{eq:TimeTransformedGreenFnc}) is verified
  by showing that the asserted expression for the transformed Green's
  function $\mat{\bar G}'(z_1,z_2)$ together
  Eq.~(\ref{eq:TtransformedSE}) satisfies the equation of motion for
  transformed one-particle parameters,
  Eq.~(\ref{eq:timetrans1partpar}).  A proof for this can be found in
  appendix \ref{sec:energy-app}.

  The first-order variations of $\mat \Sigma'$ induced by this
  transformation, $\delta\mat \Sigma' =
  \left.\delta\mat\Sigma'(\theta_1,\theta_2) /
    \delta\theta\right|_{\theta=t}\circ\delta\theta$, are given by:
  \begin{eqnarray}
    \delta\Sigma'(2,1) &=& \int_{\mc C} dz\,
    \Big[
    \frac{3}{4} \Sigma'(2,1) \left( \frac{\partial}{\partial z_2} \delta(z_2-z) \right) 
    + \frac{3}{4} \Sigma'(2,1) \left( \frac{\partial}{\partial z_1} \delta(z_1-z) \right) 
    \nonumber \\
    &+&
    \left( \frac{\partial}{\partial z_2} \Sigma'(2,1) \right) \delta(z_2-z) 
    + \left(\frac{\partial}{\partial z_1} \Sigma'(2,1) \right) \delta(z_1-z) 
    \Big] \delta\theta(t)
    \,.
  \end{eqnarray}
  Inserting this into the Euler equation~(\ref{eq:EulerEqn}),
  integrating by parts and exploiting the $\delta$-functions, we are
  left with:
  \begin{eqnarray}
    0
    &=&
    \left.\beta \delta \fcl \Omega_{\mat T,\mat U}[\mat \Sigma_{\mat
        \lambda',\mat U}] \right|_{\mat\lambda'=\mat\lambda'_{\rm opt}}  
    \nonumber \\
    &=&
    \int_{\mc C} dz_1\, \left[ - \frac{3}{4} \sum_{\alpha_1}\int d2\,
      \frac{\partial}{\partial z_1} \bigg( \Sigma_{\mat \lambda',\mat
        U}(1,2)G^{\rm SFT}(2,1^+)
      + G^{\rm SFT}(1,2)\Sigma_{\mat \lambda',\mat U}(2,1^+) \bigg) \right.  
    \nonumber \\
    &+& 
    \sum_{\alpha_1}\int d2\,\left( \left( \frac{\partial}{\partial
          z_1} \Sigma_{\mat \lambda',\mat U}(1,2)\right)G^{\rm
        SFT}(2,1^+) + G^{\rm SFT}(1,2) \frac{\partial}{\partial z_1}
      \Sigma_{\mat \lambda',\mat U}(2,1^+)
    \right)  
    \nonumber \\
    &+& 
    \frac{3}{4} \sum_{\alpha_1}\int d2\,
    \frac{\partial}{\partial z_1} \bigg( \Sigma_{\mat \lambda',\mat
      U}(1,2)G_{\mat\lambda',\mat U}(2,1^+) + G_{\mat\lambda',\mat
      U}(1,2)\Sigma_{\mat \lambda',\mat U}(2,1^+)
    \bigg)  
    \nonumber \\
    &-& 
    \left. 
      \sum_{\alpha_1}\int d2\,\left(
        \left( \frac{\partial}{\partial z_1} \Sigma_{\mat \lambda',\mat
            U}(1,2)\right)G_{\mat\lambda',\mat U}(2,1^+) +
        G_{\mat\lambda',\mat U}(1,2) \frac{\partial}{\partial z_1}
        \Sigma_{\mat \lambda',\mat U}(2,1^+) \right) \right]_{\mat\lambda'=\mat\lambda'_{\rm opt}} \delta
    \theta(z_1) \; .  \label{eq:econ}
  \end{eqnarray}
\end{widetext}
At the stationary point, this holds for all variations
$\delta\theta(z_1)$.  Hence, the term in the square brackets must
vanish.  We assume that the energy-balance relation is satisfied
\emph{in the reference system} as expressed by
Eq.~(\ref{eq:energycons_condition}), with $\mat \Sigma \equiv \mat
\Sigma_{\mat T, \mat U}$ replaced by $\mat \Sigma_{\mat \lambda',\mat
  U}$ and with $\mat G \equiv \mat G_{\mat T, \mat U}$ replaced by
$\mat G_{\mat \lambda',\mat U}$.  This implies that the last two terms
in Eq.~(\ref{eq:econ}) vanish and therewith the first two terms in the
square bracket must vanish which is just equivalent with total-energy
balance within the SFT.  We conclude that within the SFT the
energy-conservation law is proliferated from the reference system to
the original system, if stationarity of the self-energy functional
under the variations defined by Eq.~(\ref{eq:timetrans1partpar}) can
be enforced.

% ----------------------------------------------------
\subsection{Discussion}
\label{sec:discussion}

However, there are two important points that need further discussion.
First, we recall that the interaction parameters must be assumed as
time independent, $\mat U=\mbox{const.}$, to show that the
nonequilibrium SFT respects conservation of energy.  In case of a
time-dependent interaction $\mat U(t)$ (and assuming the one-particle
parameters as constant for a moment), the energy-balance relation will
involve a two-particle correlation function,
\begin{equation}
  \frac{\partial E_{\rm tot}(t)}{\partial t} 
  =
  \sum_{\alpha\beta\gamma\delta}\frac{\partial
    U_{\alpha\beta\delta\gamma}(t)}{\partial t}
  \ev<\cc{\alpha}(t)\cc{\beta}(t)\ac{\gamma}(t)\ac{\delta}(t)>
  \,, \label{eq:dtEtotTconst}
\end{equation}
which cannot (easily) be expressed in terms $\mat\Sigma$ and $\mat G$.
Therefore, without further approximations, it is impossible to set up
(and prove) an energy balance equation within SFT in this case.

An exception worth mentioning is a time dependence of the simple form
$U_{\alpha\beta\delta\gamma}(t) = \kappa(t)
U_{\alpha\beta\delta\gamma}$ where we furthermore assume $\kappa(t) =
\dot\varphi^{-1}(t)$ with $\dot\varphi
(t)\equiv \partial_t\varphi(t)\neq 0$.  In this case, the time
dependence can be shifted to the one-particle parameters by a
transformation of the time scale: $\op{H}(t) \mapsto \op{\widetilde
  H}(t) = \dot\varphi(t) \op H(\varphi(t))$ and $| \widetilde \psi(t)
\rangle = | \psi(\varphi(t)) \rangle$ which leaves the Schr\"odinger
equation form invariant:
\begin{equation}
  (i \partial_t - \op{\widetilde H}(t)) | \widetilde \psi(t) \rangle
  =
  \dot\varphi(t)
  (i \partial_\varphi - \op{H}(\varphi)) | \psi(\varphi) \rangle
  = 0 \: .
\end{equation}

The second point to be discussed is that according to the presence of
the contour derivative $\partial_{z}$ in the transformation law Eq.\
(\ref{eq:timetrans1partpar}), time-non-local one-particle parameters
of the reference system are generated by a generic transformation of
the time scale $\theta(z)$.  Within the present (Hamiltonian)
formalism, time-non-local parameters $\mat \lambda'(z_{1},z_{2})$ must
be generated effectively by considering additional bath degrees of
freedom in the reference system, i.e., $\mat \lambda'(z_{1},z_{2})$
must be understood as a corresponding hybridization function
\begin{equation}
  \mat \lambda'(z_{1},z_{2})
  =
  \mat V'(z_{1}) 
  \mat G'_{0}(z_{1},z_{2})
  \mat V'(z_{2}) 
\end{equation}
where $\mat G'_{0}$ is the non-interacting bath Green's function and
$\mat V'$ the hybridization matrix element.  However, a time-non-local
term of the form $\partial_{z}$ can presumably not be represented with
the help of a \emph{finite} number of bath degrees of freedom (see
also Ref.\ \onlinecite{gramsch2013} for a discussion).  On the other
hand, with the consideration of a \emph{continuum} of bath sites one
is essentially restricted to DMFT or to cellular DMFT as
approximations that can be constructed within the SFT framework.  This
conflicts with the original intention to construct variational and
consistent approximations using reference systems with a few degrees
of freedom only which are accessible to an exact-diagonalization
technique.

However, the argument can also be turned by stating that the degree to
which energy conservation is violated within an SFT-based
approximation can be controlled systematically by increasing the
number of variational degrees of freedom in the reference system.
Adding bath degrees of freedom, for example, is expected to
substantially improve the degree to which energy conservation is
respected.  Furthermore, the analysis in Sec.\ \ref{sec:energy} shows
that a substantial violation of energy conservation should not
expected {\em for short times}.  Here, the system's dynamics is
dominated by high-energy excitations and is thus only weakly affected
by a discrete level structure.

Another option is to enforce energy conservation.  As the SFT is a
variational approach, energy conservation can easily be imposed as an
additional constraint that is used to fix the time-dependence of one
of the variational parameters.  This represents an {\em ad hoc} but
physically motivated modification of the original theory by which the
search for optimal values of the remaining variational parameters is
confined to a subspace where $E_{\rm tot} = \mbox{const}$.  Here,
$E_{\rm tot}=E_{\rm tot}[\mat \lambda'](z)$ is given by Eqs.\
(\ref{eq:Ekin}) and (\ref{eq:Epot}) with $\mat G$ and $\mat \Sigma$
replaced by $\mat G^{{\rm SFT}}$ and $\mat \Sigma_{\mat \lambda',\mat
  U}$.  The SFT variational principle, Eq.\ (\ref{eq:EulerEqn}), is
replaced by:
\begin{equation}
  \label{eq:eex1}
  E_{\rm tot}[\mat \lambda'](z) - \mbox{\rm const.} = 0
\end{equation}
and
\begin{equation}
  \label{eq:eex2}
  \frac{\delta}{\delta \mat \lambda'(z)} 
  \Big(
  \fcl\Omega_{\mat T,\mat U}[\mat\Sigma_{\mat \lambda',\mat U}]
  - \int_{\mc K} dz' \xi (z') E_{\rm tot}[\mat \lambda'](z') 
  \Big)
  =
  0
  \, ,
\end{equation}
where $\xi(z)$ is a Lagrange multiplier on the Keldysh branch $\mc K$.
Alternatively, for driven systems with an explicitly time-dependent
Hamiltonian, one may impose Eq.\ (\ref{eq:energycons_condition}),
again formulated in terms of $\mat G^{{\rm SFT}}$ and $\mat
\Sigma_{\mat \lambda',\mat U}$, as a constraint.  Again, variations in
the {\rm transverse} direction must be considered (i.e., $\mat
\lambda_{+}(t) = - \mat \lambda_{-}(t)$), followed by an evaluation on
the {\rm physical} manifold (i.e., $\mat \lambda_{+}(t) = \mat
\lambda_{-}(t)$, $\xi_{+}=\xi_{-}$), as discussed in Sec.\
\ref{sec:triv-nontriv-var}.  Furthermore, Eqs.\ (\ref{eq:eex1}) and
(\ref{eq:eex2}) have an inherent causal structure analogous to the
full SFT equations and may thus be solved by a similar propagation
algorithm as discussed in Sec.~\ref{sec:numerics}.  An overall
time-dependent scaling of the hopping parameters may be considered as
a variational parameter taken to satisfy the constraint but there is
no obvious optimal choice.

% ----------------------------------------------------
\section{Conclusions}
\label{sec:conclusions-outlook}

Self-energy functional theory (SFT) addresses the problem of strongly
correlated fermions with local Hubbard-type interactions on a
low-dimensional lattice.  One of the main advantages of the standard
equilibrium SFT is that it unifies and also extends different
approximations within a single theoretical framework.  This comprises
``two-site'' approximations \cite{potthoff2001b,potthoff2003} and the
linearized DMFT, \cite{bulla2000} dynamical impurity approximations
(DIA), \cite{potthoff2003,pozgajcic2004,eckstein2007} but also
dynamical mean-field theory (DMFT) and its cluster extensions, i.e.,
the cellular DMFT (C-DMFT)
\cite{kotliar2001,lichtenstein2000,potthoff2003c} as well as the
dynamical cluster approximation (DCA)\cite{hettler1998} (see also
Ref.\ \onlinecite{potthoff2007} for deriving the DCA within SFT), and
finally the cluster-perturbation theory (CPT)
\cite{gros1993,senechal2000} and its variational extension, the
variational cluster approach (VCA). \cite{dahnken2004} The SFT has
been extended into several directions, e.g., to systems with non-local
interactions, \cite{tong2005} to disordered \cite{potthoff2007} and to
bosonic systems. \cite{koller2006,arrigoni2011}

The present study has shown how to generalize the SFT and the
different approximations that can be constructed within the SFT to the
general nonequilibrium case.  This nonequilibrium SFT addresses
problems of transient real-time dynamics in lattice-fermion systems
far from equilibrium.  It provides approximations to describe the
dynamics of single-particle observables in a state that evolves from
an initial thermal state after a sudden quench or after an arbitrarily
time-dependent and strong perturbation.  As for the equilibrium
theory, the approximations generated are non-perturbative, consistent
in itself and can be improved systematically.  In fact, the
nonequilibrium SFT reduces to the equilibrium approach in case of an
equilibrium setup, and it comprises the equilibrium SFT which
describes the initial equilibrium state from which the subsequent
final-state dynamics evolves.  The same holds for each of the
different approximations.
  
Essentially, the main starting point for the nonequilibrium
generalization is to reformulate the entire theory in terms of the
one-particle Green's functions and the self-energy on the
Keldysh-Matsubara contour in the complex time plane.  While the basic
structure of the theory remains unchanged in this way, a much more
general approach is gained which exhibits several important aspects
that have no counterpart in the equilibrium formalism:

The first essential and important difference as compared to
equilibrium SFT consists in the fact that the Euler equation that
fixes the variational parameters results from ``transverse''
variations that involve trial self-energies away from the ``physical''
manifold while stationarity with respect to ``physical'' variations
turns out to be trivial.  Another point concerns the functional
$\fcl\Omega[\mat \Sigma]$ itself.  Evaluating the self-energy
functional at the (physical) stationary point, yields the grand
potential of the initial thermal state.  The value of the functional
thereby has a clear physical meaning which may be used to decide
between several solutions of the Euler equation.

It is remarkable that even the most simple approximations, such as the
nonequilibrium variant of the two-site DIA, can be shown to respect
the conservation laws resulting from the U(1) and SU(2) symmetries of
the original Hamiltonian.  This demonstrates that there is a class of
approximations that are ``conserving'' in the sense of Baym and
Kadanoff but non-perturbative at the same time -- apart from the
nonequilibrium DMFT, which can be understood as a $\Phi$-derivable
diagrammatic technique {\em and} as an approximation within the
nonequilibrium SFT framework.

As the nonequilibrium SFT represents a variational approach that is
based on one-particle quantities, it is not surprising that
complications show up in the context of total energy conservation.
Energy conservation can be ensured with the help of time-non-local
variational parameters or can be enforced by means of a constrained
variation -- as an {\em ad hoc} but physically motivated alternative.
We expect, however, that there is no substantial violation of
total-energy conservation in the short-time domain anyway.

Finally, the nonequilibrium SFT has an inherently causal structure,
i.e., approximations do respect the physical causality principle.
This not only is satisfying fundamentally but also important for the
numerical implementation of the theory. A time-propagation algorithm
has been proposed here which requires the exact computation of one-
and more-particle time-dependent correlation functions for the
reference system that specifies the approximation.
  
While the practical usefulness and the reliability of such
approximations has to be awaited, we do not see severe obstacles for
an implementation using reference systems with a small number of
degrees of freedom.  A very simple non-variational variant of the
nonequilibrium VCA has been implemented
already. \cite{potthoff2011c,balzer2012,jurgenowski2013} This
essentially consists in the numerical solution of the CPT equation
(\ref{eq:CPTequation}).  From the computational point of view, we
expect that the CPT equation also represents the bottleneck in case of
a fully variational NE-VCA.

Clearly, the implementation of cluster and of impurity approximations
is more involved compared to the direct mapping of the DMFT
hybridization function to a single-impurity Anderson
model,\cite{gramsch2013} but the many favorable properties of the
NE-SFT make it a very promising way to employ an exact-diagonalization
solver in the context of nonequilibrium dynamical mean-field or
cluster mean-field approaches. Work along these lines is in progress.

% ----------------------------------------------------

\acknowledgments

We would like to thank Philipp Werner for instructive discussions.
Support of this work by the Deutsche Forschungsgemeinschaft within the
Sonderforschungsbereich 925 (project B5) and by the excellence cluster
``The Hamburg Centre for Ultrafast Imaging - Structure, Dynamics and
Control of Matter at the Atomic Scale'' is gratefully acknowledged.
EA acknowledges support by the Austrian Science Fund (FWF) F4103-N13
and P24081-N16.

% ----------------------------------------------------

\appendix

% ----------------------------------------------------
\section{Analytical functions of contour functions}
\label{sec:analyt-funct-cont}

Analytical functions of contour functions $X(z,z')$ are formally
defined as
\begin{equation}
  \label{eq:AnalytFctContFct}
  f (\mat X ) = \sum_n \frac{f^{(n)}(0)}{n!} \mat X^{\circ n} \,,
\end{equation}
where the notations $\mat X^{\circ n} = \underbrace{\mat X \circ \dots
  \circ \mat X}_{n \ \rm times}$ and $\mat X^{\circ 0} = \mat 1$ are
used.  We immediately have
\begin{equation}
  \label{eq:Tr(f(X))_contour}
  \frac{\delta \Tr(f(\mat X))}{\delta \mat X(1,2)} = f'(\mat X)(2,1^+)\,.
\end{equation}
By setting $\hbar$ to one, time is measured in units $1/$energy, and
hence the contour integration carries the unit $1/$energy, too.  Therefore,
for a meaningful definition of $f$ via Eq.~(\ref{eq:AnalytFctContFct}),
its argument $\mat X$ must have energy units.  This ensures that each
${\circ}$-power of $\mat X$ has the same unit.

With the trivial inverse Green's function
\begin{equation}
  G^{-1}_{\varepsilon_0,0;\alpha\alpha'}(z,z') 
  =
  \delta_{\alpha\alpha'}\delta_{\mc C}(z,z') 
  \left( 
    i \partial_{z'} + \mu - \varepsilon_{0} 
  \right) \,,
\end{equation}
the term $\mat G_{\varepsilon_0,0}^{-1}\circ \mat G_{\mat T,\mat U}$
carries energy units, and the principal branch of the logarithm $\ln
\left( \mat G_{\varepsilon_0,0}^{-1}\circ \mat G_{\mat T,\mat U}
\right)$ is well defined for any $\varepsilon_{0}$.  For
$\varepsilon_{0} \ra \infty$, it represents a regularization of the
ill-defined expression $\ln \mat G_{\mat T,\mat U}$.  In particular,
we find that this is related to the grand potential,
\begin{equation}
  \Omega_{\mat T,0} = \frac{1}{\beta}\Tr\ln \left( \mat G_{\varepsilon_0,0}^{-1}\circ \mat
    G_{\mat T,\mat 0} \right) \: ,
\end{equation}
in the non-interacting case (see also Eq.~\ref{eq:GrandPotLWfcl}).

% ----------------------------------------------------
\section{Dependence of the Green's function on the one-particle
  parameters}
\label{sec:phys}

To exhibit the full $\mat \lambda'$-dependence of the Green's function
of the reference system $\mat G_{\mat \lambda',\mat U}(z_{1},z_{2})$,
one may switch to an ``inverted'' interaction picture where the roles
of the ``free'' and the ``interacting'' part are interchanged.  With
this choice, all expectation values and time dependencies are due to
$\op H_{0,\mat U}$ whereas all one-particle terms of the Hamiltonian
enter via the S-matrix only.  Therewith, analogously to
Eq.~(\ref{eq:IntPicNEGF}), the Green's function can be written as:
\begin{equation}
  \label{eq:RevIntPicNEGF}
  i  G_{\mat \lambda',\mat U;\alpha_{1}\alpha_{2}}(z_{1},z_{2}) 
  = 
  \frac{\ev< \ord_{\mc C} e^{-i\int_{\mc C} d z \, \gop H_{\mat \lambda',0}(z)}
    \ac{\alpha_{1}}(z_{1})\cc{\alpha_{2}}(z_{2}) >_{0,\mat U}}{\ev< \ord_{\mc C} e^{-i\int_{\mc
        C} d z \, \gop H_{\mat \lambda',0}(z)} >_{0,\mat U}} \, .
\end{equation}
Here, one can directly read off the functional derivative with respect
to $\mat \lambda'$:
\begin{multline}
  \label{eq:dGdlambda}
  \frac{\delta G_{\mat\lambda',\mat U}(1,2)}{\delta
    \lambda'_{\alpha_3\alpha_4}(z_3)} = G_{\mat\lambda',\mat U}(1,2)
  \left. G_{\mat\lambda',\mat U}(4,3^+) \right|_{z_4=z_3} \\ -
  \left. G^{(2)}_{\mat\lambda',\mat U}(1,4,3^+,2) \right|_{z_4=z_3}
  \,,
\end{multline}
where
\begin{equation}
  \label{eq:TwoPartGreensFnc}
  G^{(2)}_{\mat \lambda',\mat U}(1,2,3,4) = (-i)^2\ev<\ord_{\mc C}\ac{}(1)\ac{}(2)\cc{}(3)\cc{}(4)>
\end{equation}
is the two-particle Green's function of the reference system.

\begin{widetext}

  \section{Time-independence of the Euler equation in the equilibrium
    case}
  \label{sec:time-indep-euler}
 
  In the following, we show the time-independence of the Euler
  equation on the Keldysh contour in the equilibrium case, i.e.:
  \begin{equation}
    \label{eq:keltindep}
    \left.\frac{\delta\fcl\Omega_{\mat T,\mat
          U}[\mat\Sigma_{\mat \lambda',\mat U}]}{\delta\lambda'_{\alpha_1\alpha_2}(z_1)}\right|_{\mat\lambda'_{\rm
        opt}=\rm const} = \left.\frac{\partial\fcl\Omega_{\mat T,\mat
          U}[\mat\Sigma_{\mat \lambda',\mat U}]}{\partial\lambda'_{\alpha_1\alpha_2}}\right|_{\mat\lambda'_{\rm
        opt}=\rm const} \,, 
  \end{equation}
  where $z_1 = t_1$ is the physically largest time on the Keldysh
  contour.  To this end, we start with Eq.~(\ref{eq:EulerEqnReform})
  and make all contour integrations explicit: {\allowdisplaybreaks
    \begin{align}
      -&\beta \frac{\delta\fcl\Omega_{\mat T,\mat
          U}[\mat\Sigma_{\mat \lambda',\mat U}]}{\delta\lambda'_{\alpha_1\alpha_2}(t_1)} = \notag\\
      =& \int d 3\int d 4\, \left. Y_{\mat\lambda',\mat T,\mat U}(4,3)
        L_{\mat\lambda',\mat U}(3,2,1^+,4)\right|_{z_2=z_1} \notag\\%
      =&\phantom{+}\; \sum_{\alpha_3\alpha_4}\int d z_3\,
      Y_{43}^\delta(z_3) L_{3214}(z_3,t_1,t_1^+,z_3^+) \label{eq:delta}\\
      &+\sum_{\alpha_3\alpha_4}\int_{t_0}^{t_1} d t_3 \int_{t_0}^{t_3} d t_4\, Y^<_{43}(t_4,t_3) L^{3>}_{3214}(t_3,t_1,t_1^+,t_4) \label{eq:I.1}\\
      &+\sum_{\alpha_3\alpha_4}\int_{t_0}^{t_1} d t_3 \int_{t_3}^{t_1} d t_4\, Y^>_{43}(t_4,t_3) L^{3<}_{3214}(t_3,t_1,t_1^+,t_4) \label{eq:II.1}\\
      &-\sum_{\alpha_3\alpha_4}\int_{t_0}^{t_1} d t_3 \int_{t_0}^{t_1} d t_4\, Y^<_{43}(t_4,t_3) L^{2>}_{3214}(t_3,t_1,t_1^+,t_4) \label{eq:III.1}\\
      &-\sum_{\alpha_3\alpha_4}\int_{t_0}^{t_1} d t_3 \int_{t_0}^{t_1} d t_4\, Y^>_{43}(t_4,t_3) L^{2<}_{3214}(t_3,t_1,t_1^+,t_4) \label{eq:III.2}\\
      &+\sum_{\alpha_3\alpha_4}\int_{t_0}^{t_1} d t_3 \int_{t_3}^{t_1} d t_4\, Y^<_{43}(t_4,t_3) L^{1>}_{3214}(t_3,t_1,t_1^+,t_4) \label{eq:II.2}\\
      &+\sum_{\alpha_3\alpha_4}\int_{t_0}^{t_1} d t_3 \int_{t_0}^{t_3} d t_4\, Y^>_{43}(t_4,t_3) L^{1<}_{3214}(t_3,t_1,t_1^+,t_4) \label{eq:I.2}\\
      &-i \sum_{\alpha_3\alpha_4}\int_{t_0}^{t_1} d t_3 \int_0^\beta d\tau_4\, Y^>_{43}(t_0-i\tau_4,t_3) L^{2<}_{3214}(t_3,t_1,t_1^+,t_0-i\tau_4) \label{eq:IV.1}\\
      &+i \sum_{\alpha_3\alpha_4}\int_{t_0}^{t_1} d t_3 \int_0^\beta d\tau_4\, Y^>_{43}(t_0-i\tau_4,t_3) L^{1<}_{3214}(t_3,t_1,t_1^+,t_0-i\tau_4) \label{eq:IV.2}\\
      &-i \sum_{\alpha_3\alpha_4}\int_0^\beta d\tau_3 \int_{t_0}^{t_1} d t_4\, Y^<_{43}(t_4,t_0-i\tau_3) L^{2>}_{3214}(t_0-i\tau_3,t_1,t_1^+,t_4) \label{eq:V.1}\\
      &+i \sum_{\alpha_3\alpha_4}\int_0^\beta d\tau_3 \int_{t_0}^{t_1} d t_4\, Y^<_{43}(t_4,t_0-i\tau_3) L^{1>}_{3214}(t_0-i\tau_3,t_1,t_1^+,t_4) \label{eq:V.2}\\
      &+(-i)^2 \sum_{\alpha_3\alpha_4}\int_0^\beta d\tau_3 \int_0^{\tau_3} d\tau_4 \,  Y^<_{43}(t_0-i\tau_4,t_0-i\tau_3) L^{1>}_{3214}(t_0-i\tau_3,t_1,t_1^+,t_0-i\tau_4) \label{eq:M.1}\\
      &+(-i)^2 \sum_{\alpha_3\alpha_4}\int_0^\beta d\tau_3
      \int_{\tau_3}^\beta d\tau_4 \, Y^>_{43}(t_0-i\tau_4,t_0-i\tau_3)
      L^{1<}_{3214}(t_0-i\tau_3,t_1,t_1^+,t_0-i\tau_4)
      \,.\label{eq:M.2}
    \end{align}
  } Here we have split up the $T$-matrix $\mat Y$ into a singular,
  lesser and greater part:
  \begin{align}
    \mat Y_{\mat\lambda',\mat T,\mat U}(z_1,z_2) &= \mat V(z_1)
    \delta_{\mc C}(z_1,z_2) + \mat V(z_1) \mat G^{\rm
      SFT}(z_1,z_2) \mat V(z_2) \\
    &=: \mat Y_{\mat\lambda',\mat T,\mat U}^\delta(z_1) \delta_{\mc
      C}(z_1,z_2) + \Theta_{\mc C}(z_1,z_2)\mat Y_{\mat\lambda',\mat
      T,\mat U}^>(z_1,z_2) + \Theta_{\mc C}(z_2,z_1) \mat
    Y_{\mat\lambda',\mat T,\mat U}^<(z_1,z_2) \,, \label{eq:Ysplit}
  \end{align}
  where $\Theta_{\mc C}(z,z')$ denotes the Heaviside step function on
  the contour.  For the two-particle vertex function $\mat
  L_{\mat\lambda',\mat U}$ (Eq.~(\ref{eq:4PointVertexFnc})), the
  notation $L^{i\gtrless}(z_3,z_1,z_1^+,z_4)$ indicates that $z_1$ is
  the $i$-th time on the contour and that $z_3$ is a later/earlier
  contour-time than $z_4$. Note that we write $L^{i\gtrless}_{3214}$
  which is short for
  $L^{i\gtrless}_{\alpha_3\alpha_2\alpha_1\alpha_4}$ and that the
  indexing with the parameters $\mat\lambda'$ and $\mat U$ has been
  suppressed for brevity.

  To evaluate the above integrations in equilibrium, we express the
  $T$-matrix via its spectral representation\cite{abrikosov1975} with
  the respective spectral function $\mat A^Y$:
  \begin{equation}
    \mat Y^\gtrless(z_1,z_2) = i\int d\omega\, e^{-i\omega(z_1-z_2)}
    f^\gtrless(\omega) \mat A^Y(\omega)\,,
  \end{equation}
  with $f^<(\omega) = f(\omega)$ and $f^>(\omega) = f(\omega) - 1$ and
  where $f(\omega)$ is the Fermi function. This also implies
  \begin{equation}
    \label{eq:gtrlessRel}
    f^>(\omega)  = - e^{\omega\beta}f^<(\omega) \,.
  \end{equation}

  For the two-particle vertex function we choose the Lehmann
  representation by inserting the completeness relation $\mat 1 =
  \sum_m\ket|m>\bra<m|$ between all operators.  We find:
  \begin{align}
    L^{3>}(3,2,1^+,4)_{z_2=z_1} =\sum_{mnkl}\Big[& -M_{3214}^{mnkl}
    e^{-\beta E_m}e^{-\beta
      E_k}e^{i(E_m+E_k-E_n-E_l)z_1}e^{i(E_n-E_m)z_3}e^{i(E_l-E_k)z_4}\notag\\ %
    & +\tilde M_{3214}^{mnkl} e^{-\beta E_m}e^{-\beta
      E_k}e^{i(E_m-E_n)z_3}e^{i(E_n-E_m)z_4}\notag\\ %
    & -N_{3214}^{mnkl} e^{-\beta
      E_m}e^{i(E_m-E_k)z_1}e^{i(E_k-E_l)z_3}e^{i(E_l-E_m)z_4}\Big]\,,\label{eq:L3>}\\ %
    L^{3<}(3,2,1^+,4)_{z_2=z_1} =\sum_{mnkl}\Big[&%
    -M_{3214}^{mnkl} e^{-\beta E_m}e^{-\beta
      E_k}e^{i(E_m+E_k-E_n-E_l)z_1}e^{i(E_n-E_m)z_3}e^{i(E_l-E_k)z_4}\notag\\ %
    & -\tilde M_{3214}^{mnkl} e^{-\beta E_n}e^{-\beta
      E_k}e^{i(E_m-E_n)z_3}e^{i(E_n-E_m)z_4}\notag\\ %
    & +\tilde N_{3214}^{mnkl} e^{-\beta
      E_m}e^{i(E_m-E_k)z_1}e^{i(E_l-E_m)z_3}e^{i(E_k-E_l)z_4}\Big]
    \,,\label{eq:L3<}% \\ %
  \end{align}
  and similar expressions for $L^{1\gtrless}$ and $L^{2\gtrless}$. For
  the amplitudes we used the short-hand notations:
  \begin{align}
    M_{3214}^{mnkl}&=\frac{(-i)^2}{Z^2}\braket<m|\cc{1}|n>\braket<n|\ac{3}|m>\braket<k|\ac{2}|l>\braket<l|\cc{4}|k>\,,\label{eq:M}\\
    \tilde M_{3214}^{mnkl}&=\frac{(-i)^2}{Z^2}\braket<m|\ac{3}|n>\braket<n|\cc{4}|m>\braket<k|\cc{1}|l>\braket<l|\ac{2}|k>\,,\label{eq:Mtilde}\\
    N_{3214}^{mnkl}&=\frac{(-i)^2}{Z^2}\braket<m|\cc{1}|n>\braket<l|\cc{4}|m>\braket<n|\ac{2}|k>\braket<k|\ac{3}|l>\,,\label{eq:N}\\
    \tilde
    N_{3214}^{mnkl}&=\frac{(-i)^2}{Z^2}\braket<m|\cc{1}|n>\braket<k|\cc{4}|l>\braket<n|\ac{2}|k>\braket<l|\ac{3}|m>\,.\label{eq:Ntilde}
  \end{align}

  Let us first focus on those terms involving only greater and lesser
  parts of $\mat Y_{\mat\lambda',\mat T,\mat U}$
  (Eqs.~\ref{eq:I.1}~-~\ref{eq:M.2}) and evaluate them for each
  amplitude (Eqs.~\ref{eq:M}~-~\ref{eq:Ntilde}) separately.  For this
  purpose we write all summands~(\ref{eq:I.1})~-~(\ref{eq:M.2}) in the
  compact form
  \begin{equation}
    i \sum_{\alpha_3\alpha_4}\sum_{mnkl}\sum_{X}\int d\omega \,
    f(\omega) A^Y_{43}(\omega)
    X_{3214}^{mnkl} \mc R_{mnkl}^X(\omega,t_1) \,.
  \end{equation}
  To this end, we have made use of Eq.~(\ref{eq:gtrlessRel}) and
  factored out all common terms for each combination of amplitudes
  $X_{3214}^{mnkl}$, where $X$ stands for $M,\tilde M,N,$ or $\tilde
  N$. The remaining exponential factors, resulting from the
  time-evolution operator and the density matrix when introducing the
  Lehmann representation, and the two time integrations along the
  different branches are collected in the term $\mc
  R_{mnkl}^X(\omega,t_1)$ for each $X$. As an example, we give an
  expression for $\mc R_{mnkl}^N(\omega,t_1)$ in the following and tag
  each summand according to its origin in the above expression for
  $-\beta \delta\fcl\Omega_{\mat T,\mat U}[\mat\Sigma_{\mat
    \lambda',\mat U}] / \delta\lambda'_{\alpha_1\alpha_2}(t_1)$:
  \begin{align*}
    &\mc R_{mnkl}^N(\omega,t_1) =\\
    &\;- e^{-\beta E_m} \left( -\mc I_{c'}^N + \mc I_{a'}^N + \mc
      I_{b_1}^N - \mc I_{a}^N \right) \tag{{\rm from} \ref{eq:I.1}}\\
    &\;+ e^{-\beta (E_l-\omega)} \left( -\mc I_{c}^N + \mc I_{b_1}^N + \mc I_{b_2}^N - \mc I_{a}^N \right) \tag{{\rm from} \ref{eq:III.2}}\\
    &\;- e^{-\beta E_k} \left( -\mc I_{c}^N + \mc I_{b_2}^N + \mc I_{c'}^N - \mc I_{a'}^N \right) \tag{{\rm from} \ref{eq:II.2}}\\
    &\;+  \left( e^{-\beta E_m} - e^{-\beta (E_l-\omega)} \right) \left( \mc I_{b_1}^N - \mc I_{a}^N \right) \tag{{\rm from} \ref{eq:IV.1}}\\
    &\;+  \left( e^{-\beta E_k} - e^{-\beta(E_l-\omega)} \right) \left( \mc I_{b_2}^N - \mc I_{a}^N \right) \tag{{\rm from} \ref{eq:V.2}}\\
    &\;+ \left[ \left( e^{-\beta E_m} - e^{-\beta E_k} \right)\mc
      I_{a'}^N + \left( e^{-\beta E_k} + e^{-\beta(E_l-\omega)}
      \right)\mc I_{a}^N \right] \,. \tag{{\rm from} \ref{eq:M.1}}
  \end{align*}
  Here, the results of the different integrals are given by:
  \begin{alignat}{3}
    \mc I_{a}^N &:= \mc I_{a,mnkl}^N(\omega,t_1) &\,=\,& \frac{1}{E_l-E_m-\omega}\frac{1}{E_k-E_l+\omega} e^{i(E_m-E_k)(t_1-t_0)} \,,\\
    \mc I_{a'}^N &:= \mc I_{a',mnkl}^N(\omega,t_1) &\,=\,& \frac{1}{E_l-E_m-\omega}\frac{1}{E_k-E_m} e^{i(E_m-E_k)(t_1-t_0)} \,,\\
    \mc I_{b_1}^N &:= \mc I_{b_1,mnkl}^N(\omega,t_1) &\,=\,& \frac{1}{E_l-E_m-\omega}\frac{1}{E_k-E_l+\omega} e^{i(E_m-E_l+\omega)(t_1-t_0)} \,,\\
    \mc I_{b_2}^N &:= \mc I_{b_2,mnkl}^N(\omega,t_1) &\,=\,& \frac{1}{E_l-E_m-\omega}\frac{1}{E_k-E_l+\omega} e^{i(E_l-E_k-\omega)(t_1-t_0)} \,,\\
    \mc I_{c}^N &:= \mc I_{c,mnkl}^N(\omega) &\,=\,& \frac{1}{E_l-E_m-\omega}\frac{1}{E_k-E_l+\omega} \,,\\
    \mc I_{c'}^N &:= \mc I_{c',mnkl}^N(\omega) &\,=\,&
    \frac{1}{E_l-E_m-\omega}\frac{1}{E_k-E_m} \,.
  \end{alignat}
  By collecting prefactors, we find that all explicitly
  $t_1$-dependent parts drop out and that only those containing $\mc
  I_{c}^N$ and $\mc I_{c'}^N$ contribute. Analogous calculations lead
  to the same result for $M$, $\tilde M$ and $\tilde N$, and we thus
  conclude:
  \begin{equation}
    \mc R_{mnkl}^X(\omega,t_1) = \mc R_{mnkl}^X(\omega) \quad\forall X\,.
  \end{equation}

  The singular part $\sum_{\alpha_3\alpha_4}\int d z_3\,
  Y_{43}^\delta(z_3) L_{3214}(z_3,t_1,t_1^+,z_3^+)$
  (Eq.~\ref{eq:delta}) is evaluated straightforwardly and also turns
  out to be independent of the time $t_1$. This completes the proof of
  Eq.\ (\ref{eq:keltindep}).

  % ----------------------------------------------------

  \section{Form invariance of the equation of motion under gauge
    transformations}
  \label{sec:auxcalc-cons}

  Here, we show the form invariance of the equation of motion for the
  Green's function under the gauge transformations Eq.\
  (\ref{eq:gaugetrans1partpar}).  To verify the transformed equation
  of motion, Eq.~(\ref{eq:EOMRefTrans}), we first compute the
  left-hand side:
  \begin{equation}
    i\partial_{z_1} \mat{\bar G}'(z_1,z_2) = e^{i\mat \chi(z_1)} i\partial_{z_1}
    \mat G'(z_1,z_2) e^{-i\mat \chi(z_2)} - (\partial_{z_1}\mat\chi(z_1))
    e^{i\mat \chi(z_1)} \mat G'(z_1,z_2) e^{-i\mat \chi(z_2)} \,. \label{eq:tdiffTransGreen1part}
  \end{equation}
  To treat the second term on the right-hand side of Eq.\
  (\ref{eq:EOMRefTrans}), we distinguish between (spatially)
  diagonal and off-diagonal parts of the one-particle parameters and
  apply the respective transformation laws,
  Eq.~(\ref{eq:gaugetrans1partpar}).  This yields:
  \begin{eqnarray}
    \mat{\bar \lambda}'(z_1)\mat{\bar G}'(z_1,z_2) &=& \mat{\bar
      \varepsilon}'(z_1)\mat{\bar G}'(z_1,z_2) + \mat{\bar
      T}'(z_1)\mat{\bar G}'(z_1,z_2) 
    \nonumber \\
    &=& \mat \varepsilon'(z_1) e^{i\mat \chi(z_1)} \mat G'(z_1,z_2)
    e^{-i\mat \chi(z_2)} + e^{i\mat \chi(z_1)} \mat T'(z_{1}) e^{-i\mat
      \chi(z_1)} e^{i\mat \chi(z_1)} \mat G'(z_1,z_2) e^{-i\mat
      \chi(z_2)} 
    \nonumber \\ 
    &-& (\partial_{z_1}\mat\chi(z_1))
    e^{i\mat \chi(z_1)} \mat
    G'(z_1,z_2) e^{-i\mat \chi(z_2)} 
    \nonumber  \\
    &=& e^{i\mat \chi(z_1)} \mat \lambda'(z_1) \mat G'(z_1,z_2)
    e^{-i\mat \chi(z_2)} - (\partial_{z_1}\mat\chi(z_1)) e^{i\mat
      \chi(z_1)} \mat G'(z_1,z_2) e^{-i\mat \chi(z_2)}
    \,. \label{eq:lamGtrans}
  \end{eqnarray}
  In the last step we made use of the commutativity of
  $\mat\varepsilon'$ and $\mat\chi$.  The second terms in
  Eq.~(\ref{eq:tdiffTransGreen1part}) and in Eq.~(\ref{eq:lamGtrans})
  cancel each other.  Finally, we have $(\mat{\bar
    \Sigma}'\circ\mat{\bar G}')(z_1,z_2) = e^{i\mat \chi(z_1)} (\mat
  \Sigma'\circ\mat G')(z_1,z_2) e^{-i\mat \chi(z_2)}$ and $\delta_{\mc
    C}(z_1,z_2) = e^{i\mat \chi(z_1)}\delta_{\mc C}(z_1,z_2)e^{-i\mat
    \chi(z_2)}$.  Thus, we conclude that the transformed equation of
  motion is solved by the transformed Green's function and self-energy
  if the original one was solved by the original quantities.
  
% ----------------------------------------------------
  \section{Form-invariance of the equation of motion under
    transformations of the time scale}
  \label{sec:energy-app}

  Here, we show the form invariance of the equation of motion for the
  Green's function under the transformations of the time scale, Eq.\
  (\ref{eq:timetrans1partpar}).  To verify the transformed equation of
  motion,
  \begin{equation}
    \label{eq:eom1}
    i\partial_{z_1} \mat{\bar G}'(z_1,z_2) =
    \delta_{\mc C}(z_1,z_2) + \mat{\bar
      \lambda}'(z_1)\mat{\bar G}'(z_1,z_2)  
    + (\mat{\bar \Sigma}'\circ\mat{\bar G}')(z_1,z_2) \: ,
  \end{equation}
  we first compute the left-hand side:
  \begin{eqnarray}
    i\partial_{z_1} \mat{\bar G'}(z_1,z_2) &=& i \partial_{z_1} \left(
      \dot\theta_1^{1/4} \mat G'(\theta_1,\theta_2) \dot\theta_2^{1/4}
    \right) = \frac{i}{4} \dot\theta_1^{-3/4} \ddot\theta_1 \mat
    G'(\theta_1,\theta_2) \dot\theta_2^{1/4} + i
    \dot\theta_1^{5/4} \partial_{\theta_1} \mat G'(\theta_1,\theta_2)
    \dot\theta_2^{1/4} 
    \nonumber \\
    &=& \dot\theta_1^{3/4} \left( \frac{i}{4} \dot\theta_1^{-3/2}
      \ddot\theta_1 \mat G'(\theta_1,\theta_2) + i
      \dot\theta_1^{1/2} \partial_{\theta_1}\mat
      G'(\theta_1,\theta_2) \right) \dot\theta_2^{1/4} \, .
  \end{eqnarray}
  With Eq.~(\ref{eq:timetrans1partpar}) we find:
  \begin{eqnarray}
    \mat{\bar \lambda}'(z_1) \mat{\bar G'}(z_1,z_2) &=& \left( i ( 1 -
      \dot\theta_1^{-1/2}) \partial_z + \frac{i}{4}
      \dot\theta_1^{-3/2}\ddot\theta_1 + \dot\theta_1^{1/2}
      \mat\lambda'(\theta_1) \right) \dot\theta_1^{1/4} \mat
    G'(\theta_1,\theta_2)
    \dot\theta_2^{1/4} 
    \nonumber \\
    &=& \left( \frac{i}{4} ( 1 - \dot\theta_1^{-1/2})
      \dot\theta_1^{-3/4}\ddot\theta_1 + \frac{i}{4}
      \dot\theta_1^{-5/4}\ddot\theta_1 + \dot\theta_1^{3/4}
      \mat\lambda'(\theta_1) + i ( 1 - \dot\theta_1^{-1/2})
      \dot\theta_1^{5/4} \partial_{\theta_1} \right) \mat G'(\theta_1,\theta_2)
    \dot\theta_2^{1/4} 
    \nonumber \\
    &=& \dot\theta_1^{3/4} \left( \frac{i}{4} \dot\theta_1^{-3/2}
      \ddot\theta_1 \mat G'(\theta_1,\theta_2) + i
      \dot\theta_1^{1/2} \partial_{\theta_1}\mat G'(\theta_1,\theta_2)
    \right) \dot\theta_2^{1/4} - \dot\theta_1^{3/4} \Big(
    i \partial_{\theta_1}\mat G'(\theta_1,\theta_2) -
    \mat\lambda'(\theta_1) \mat
    G'(\theta_1,\theta_2)  \Big) \dot\theta_2^{1/4} \,.
    \nonumber \\
  \end{eqnarray}
  Combining both equations leaves us with the following expression:
  \begin{equation}
    i\partial_{z_1} \mat{\bar G'}(z_1,z_2) - \mat{\bar
      \lambda}'(z_1) \mat{\bar G'}(z_1,z_2) = \dot\theta_1^{3/4} \Big(
    i \partial_{\theta_1}\mat G'(\theta_1,\theta_2) -
    \mat\lambda'(\theta_1) \mat
    G'(\theta_1,\theta_2)  \Big) \dot\theta_2^{1/4} \,.
  \end{equation}
  Furthermore, using the substitution rule, we find both, $(\mat{\bar
    \Sigma}'\circ\mat{\bar G}')(z_1,z_2) = \dot\theta_1^{3/4}
  (\mat\Sigma'\circ\mat G')(\theta_1,\theta_2) \dot\theta_2^{1/4}$ and
  $\delta(z_1,z_2) = \dot\theta_1 \delta(\theta_1,\theta_2) =
  \dot\theta_1^{3/4} \delta(\theta_1,\theta_2) \dot\theta_2^{1/4}$.
  Thus, assembling all parts completes the proof.

\end{widetext}

\def\BibitemShut#1{}


\begin{thebibliography}{10}%
\makeatletter
\providecommand \@ifxundefined [1]{%
 \ifx #1\undefined \expandafter \@firstoftwo
 \else \expandafter \@secondoftwo
\fi
}%
\providecommand \@ifnum [1]{%
 \ifnum #1\expandafter \@firstoftwo
 \else \expandafter \@secondoftwo
\fi
}%
\providecommand \enquote [1]{``#1''}%
\providecommand \bibnamefont  [1]{#1}%
\providecommand \bibfnamefont [1]{#1}%
\providecommand \citenamefont [1]{#1}%
\providecommand\href[0]{\@sanitize\@href}%
\providecommand\@href[1]{\endgroup\@@startlink{#1}\endgroup\@@href}%
\providecommand\@@href[1]{#1\@@endlink}%
\providecommand \@sanitize [0]{\begingroup\catcode`\&12\catcode`\#12\relax}%
\@ifxundefined \pdfoutput {\@firstoftwo}{%
 \@ifnum{\z@=\pdfoutput}{\@firstoftwo}{\@secondoftwo}%
}{%
 \providecommand\@@startlink[1]{\leavevmode}%
 \providecommand\@@endlink[0]{}%
}{%
 \providecommand\@@startlink[1]{%
  \leavevmode
  \pdfstartlink
   attr{/Border[0 0 1 ]/H/I/C[0 1 1]}%
   user{/Subtype/Link/A<</Type/Action/S/URI/URI(#1)>>}%
  \relax
 }%
 \providecommand\@@endlink[0]{\pdfendlink}%
}%
\providecommand \url  [0]{\begingroup\@sanitize \@url }%
\providecommand \@url [1]{\endgroup\@href {#1}{\urlprefix}}%
\providecommand \urlprefix [0]{URL }%
\providecommand \Eprint[0]{\href }%
\@ifxundefined \urlstyle {%
  \providecommand \doi [1]{doi:\discretionary{}{}{}#1}%
}{%
  \providecommand \doi [0]{doi:\discretionary{}{}{}\begingroup
  \urlstyle{rm}\Url }%
}%
\providecommand \doibase [0]{http://dx.doi.org/}%
\providecommand \Doi[1]{\href{\doibase#1}}%
\providecommand \bibAnnote [3]{%
  \BibitemShut{#1}%
  \begin{quotation}\noindent
    \textsc{Key:}\ #2\\\textsc{Annotation:}\ #3%
  \end{quotation}%
}%
\providecommand \bibAnnoteFile [2]{%
  \IfFileExists{#2}{\bibAnnote {#1} {#2} {\input{#2}}}{}%
}%
\providecommand \typeout [0]{\immediate \write \m@ne }%
\providecommand \selectlanguage [0]{\@gobble}%
\providecommand \bibinfo [0]{\@secondoftwo}%
\providecommand \bibfield [0]{\@secondoftwo}%
\providecommand \translation [1]{[#1]}%
\providecommand \BibitemOpen[0]{}%
\providecommand \bibitemStop [0]{}%
\providecommand \bibitemNoStop [0]{.\EOS\space}%
\providecommand \EOS [0]{\spacefactor3000\relax}%
\providecommand \BibitemShut [1]{\csname bibitem#1\endcsname}%
%</preamble>
\bibitem{kinoshita2006}%
  \BibitemOpen
  \bibfield{author}{%
  \bibinfo {author} {\bibfnamefont{T.}~\bibnamefont{Kinoshita}}, \bibinfo
  {author} {\bibfnamefont{T.}~\bibnamefont{Wenger}},\ and\ \bibinfo {author}
  {\bibfnamefont{D.~S.}\ \bibnamefont{Weiss}},\ }%
  \bibfield{journal}{%
  \Doi{10.1038/nature04693}{\bibinfo {journal} {Nature}}\ }%
  \textbf{\bibinfo {volume} {440}},\ \bibinfo {pages} {900} (\bibinfo {year}
  {2006})%
  \bibAnnoteFile{NoStop}{kinoshita2006}%
\bibitem{mitra2006}%
  \BibitemOpen
  \bibfield{author}{%
  \bibinfo {author} {\bibfnamefont{A.}~\bibnamefont{Mitra}}, \bibinfo {author}
  {\bibfnamefont{S.}~\bibnamefont{Takei}}, \bibinfo {author}
  {\bibfnamefont{Y.~B.}\ \bibnamefont{Kim}},\ and\ \bibinfo {author}
  {\bibfnamefont{A.~J.}\ \bibnamefont{Millis}},\ }%
  \bibfield{journal}{%
  \Doi{10.1103/PhysRevLett.97.236808}{\bibinfo {journal} {Phys. Rev. Lett.}}\
  }%
  \textbf{\bibinfo {volume} {97}},\ \bibinfo {eid} {236808} (\bibinfo {year}
  {2006}),\
  \Eprint{http://arxiv.org/abs/arXiv:cond-mat/0607256}{arXiv:cond-mat/0607256}%
  \bibAnnoteFile{NoStop}{mitra2006}%
\bibitem{diehl2010}%
  \BibitemOpen
  \bibfield{author}{%
  \bibinfo {author} {\bibfnamefont{S.}~\bibnamefont{Diehl}}, \bibinfo {author}
  {\bibfnamefont{A.}~\bibnamefont{Tomadin}}, \bibinfo {author}
  {\bibfnamefont{A.}~\bibnamefont{Micheli}}, \bibinfo {author}
  {\bibfnamefont{R.}~\bibnamefont{Fazio}},\ and\ \bibinfo {author}
  {\bibfnamefont{P.}~\bibnamefont{Zoller}},\ }%
  \bibfield{journal}{%
  \Doi{10.1103/PhysRevLett.105.015702}{\bibinfo {journal} {Phys. Rev. Lett.}}\
  }%
  \textbf{\bibinfo {volume} {105}},\ \bibinfo {eid} {015702} (\bibinfo {year}
  {2010}),\ \Eprint{http://arxiv.org/abs/1003.2071}{arXiv:1003.2071
  [cond-mat.quant-gas]}%
  \bibAnnoteFile{NoStop}{diehl2010}%
\bibitem{leggett1987}%
  \BibitemOpen
  \bibfield{author}{%
  \bibinfo {author} {\bibfnamefont{A.~J.}\ \bibnamefont{Leggett}}, \bibinfo
  {author} {\bibfnamefont{S.}~\bibnamefont{Chakravarty}}, \bibinfo {author}
  {\bibfnamefont{A.~T.}\ \bibnamefont{Dorsey}}, \bibinfo {author}
  {\bibfnamefont{M.~P.~A.}\ \bibnamefont{Fisher}}, \bibinfo {author}
  {\bibfnamefont{A.}~\bibnamefont{Garg}},\ and\ \bibinfo {author}
  {\bibfnamefont{W.}~\bibnamefont{Zwerger}},\ }%
  \bibfield{journal}{%
  \Doi{10.1103/RevModPhys.59.1}{\bibinfo {journal} {Rev. Mod. Phys.}}\ }%
  \textbf{\bibinfo {volume} {59}},\ \bibinfo {pages} {1} (\bibinfo {year}
  {1987})%
  \bibAnnoteFile{NoStop}{leggett1987}%
\bibitem{iwai2003}%
  \BibitemOpen
  \bibfield{author}{%
  \bibinfo {author} {\bibfnamefont{S.}~\bibnamefont{Iwai}}, \bibinfo {author}
  {\bibfnamefont{M.}~\bibnamefont{Ono}}, \bibinfo {author}
  {\bibfnamefont{A.}~\bibnamefont{Maeda}}, \bibinfo {author}
  {\bibfnamefont{H.}~\bibnamefont{Matsuzaki}}, \bibinfo {author}
  {\bibfnamefont{H.}~\bibnamefont{Kishida}}, \bibinfo {author}
  {\bibfnamefont{H.}~\bibnamefont{Okamoto}},\ and\ \bibinfo {author}
  {\bibfnamefont{Y.}~\bibnamefont{Tokura}},\ }%
  \bibfield{journal}{%
  \Doi{10.1103/PhysRevLett.91.057401}{\bibinfo {journal} {Phys. Rev. Lett.}}\
  }%
  \textbf{\bibinfo {volume} {91}},\ \bibinfo {pages} {057401} (\bibinfo {year}
  {2003})%
  \bibAnnoteFile{NoStop}{iwai2003}%
\bibitem{perfetti2006}%
  \BibitemOpen
  \bibfield{author}{%
  \bibinfo {author} {\bibfnamefont{L.}~\bibnamefont{Perfetti}}, \bibinfo
  {author} {\bibfnamefont{P.~A.}\ \bibnamefont{Loukakos}}, \bibinfo {author}
  {\bibfnamefont{M.}~\bibnamefont{Lisowski}}, \bibinfo {author}
  {\bibfnamefont{U.}~\bibnamefont{Bovensiepen}}, \bibinfo {author}
  {\bibfnamefont{H.}~\bibnamefont{Berger}}, \bibinfo {author}
  {\bibfnamefont{S.}~\bibnamefont{Biermann}}, \bibinfo {author}
  {\bibfnamefont{P.~S.}\ \bibnamefont{Cornaglia}}, \bibinfo {author}
  {\bibfnamefont{A.}~\bibnamefont{Georges}},\ and\ \bibinfo {author}
  {\bibfnamefont{M.}~\bibnamefont{Wolf}},\ }%
  \bibfield{journal}{%
  \Doi{10.1103/PhysRevLett.97.067402}{\bibinfo {journal} {Phys. Rev. Lett.}}\
  }%
  \textbf{\bibinfo {volume} {97}},\ \bibinfo {pages} {067402} (\bibinfo {year}
  {2006})%
  \bibAnnoteFile{NoStop}{perfetti2006}%
\bibitem{wall2009}%
  \BibitemOpen
  \bibfield{author}{%
  \bibinfo {author} {\bibfnamefont{S.}~\bibnamefont{Wall}}, \bibinfo {author}
  {\bibfnamefont{D.}~\bibnamefont{Prabhakaran}}, \bibinfo {author}
  {\bibfnamefont{A.~T.}\ \bibnamefont{Boothroyd}},\ and\ \bibinfo {author}
  {\bibfnamefont{A.}~\bibnamefont{Cavalleri}},\ }%
  \bibfield{journal}{%
  \Doi{10.1103/PhysRevLett.103.097402}{\bibinfo {journal} {Phys. Rev. Lett.}}\
  }%
  \textbf{\bibinfo {volume} {103}},\ \bibinfo {pages} {097402} (\bibinfo {year}
  {2009})%
  \bibAnnoteFile{NoStop}{wall2009}%
\bibitem{jaksch1998}%
  \BibitemOpen
  \bibfield{author}{%
  \bibinfo {author} {\bibfnamefont{D.}~\bibnamefont{Jaksch}}, \bibinfo {author}
  {\bibfnamefont{C.}~\bibnamefont{Bruder}}, \bibinfo {author}
  {\bibfnamefont{J.~I.}\ \bibnamefont{Cirac}}, \bibinfo {author}
  {\bibfnamefont{C.~W.}\ \bibnamefont{Gardiner}},\ and\ \bibinfo {author}
  {\bibfnamefont{P.}~\bibnamefont{Zoller}},\ }%
  \bibfield{journal}{%
  \Doi{10.1103/PhysRevLett.81.3108}{\bibinfo {journal} {Phys. Rev. Lett.}}\ }%
  \textbf{\bibinfo {volume} {81}},\ \bibinfo {pages} {3108} (\bibinfo {year}
  {1998}),\
  \Eprint{http://arxiv.org/abs/arXiv:cond-mat/9805329}{arXiv:cond-mat/9805329}%
  \bibAnnoteFile{NoStop}{jaksch1998}%
\bibitem{bloch2008}%
  \BibitemOpen
  \bibfield{author}{%
  \bibinfo {author} {\bibfnamefont{I.}~\bibnamefont{Bloch}}, \bibinfo {author}
  {\bibfnamefont{J.}~\bibnamefont{Dalibard}},\ and\ \bibinfo {author}
  {\bibfnamefont{W.}~\bibnamefont{Zwerger}},\ }%
  \bibfield{journal}{%
  \Doi{10.1103/RevModPhys.80.885}{\bibinfo {journal} {Rev. Mod. Phys.}}\ }%
  \textbf{\bibinfo {volume} {80}},\ \bibinfo {pages} {885} (\bibinfo {year}
  {2008}),\ \Eprint{http://arxiv.org/abs/0704.3011}{arXiv:0704.3011}%
  \bibAnnoteFile{NoStop}{bloch2008}%
\bibitem{strohmaier2010}%
  \BibitemOpen
  \bibfield{author}{%
  \bibinfo {author} {\bibfnamefont{N.}~\bibnamefont{Strohmaier}}, \bibinfo
  {author} {\bibfnamefont{D.}~\bibnamefont{Greif}}, \bibinfo {author}
  {\bibfnamefont{R.}~\bibnamefont{J\"ordens}}, \bibinfo {author}
  {\bibfnamefont{L.}~\bibnamefont{Tarruell}}, \bibinfo {author}
  {\bibfnamefont{H.}~\bibnamefont{Moritz}}, \bibinfo {author}
  {\bibfnamefont{T.}~\bibnamefont{Esslinger}}, \bibinfo {author}
  {\bibfnamefont{R.}~\bibnamefont{Sensarma}}, \bibinfo {author}
  {\bibfnamefont{D.}~\bibnamefont{Pekker}}, \bibinfo {author}
  {\bibfnamefont{E.}~\bibnamefont{Altman}},\ and\ \bibinfo {author}
  {\bibfnamefont{E.}~\bibnamefont{Demler}},\ }%
  \bibfield{journal}{%
  \Doi{10.1103/PhysRevLett.104.080401}{\bibinfo {journal} {Phys. Rev. Lett.}}\
  }%
  \textbf{\bibinfo {volume} {104}},\ \bibinfo {pages} {080401} (\bibinfo {year}
  {2010}),\ \Eprint{http://arxiv.org/abs/0905.2963}{arXiv:0905.2963
  [cond-mat.quant-gas]}%
  \bibAnnoteFile{NoStop}{strohmaier2010}%
\bibitem{gutzwiller1963}%
  \BibitemOpen
  \bibfield{author}{%
  \bibinfo {author} {\bibfnamefont{M.~C.}\ \bibnamefont{Gutzwiller}},\ }%
  \bibfield{journal}{%
  \Doi{10.1103/PhysRevLett.10.159}{\bibinfo {journal} {Phys. Rev. Lett.}}\ }%
  \textbf{\bibinfo {volume} {10}},\ \bibinfo {pages} {159} (\bibinfo {year}
  {1963})%
  \bibAnnoteFile{NoStop}{gutzwiller1963}%
\bibitem{hubbard1963}%
  \BibitemOpen
  \bibfield{author}{%
  \bibinfo {author} {\bibfnamefont{J.}~\bibnamefont{Hubbard}},\ }%
  \bibfield{journal}{%
  \Doi{10.1098/rspa.1963.0204}{\bibinfo {journal} {Proc. R. Soc. Lond. A}}\ }%
  \textbf{\bibinfo {volume} {276}},\ \bibinfo {pages} {238} (\bibinfo {year}
  {1963})%
  \bibAnnoteFile{NoStop}{hubbard1963}%
\bibitem{kanamori1963}%
  \BibitemOpen
  \bibfield{author}{%
  \bibinfo {author} {\bibfnamefont{J.}~\bibnamefont{Kanamori}},\ }%
  \bibfield{journal}{%
  \Doi{10.1143/PTP.30.275}{\bibinfo {journal} {Prog. Theor. Phys.}}\ }%
  \textbf{\bibinfo {volume} {30}},\ \bibinfo {pages} {275} (\bibinfo {year}
  {1963})%
  \bibAnnoteFile{NoStop}{kanamori1963}%
\bibitem{weiss1907}%
  \BibitemOpen
  \bibfield{author}{%
  \bibinfo {author} {\bibfnamefont{P.}~\bibnamefont{Weiss}},\ }%
  \bibfield{journal}{%
  \Doi{10.1051/jphystap:019070060066100}{\bibinfo {journal} {J. Phys. Theor.
  Appl.}}\ }%
  \textbf{\bibinfo {volume} {6}},\ \bibinfo {pages} {661} (\bibinfo {year}
  {1907})%
  \bibAnnoteFile{NoStop}{weiss1907}%
\bibitem{metzner1989}%
  \BibitemOpen
  \bibfield{author}{%
  \bibinfo {author} {\bibfnamefont{W.}~\bibnamefont{Metzner}}\ and\ \bibinfo
  {author} {\bibfnamefont{D.}~\bibnamefont{Vollhardt}},\ }%
  \bibfield{journal}{%
  \Doi{10.1103/PhysRevLett.62.324}{\bibinfo {journal} {Phys. Rev. Lett.}}\ }%
  \textbf{\bibinfo {volume} {62}},\ \bibinfo {pages} {324} (\bibinfo {year}
  {1989})%
  \bibAnnoteFile{NoStop}{metzner1989}%
\bibitem{georges1992}%
  \BibitemOpen
  \bibfield{author}{%
  \bibinfo {author} {\bibfnamefont{A.}~\bibnamefont{Georges}}\ and\ \bibinfo
  {author} {\bibfnamefont{G.}~\bibnamefont{Kotliar}},\ }%
  \bibfield{journal}{%
  \Doi{10.1103/PhysRevB.45.6479}{\bibinfo {journal} {Phys. Rev. B}}\ }%
  \textbf{\bibinfo {volume} {45}},\ \bibinfo {pages} {6479} (\bibinfo {year}
  {1992})%
  \bibAnnoteFile{NoStop}{georges1992}%
\bibitem{jarrell1992}%
  \BibitemOpen
  \bibfield{author}{%
  \bibinfo {author} {\bibfnamefont{M.}~\bibnamefont{Jarrell}},\ }%
  \bibfield{journal}{%
  \Doi{10.1103/PhysRevLett.69.168}{\bibinfo {journal} {Phys. Rev. Lett.}}\ }%
  \textbf{\bibinfo {volume} {69}},\ \bibinfo {pages} {168} (\bibinfo {year}
  {1992})%
  \bibAnnoteFile{NoStop}{jarrell1992}%
\bibitem{georges1996}%
  \BibitemOpen
  \bibfield{author}{%
  \bibinfo {author} {\bibfnamefont{A.}~\bibnamefont{Georges}}, \bibinfo
  {author} {\bibfnamefont{G.}~\bibnamefont{Kotliar}}, \bibinfo {author}
  {\bibfnamefont{W.}~\bibnamefont{Krauth}},\ and\ \bibinfo {author}
  {\bibfnamefont{M.~J.}\ \bibnamefont{Rozenberg}},\ }%
  \bibfield{journal}{%
  \Doi{10.1103/RevModPhys.68.13}{\bibinfo {journal} {Rev. Mod. Phys.}}\ }%
  \textbf{\bibinfo {volume} {68}},\ \bibinfo {pages} {13} (\bibinfo {year}
  {1996})%
  \bibAnnoteFile{NoStop}{georges1996}%
\bibitem{freericks2006}%
  \BibitemOpen
  \bibfield{author}{%
  \bibinfo {author} {\bibfnamefont{J.~K.}\ \bibnamefont{Freericks}}, \bibinfo
  {author} {\bibfnamefont{V.~M.}\ \bibnamefont{Turkowski}},\ and\ \bibinfo
  {author} {\bibfnamefont{V.}~\bibnamefont{Zlati\'c}},\ }%
  \bibfield{journal}{%
  \Doi{10.1103/PhysRevLett.97.266408}{\bibinfo {journal} {Phys. Rev. Lett.}}\
  }%
  \textbf{\bibinfo {volume} {97}},\ \bibinfo {eid} {266408} (\bibinfo {year}
  {2006}),\
  \Eprint{http://arxiv.org/abs/arXiv:cond-mat/0607053}{arXiv:cond-mat/0607053}%
  \bibAnnoteFile{NoStop}{freericks2006}%
\bibitem{schmidt2002}%
  \BibitemOpen
  \bibfield{author}{%
  \bibinfo {author} {\bibfnamefont{P.}~\bibnamefont{Schmidt}}\ and\ \bibinfo
  {author} {\bibfnamefont{H.}~\bibnamefont{Monien}},\ }%
  \bibfield{journal}{%
  \bibinfo {journal} {ArXiv e-prints \;}}%
   (\bibinfo {year} {2002}),\
  \Eprint{http://arxiv.org/abs/arXiv:cond-mat/0202046}{arXiv:cond-mat/0202046}%
  \bibAnnoteFile{NoStop}{schmidt2002}%
\bibitem{eckstein2009b}%
  \BibitemOpen
  \bibfield{author}{%
  \bibinfo {author} {\bibfnamefont{M.}~\bibnamefont{Eckstein}}, \bibinfo
  {author} {\bibfnamefont{M.}~\bibnamefont{Kollar}},\ and\ \bibinfo {author}
  {\bibfnamefont{P.}~\bibnamefont{Werner}},\ }%
  \bibfield{journal}{%
  \Doi{10.1103/PhysRevLett.103.056403}{\bibinfo {journal} {Phys. Rev. Lett.}}\
  }%
  \textbf{\bibinfo {volume} {103}},\ \bibinfo {eid} {056403} (\bibinfo {year}
  {2009}),\ \Eprint{http://arxiv.org/abs/0904.0976}{arXiv:0904.0976
  [cond-mat.str-el]}%
  \bibAnnoteFile{NoStop}{eckstein2009b}%
\bibitem{tsuji2009}%
  \BibitemOpen
  \bibfield{author}{%
  \bibinfo {author} {\bibfnamefont{N.}~\bibnamefont{Tsuji}}, \bibinfo {author}
  {\bibfnamefont{T.}~\bibnamefont{Oka}},\ and\ \bibinfo {author}
  {\bibfnamefont{H.}~\bibnamefont{Aoki}},\ }%
  \bibfield{journal}{%
  \Doi{10.1103/PhysRevLett.103.047403}{\bibinfo {journal} {Phys. Rev. Lett.}}\
  }%
  \textbf{\bibinfo {volume} {103}},\ \bibinfo {pages} {047403} (\bibinfo {year}
  {2009}),\ \Eprint{http://arxiv.org/abs/0903.2332}{arXiv:0903.2332
  [cond-mat.str-el]}%
  \bibAnnoteFile{NoStop}{tsuji2009}%
\bibitem{eckstein2010}%
  \BibitemOpen
  \bibfield{author}{%
  \bibinfo {author} {\bibfnamefont{M.}~\bibnamefont{Eckstein}}\ and\ \bibinfo
  {author} {\bibfnamefont{P.}~\bibnamefont{Werner}},\ }%
  \bibfield{journal}{%
  \Doi{10.1103/PhysRevB.82.115115}{\bibinfo {journal} {Phys. Rev. B}}\ }%
  \textbf{\bibinfo {volume} {82}},\ \bibinfo {eid} {115115} (\bibinfo {year}
  {2010}),\ \Eprint{http://arxiv.org/abs/1005.1872}{arXiv:1005.1872
  [cond-mat.str-el]}%
  \bibAnnoteFile{NoStop}{eckstein2010}%
\bibitem{werner2012}%
  \BibitemOpen
  \bibfield{author}{%
  \bibinfo {author} {\bibfnamefont{P.}~\bibnamefont{Werner}}\ and\ \bibinfo
  {author} {\bibfnamefont{M.}~\bibnamefont{Eckstein}},\ }%
  \bibfield{journal}{%
  \Doi{10.1103/PhysRevB.86.045119}{\bibinfo {journal} {Phys. Rev. B}}\ }%
  \textbf{\bibinfo {volume} {86}},\ \bibinfo {eid} {045119} (\bibinfo {year}
  {2012}),\ \Eprint{http://arxiv.org/abs/1204.5418}{arXiv:1204.5418
  [cond-mat.str-el]}%
  \bibAnnoteFile{NoStop}{werner2012}%
\bibitem{amaricci2012}%
  \BibitemOpen
  \bibfield{author}{%
  \bibinfo {author} {\bibfnamefont{A.}~\bibnamefont{Amaricci}}, \bibinfo
  {author} {\bibfnamefont{C.}~\bibnamefont{Weber}}, \bibinfo {author}
  {\bibfnamefont{M.}~\bibnamefont{Capone}},\ and\ \bibinfo {author}
  {\bibfnamefont{G.}~\bibnamefont{Kotliar}},\ }%
  \bibfield{journal}{%
  \Doi{10.1103/PhysRevB.86.085110}{\bibinfo {journal} {Phys. Rev. B}}\ }%
  \textbf{\bibinfo {volume} {86}},\ \bibinfo {pages} {085110} (\bibinfo {year}
  {2012}),\ \Eprint{http://arxiv.org/abs/1106.3483}{arXiv:1106.3483
  [cond-mat.str-el]}%
  \bibAnnoteFile{NoStop}{amaricci2012}%
\bibitem{gull2011}%
  \BibitemOpen
  \bibfield{author}{%
  \bibinfo {author} {\bibfnamefont{E.}~\bibnamefont{Gull}}, \bibinfo {author}
  {\bibfnamefont{A.~J.}\ \bibnamefont{Millis}}, \bibinfo {author}
  {\bibfnamefont{A.~I.}\ \bibnamefont{Lichtenstein}}, \bibinfo {author}
  {\bibfnamefont{A.~N.}\ \bibnamefont{Rubtsov}}, \bibinfo {author}
  {\bibfnamefont{M.}~\bibnamefont{Troyer}},\ and\ \bibinfo {author}
  {\bibfnamefont{P.}~\bibnamefont{Werner}},\ }%
  \bibfield{journal}{%
  \Doi{10.1103/RevModPhys.83.349}{\bibinfo {journal} {Rev. Mod. Phys.}}\ }%
  \textbf{\bibinfo {volume} {83}},\ \bibinfo {pages} {349} (\bibinfo {year}
  {2011}),\ \Eprint{http://arxiv.org/abs/1012.4474}{arXiv:1012.4474
  [cond-mat.str-el]}%
  \bibAnnoteFile{NoStop}{gull2011}%
\bibitem{caffarel1994}%
  \BibitemOpen
  \bibfield{author}{%
  \bibinfo {author} {\bibfnamefont{M.}~\bibnamefont{Caffarel}}\ and\ \bibinfo
  {author} {\bibfnamefont{W.}~\bibnamefont{Krauth}},\ }%
  \bibfield{journal}{%
  \Doi{10.1103/PhysRevLett.72.1545}{\bibinfo {journal} {Phys. Rev. Lett.}}\ }%
  \textbf{\bibinfo {volume} {72}},\ \bibinfo {pages} {1545} (\bibinfo {year}
  {1994})%
  \bibAnnoteFile{NoStop}{caffarel1994}%
\bibitem{liebsch2012}%
  \BibitemOpen
  \bibfield{author}{%
  \bibinfo {author} {\bibfnamefont{A.}~\bibnamefont{Liebsch}}\ and\ \bibinfo
  {author} {\bibfnamefont{H.}~\bibnamefont{Ishida}},\ }%
  \bibfield{journal}{%
  {\bibinfo {journal} {J. Phys.: Condens. Matter}}\ }%
  \textbf{\bibinfo {volume} {24}},\ \bibinfo {pages} {053201} (\bibinfo {year}
  {2012})%
  \bibAnnoteFile{NoStop}{caffarel1994}%
\bibitem{potthoff2003}%
  \BibitemOpen
  \bibfield{author}{%
  \bibinfo {author} {\bibfnamefont{M.}~\bibnamefont{Potthoff}},\ }%
  \bibfield{journal}{%
  \Doi{10.1140/epjb/e2003-00121-8}{\bibinfo {journal} {Eur. Phys. J. B}}\ }%
  \textbf{\bibinfo {volume} {32}},\ \bibinfo {pages} {429} (\bibinfo {year}
  {2003}),\
  \Eprint{http://arxiv.org/abs/arXiv:cond-mat/0301137}{arXiv:cond-mat/0301137}%
  \bibAnnoteFile{NoStop}{potthoff2003}%
\bibitem{potthoff2003c}%
  \BibitemOpen
  \bibfield{author}{%
  \bibinfo {author} {\bibfnamefont{M.}~\bibnamefont{Potthoff}}, \bibinfo
  {author} {\bibfnamefont{M.}~\bibnamefont{Aichhorn}},\ and\ \bibinfo {author}
  {\bibfnamefont{C.}~\bibnamefont{Dahnken}},\ }%
  \bibfield{journal}{%
  \Doi{10.1103/PhysRevLett.91.206402}{\bibinfo {journal} {Phys. Rev. Lett.}}\
  }%
  \textbf{\bibinfo {volume} {91}},\ \bibinfo {pages} {206402} (\bibinfo {year}
  {2003}),\
  \Eprint{http://arxiv.org/abs/arXiv:cond-mat/0303136}{arXiv:cond-mat/0303136}%
  \bibAnnoteFile{NoStop}{potthoff2003c}%
\bibitem{potthoff2007}%
  \BibitemOpen
  \bibfield{author}{%
  \bibinfo {author} {\bibfnamefont{M.}~\bibnamefont{Potthoff}}\ and\ \bibinfo
  {author} {\bibfnamefont{M.}~\bibnamefont{Balzer}},\ }%
  \bibfield{journal}{%
  \Doi{10.1103/PhysRevB.75.125112}{\bibinfo {journal} {Phys. Rev. B}}\ }%
  \textbf{\bibinfo {volume} {75}},\ \bibinfo {pages} {125112} (\bibinfo {year}
  {2007}),\
  \Eprint{http://arxiv.org/abs/arXiv:cond-mat/0610217}{arXiv:cond-mat/0610217}%
  \bibAnnoteFile{NoStop}{potthoff2007}%
\bibitem{potthoff2012}%
  \BibitemOpen
  \bibfield{author}{%
  \bibinfo {author} {\bibfnamefont{M.}~\bibnamefont{Potthoff}},\ }%
  in\ \Doi{10.1007/978-3-642-21831-6_10}{\emph{\bibinfo {booktitle} {Strongly
  Correlated Systems: Theoretical Methods}}},\ \bibinfo {series} {Springer
  Series in Solid-State Sciences}, Vol.\ \bibinfo {volume} {171},\ \bibinfo
  {editor} {edited by\ \bibinfo {editor}
  {\bibfnamefont{A.}~\bibnamefont{Avella}}\ and\ \bibinfo {editor}
  {\bibfnamefont{F.}~\bibnamefont{Mancini}}}\ (\bibinfo {publisher}
  {Springer},\ \bibinfo {address} {Berlin},\ \bibinfo {year} {2012})\ pp.\
  \bibinfo {pages} {303--339},\
  \Eprint{http://arxiv.org/abs/1108.2183}{arXiv:1108.2183 [cond-mat.str-el]}%
  \bibAnnoteFile{NoStop}{potthoff2012}%
\bibitem{pozgajcic2004}%
  \BibitemOpen
  \bibfield{author}{%
  \bibinfo {author} {\bibfnamefont{K.}~\bibnamefont{Pozgajcic}},\ }%
  \bibfield{journal}{%
  \bibinfo {journal} {ArXiv e-prints \;}}%
   (\bibinfo {year} {2004}),\
  \Eprint{http://arxiv.org/abs/arXiv:cond-mat/0407172}{arXiv:cond-mat/0407172}%
  \bibAnnoteFile{NoStop}{pozgajcic2004}%
\bibitem{eckstein2007}%
  \BibitemOpen
  \bibfield{author}{%
  \bibinfo {author} {\bibfnamefont{M.}~\bibnamefont{Eckstein}}, \bibinfo
  {author} {\bibfnamefont{M.}~\bibnamefont{Kollar}}, \bibinfo {author}
  {\bibfnamefont{M.}~\bibnamefont{Potthoff}},\ and\ \bibinfo {author}
  {\bibfnamefont{D.}~\bibnamefont{Vollhardt}},\ }%
  \bibfield{journal}{%
  \Doi{10.1103/PhysRevB.75.125103}{\bibinfo {journal} {Phys. Rev. B}}\ }%
  \textbf{\bibinfo {volume} {75}},\ \bibinfo {eid} {125103} (\bibinfo {year}
  {2007}),\
  \Eprint{http://arxiv.org/abs/arXiv:cond-mat/0610803}{arXiv:cond-mat/0610803}%
  \bibAnnoteFile{NoStop}{eckstein2007}%
\bibitem{werner2009}%
  \BibitemOpen
  \bibfield{author}{%
  \bibinfo {author} {\bibfnamefont{P.}~\bibnamefont{Werner}}, \bibinfo {author}
  {\bibfnamefont{T.}~\bibnamefont{Oka}},\ and\ \bibinfo {author}
  {\bibfnamefont{A.~J.}\ \bibnamefont{Millis}},\ }%
  \bibfield{journal}{%
  \Doi{10.1103/PhysRevB.79.035320}{\bibinfo {journal} {Phys. Rev. B}}\ }%
  \textbf{\bibinfo {volume} {79}},\ \bibinfo {eid} {035320} (\bibinfo {year}
  {2009}),\ \Eprint{http://arxiv.org/abs/0810.2345}{arXiv:0810.2345
  [cond-mat.mes-hall]}%
  \bibAnnoteFile{NoStop}{werner2009}%
\bibitem{jung2012}%
  \BibitemOpen
  \bibfield{author}{%
  \bibinfo {author} {\bibfnamefont{C.}~\bibnamefont{Jung}}, \bibinfo {author}
  {\bibfnamefont{A.}~\bibnamefont{Lieder}}, \bibinfo {author}
  {\bibfnamefont{S.}~\bibnamefont{Brener}}, \bibinfo {author}
  {\bibfnamefont{H.}~\bibnamefont{Hafermann}}, \bibinfo {author}
  {\bibfnamefont{B.}~\bibnamefont{Baxevanis}}, \bibinfo {author}
  {\bibfnamefont{A.}~\bibnamefont{Chudnovskiy}}, \bibinfo {author}
  {\bibfnamefont{A.~N.}\ \bibnamefont{Rubtsov}}, \bibinfo {author}
  {\bibfnamefont{M.~I.}\ \bibnamefont{Katsnelson}},\ and\ \bibinfo {author}
  {\bibfnamefont{A.~I.}\ \bibnamefont{Lichtenstein}},\ }%
  \bibfield{journal}{%
  \Doi{10.1002/andp.201100045}{\bibinfo {journal} {Ann. Phys.}}\ }%
  \textbf{\bibinfo {volume} {524}},\ \bibinfo {pages} {49} (\bibinfo {year}
  {2012}),\ \Eprint{http://arxiv.org/abs/1011.3264}{arXiv:1011.3264
  [cond-mat.str-el]}%
  \bibAnnoteFile{NoStop}{jung2012}%
\bibitem{arrigoni2012}%
  \BibitemOpen
  \bibfield{author}{%
  \bibinfo {author} {\bibfnamefont{E.}~\bibnamefont{Arrigoni}}, \bibinfo
  {author} {\bibfnamefont{M.}~\bibnamefont{Knap}},\ and\ \bibinfo {author}
  {\bibfnamefont{W.}~\bibnamefont{von~der Linden}},\ }%
  \bibfield{journal}{%
  \Doi{10.1103/PhysRevLett.110.086403}{\bibinfo {journal} {Phys. Rev. Lett.}}\
  }%
  \textbf{\bibinfo {volume} {110}},\ \bibinfo {eid} {086403} (\bibinfo {year}
  {2013}),\ \Eprint{http://arxiv.org/abs/1210.4167}{arXiv:1210.4167
  [cond-mat.str-el]}%
  \bibAnnoteFile{NoStop}{arrigoni2012}%
\bibitem{gramsch2013}%
  \BibitemOpen
  \bibfield{author}{%
  \bibinfo {author} {\bibfnamefont{C.}~\bibnamefont{Gramsch}}, \bibinfo
  {author} {\bibfnamefont{K.}~\bibnamefont{Balzer}}, \bibinfo {author}
  {\bibfnamefont{M.}~\bibnamefont{Eckstein}},\ and\ \bibinfo {author}
  {\bibfnamefont{M.}~\bibnamefont{Kollar}},\ }%
  \bibfield{journal}{%
  \bibinfo {journal} {ArXiv e-prints }}%
   (\bibinfo {year} {2013}),\
  \Eprint{http://arxiv.org/abs/1306.6315}{arXiv:1306.6315 [cond-mat.str-el]}%
  \bibAnnoteFile{NoStop}{gramsch2013}%
\bibitem{potthoff2011c}%
  \BibitemOpen
  \bibfield{author}{%
  \bibinfo {author} {\bibfnamefont{M.}~\bibnamefont{Balzer}}\ and\ \bibinfo
  {author} {\bibfnamefont{M.}~\bibnamefont{Potthoff}},\ }%
  \bibfield{journal}{%
  \Doi{10.1103/PhysRevB.83.195132}{\bibinfo {journal} {Phys. Rev. B}}\ }%
  \textbf{\bibinfo {volume} {83}},\ \bibinfo {pages} {195132} (\bibinfo {year}
  {2011}),\ \Eprint{http://arxiv.org/abs/1102.3344}{arXiv:1102.3344
  [cond-mat.str-el]}%
  \bibAnnoteFile{NoStop}{potthoff2011c}%
\bibitem{balzer2012}%
  \BibitemOpen
  \bibfield{author}{%
  \bibinfo {author} {\bibfnamefont{M.}~\bibnamefont{Balzer}}, \bibinfo {author}
  {\bibfnamefont{N.}~\bibnamefont{Gdaniec}},\ and\ \bibinfo {author}
  {\bibfnamefont{M.}~\bibnamefont{Potthoff}},\ }%
  \bibfield{journal}{%
  \Doi{10.1088/0953-8984/24/3/035603}{\bibinfo {journal} {J. Phys.: Condens.
  Matter}}\ }%
  \textbf{\bibinfo {volume} {24}},\ \bibinfo {eid} {035603} (\bibinfo {year}
  {2012}),\ \Eprint{http://arxiv.org/abs/1109.1205}{arXiv:1109.1205
  [cond-mat.str-el]}%
  \bibAnnoteFile{NoStop}{balzer2012}%
\bibitem{jurgenowski2013}%
  \BibitemOpen
  \bibfield{author}{%
  \bibinfo {author} {\bibfnamefont{P.}~\bibnamefont{Jurgenowski}}\ and\
  \bibinfo {author} {\bibfnamefont{M.}~\bibnamefont{Potthoff}},\ }%
  \bibfield{journal}{%
  \Doi{10.1103/PhysRevB.87.205118}{\bibinfo {journal} {Phys. Rev. B}}\ }%
  \textbf{\bibinfo {volume} {87}},\ \bibinfo {eid} {205118} (\bibinfo {year}
  {2013}),\ \Eprint{http://arxiv.org/abs/1302.5317}{arXiv:1302.5317
  [cond-mat.str-el]}%
  \bibAnnoteFile{NoStop}{jurgenowski2013}%
\bibitem{knap2011}%
  \BibitemOpen
  \bibfield{author}{%
  \bibinfo {author} {\bibfnamefont{M.}~\bibnamefont{Knap}}, \bibinfo {author}
  {\bibfnamefont{W.}~\bibnamefont{von~der Linden}},\ and\ \bibinfo {author}
  {\bibfnamefont{E.}~\bibnamefont{Arrigoni}},\ }%
  \bibfield{journal}{%
  \Doi{10.1103/PhysRevB.84.115145}{\bibinfo {journal} {Phys. Rev. B}}\ }%
  \textbf{\bibinfo {volume} {84}},\ \bibinfo {eid} {115145} (\bibinfo {year}
  {2011}),\ \Eprint{http://arxiv.org/abs/1104.3838}{arXiv:1104.3838
  [cond-mat.str-el]}%
  \bibAnnoteFile{NoStop}{knap2011}%
\bibitem{nuss2012}%
  \BibitemOpen
  \bibfield{author}{%
  \bibinfo {author} {\bibfnamefont{M.}~\bibnamefont{Nuss}}, \bibinfo {author}
  {\bibfnamefont{C.}~\bibnamefont{Heil}}, \bibinfo {author}
  {\bibfnamefont{M.}~\bibnamefont{Ganahl}}, \bibinfo {author}
  {\bibfnamefont{M.}~\bibnamefont{Knap}}, \bibinfo {author}
  {\bibfnamefont{H.~G.}\ \bibnamefont{Evertz}}, \bibinfo {author}
  {\bibfnamefont{E.}~\bibnamefont{Arrigoni}},\ and\ \bibinfo {author}
  {\bibfnamefont{W.}~\bibnamefont{von~der Linden}},\ }%
  \bibfield{journal}{%
  \Doi{10.1103/PhysRevB.86.245119}{\bibinfo {journal} {Phys. Rev. B}}\ }%
  \textbf{\bibinfo {volume} {86}},\ \bibinfo {eid} {245119} (\bibinfo {year}
  {2012}),\ \Eprint{http://arxiv.org/abs/1207.5641}{arXiv:1207.5641
  [cond-mat.str-el]}%
  \bibAnnoteFile{NoStop}{nuss2012}%
\bibitem{kotliar2001}%
  \BibitemOpen
  \bibfield{author}{%
  \bibinfo {author} {\bibfnamefont{G.}~\bibnamefont{Kotliar}}, \bibinfo
  {author} {\bibfnamefont{S.~Y.}\ \bibnamefont{Savrasov}}, \bibinfo {author}
  {\bibfnamefont{G.}~\bibnamefont{P\'{a}lsson}},\ and\ \bibinfo {author}
  {\bibfnamefont{G.}~\bibnamefont{Biroli}},\ }%
  \bibfield{journal}{%
  \Doi{10.1103/PhysRevLett.87.186401}{\bibinfo {journal} {Phys. Rev. Lett.}}\
  }%
  \textbf{\bibinfo {volume} {87}},\ \bibinfo {pages} {186401} (\bibinfo {year}
  {2001})%
  \bibAnnoteFile{NoStop}{kotliar2001}%
\bibitem{hettler1998}%
  \BibitemOpen
  \bibfield{author}{%
  \bibinfo {author} {\bibfnamefont{M.~H.}\ \bibnamefont{Hettler}}, \bibinfo
  {author} {\bibfnamefont{A.~N.}\ \bibnamefont{Tahvildar-Zadeh}}, \bibinfo
  {author} {\bibfnamefont{M.}~\bibnamefont{Jarrell}}, \bibinfo {author}
  {\bibfnamefont{T.}~\bibnamefont{Pruschke}},\ and\ \bibinfo {author}
  {\bibfnamefont{H.~R.}\ \bibnamefont{Krishnamurthy}},\ }%
  \bibfield{journal}{%
  \Doi{10.1103/PhysRevB.58.R7475}{\bibinfo {journal} {Phys. Rev. B}}\ }%
  \textbf{\bibinfo {volume} {58}},\ \bibinfo {pages} {R7475} (\bibinfo {year}
  {1998}),\
  \Eprint{http://arxiv.org/abs/arXiv:cond-mat/9803295}{arXiv:cond-mat/9803295}%
  \bibAnnoteFile{NoStop}{hettler1998}%
\bibitem{baym1961}%
  \BibitemOpen
  \bibfield{author}{%
  \bibinfo {author} {\bibfnamefont{G.}~\bibnamefont{Baym}}\ and\ \bibinfo
  {author} {\bibfnamefont{L.~P.}\ \bibnamefont{Kadanoff}},\ }%
  \bibfield{journal}{%
  \Doi{10.1103/PhysRev.124.287}{\bibinfo {journal} {Phys. Rev.}}\ }%
  \textbf{\bibinfo {volume} {124}},\ \bibinfo {pages} {287} (\bibinfo {year}
  {1961})%
  \bibAnnoteFile{NoStop}{baym1961}%
\bibitem{baym1962}%
  \BibitemOpen
  \bibfield{author}{%
  \bibinfo {author} {\bibfnamefont{G.}~\bibnamefont{Baym}},\ }%
  \bibfield{journal}{%
  \Doi{10.1103/PhysRev.127.1391}{\bibinfo {journal} {Phys. Rev.}}\ }%
  \textbf{\bibinfo {volume} {127}},\ \bibinfo {pages} {1391} (\bibinfo {year}
  {1962})%
  \bibAnnoteFile{NoStop}{baym1962}%
\bibitem{kubo1957}%
  \BibitemOpen
  \bibfield{author}{%
  \bibinfo {author} {\bibfnamefont{R.}~\bibnamefont{Kubo}},\ }%
  \bibfield{journal}{%
  \Doi{10.1143/JPSJ.12.570}{\bibinfo {journal} {J. Phys. Soc. Jap.}}\ }%
  \textbf{\bibinfo {volume} {12}},\ \bibinfo {pages} {570} (\bibinfo {year}
  {1957})%
  \bibAnnoteFile{NoStop}{kubo1957}%
\bibitem{matsubara1955}%
  \BibitemOpen
  \bibfield{author}{%
  \bibinfo {author} {\bibfnamefont{T.}~\bibnamefont{Matsubara}},\ }%
  \bibfield{journal}{%
  \Doi{10.1143/PTP.14.351}{\bibinfo {journal} {Prog. Theor. Phys.}}\ }%
  \textbf{\bibinfo {volume} {14}},\ \bibinfo {pages} {351} (\bibinfo {year}
  {1955})%
  \bibAnnoteFile{NoStop}{matsubara1955}%
\bibitem{schwinger1961}%
  \BibitemOpen
  \bibfield{author}{%
  \bibinfo {author} {\bibfnamefont{J.}~\bibnamefont{Schwinger}},\ }%
  \bibfield{journal}{%
  \Doi{10.1063/1.1703727}{\bibinfo {journal} {Journal of Mathematical
  Physics}}\ }%
  \textbf{\bibinfo {volume} {2}},\ \bibinfo {pages} {407} (\bibinfo {year}
  {1961})%
  \bibAnnoteFile{NoStop}{schwinger1961}%
\bibitem{keldysh1965}%
  \BibitemOpen
  \bibfield{author}{%
  \bibinfo {author} {\bibfnamefont{L.}~\bibnamefont{Keldysh}},\ }%
  \bibfield{journal}{%
  \bibinfo {journal} {Sov. Phys. JETP}\ }%
  \textbf{\bibinfo {volume} {20}},\ \bibinfo {pages} {1018} (\bibinfo {year}
  {1965})%
  \bibAnnoteFile{NoStop}{keldysh1965}%
\bibitem{danielewicz1984}%
  \BibitemOpen
  \bibfield{author}{%
  \bibinfo {author} {\bibfnamefont{P.}~\bibnamefont{Danielewicz}},\ }%
  \bibfield{journal}{%
  \Doi{10.1016/0003-4916(84)90092-7}{\bibinfo {journal} {Ann. Phys.}}\ }%
  \textbf{\bibinfo {volume} {152}},\ \bibinfo {pages} {239} (\bibinfo {year}
  {1984})%
  \bibAnnoteFile{NoStop}{danielewicz1984}%
\bibitem{wagner1991}%
  \BibitemOpen
  \bibfield{author}{%
  \bibinfo {author} {\bibfnamefont{M.}~\bibnamefont{Wagner}},\ }%
  \bibfield{journal}{%
  \Doi{10.1103/PhysRevB.44.6104}{\bibinfo {journal} {Phys. Rev. B}}\ }%
  \textbf{\bibinfo {volume} {44}},\ \bibinfo {pages} {6104} (\bibinfo {year}
  {1991})%
  \bibAnnoteFile{NoStop}{wagner1991}%
\bibitem{leeuwen2006c}%
  \BibitemOpen
  \bibfield{author}{%
  \bibinfo {author} {\bibfnamefont{R.}~\bibnamefont{van Leeuwen}}, \bibinfo
  {author} {\bibfnamefont{N.~E.}\ \bibnamefont{Dahlen}}, \bibinfo {author}
  {\bibfnamefont{G.}~\bibnamefont{Stefanucci}}, \bibinfo {author}
  {\bibfnamefont{C.~O.}\ \bibnamefont{Almbladh}},\ and\ \bibinfo {author}
  {\bibfnamefont{U.}~\bibnamefont{von Barth}},\ }%
  in\ \Doi{10.1007/3-540-35426-3_3}{\emph{\bibinfo {booktitle} {Time-Dependent
  Density Functional Theory}}},\ Vol.\ \bibinfo {volume} {706},\ \bibinfo
  {editor} {edited by\ \bibinfo {editor} {\bibfnamefont{M.~A.~L.}\
  \bibnamefont{Marques}}, \bibinfo {editor} {\bibfnamefont{C.~A.}\
  \bibnamefont{Ullrich}}, \bibinfo {editor}
  {\bibfnamefont{F.}~\bibnamefont{Nogueira}}, \bibinfo {editor}
  {\bibfnamefont{A.}~\bibnamefont{Rubio}}, \bibinfo {editor}
  {\bibfnamefont{K.}~\bibnamefont{Burke}},\ and\ \bibinfo {editor}
  {\bibfnamefont{E.~K.~U.}\ \bibnamefont{Gross}}}\ (\bibinfo {publisher}
  {Springer},\ \bibinfo {address} {Berlin Heidelberg},\ \bibinfo {year}
  {2006})\ Chap.\ \bibinfo {chapter} {Lecture Notes in Physics}, pp.\ \bibinfo
  {pages} {33--59}%
  \bibAnnoteFile{NoStop}{leeuwen2006c}%
\bibitem{rammer2007}%
  \BibitemOpen
  \bibfield{author}{%
  \bibinfo {author} {\bibfnamefont{J.}~\bibnamefont{Rammer}},\ }%
  \emph{\bibinfo {title} {Quantum field theory of nonequilibrium states}}\
  (\bibinfo {publisher} {Cambridge University Press},\ \bibinfo {year} {2007})%
  \bibAnnoteFile{NoStop}{rammer2007}%
\bibitem{kamenev2011}%
  \BibitemOpen
  \bibfield{author}{%
  \bibinfo {author} {\bibfnamefont{A.}~\bibnamefont{Kamenev}},\ }%
  \emph{\bibinfo {title} {Field theory of non-equilibrium systems}}\ (\bibinfo
  {publisher} {Cambridge University Press},\ \bibinfo {address} {Cambridge, New
  York},\ \bibinfo {year} {2011})%
  \bibAnnoteFile{NoStop}{kamenev2011}%
\bibitem{luttinger1960}%
  \BibitemOpen
  \bibfield{author}{%
  \bibinfo {author} {\bibfnamefont{J.~M.}\ \bibnamefont{Luttinger}}\ and\
  \bibinfo {author} {\bibfnamefont{J.~C.}\ \bibnamefont{Ward}},\ }%
  \bibfield{journal}{%
  \Doi{10.1103/PhysRev.118.1417}{\bibinfo {journal} {Phys. Rev.}}\ }%
  \textbf{\bibinfo {volume} {118}},\ \bibinfo {pages} {1417} (\bibinfo {year}
  {1960})%
  \bibAnnoteFile{NoStop}{luttinger1960}%
\bibitem{potthoff2006b}%
  \BibitemOpen
  \bibfield{author}{%
  \bibinfo {author} {\bibfnamefont{M.}~\bibnamefont{Potthoff}},\ }%
  \bibfield{journal}{%
  \bibinfo {journal} {Condens. Mat. Phys.}\ }%
  \textbf{\bibinfo {volume} {9}},\ \bibinfo {pages} {557} (\bibinfo {year}
  {2006}),\
  \Eprint{http://arxiv.org/abs/arXiv:cond-mat/0406671}{arXiv:cond-mat/0406671}%
  \bibAnnoteFile{NoStop}{potthoff2006b}%
\bibitem{balzer2010}%
  \BibitemOpen
  \bibfield{author}{%
  \bibinfo {author} {\bibfnamefont{M.}~\bibnamefont{Balzer}}\ and\ \bibinfo
  {author} {\bibfnamefont{M.}~\bibnamefont{Potthoff}},\ }%
  \bibfield{journal}{%
  \Doi{10.1103/PhysRevB.82.174441}{\bibinfo {journal} {Phys. Rev. B}}\ }%
  \textbf{\bibinfo {volume} {82}},\ \bibinfo {eid} {174441} (\bibinfo {year}
  {2010}),\ \Eprint{http://arxiv.org/abs/1007.2517}{arXiv:1007.2517
  [cond-mat.str-el]}%
  \bibAnnoteFile{NoStop}{balzer2010}%
\bibitem{aichhorn2006b}%
  \BibitemOpen
  \bibfield{author}{%
  \bibinfo {author} {\bibfnamefont{M.}~\bibnamefont{Aichhorn}}, \bibinfo
  {author} {\bibfnamefont{E.}~\bibnamefont{Arrigoni}}, \bibinfo {author}
  {\bibfnamefont{M.}~\bibnamefont{Potthoff}},\ and\ \bibinfo {author}
  {\bibfnamefont{W.}~\bibnamefont{Hanke}},\ }%
  \bibfield{journal}{%
  \Doi{10.1103/PhysRevB.74.024508}{\bibinfo {journal} {Phys. Rev. B}}\ }%
  \textbf{\bibinfo {volume} {74}},\ \bibinfo {eid} {024508} (\bibinfo {year}
  {2006}),\
  \Eprint{http://arxiv.org/abs/arXiv:cond-mat/0511460}{arXiv:cond-mat/0511460}%
  \bibAnnoteFile{NoStop}{aichhorn2006b}%
\bibitem{aichhorn2004}%
  \BibitemOpen
  \bibfield{author}{%
  \bibinfo {author} {\bibfnamefont{M.}~\bibnamefont{Aichhorn}}, \bibinfo
  {author} {\bibfnamefont{H.~G.}\ \bibnamefont{Evertz}}, \bibinfo {author}
  {\bibfnamefont{W.}~\bibnamefont{von~der Linden}},\ and\ \bibinfo {author}
  {\bibfnamefont{M.}~\bibnamefont{Potthoff}},\ }%
  \bibfield{journal}{%
  \Doi{10.1103/PhysRevB.70.235107}{\bibinfo {journal} {Phys. Rev. B}}\ }%
  \textbf{\bibinfo {volume} {70}},\ \bibinfo {eid} {235107} (\bibinfo {year}
  {2004}),\
  \Eprint{http://arxiv.org/abs/arXiv:cond-mat/0402580}{arXiv:cond-mat/0402580}%
  \bibAnnoteFile{NoStop}{aichhorn2004}%
\bibitem{potthoff2001b}%
  \BibitemOpen
  \bibfield{author}{%
  \bibinfo {author} {\bibfnamefont{M.}~\bibnamefont{Potthoff}},\ }%
  \bibfield{journal}{%
  \Doi{10.1103/PhysRevB.64.165114}{\bibinfo {journal} {Phys. Rev. B}}\ }%
  \textbf{\bibinfo {volume} {64}},\ \bibinfo {pages} {165114} (\bibinfo {year}
  {2001}),\
  \Eprint{http://arxiv.org/abs/arXiv:cond-mat/0107502}{arXiv:cond-mat/0107502}%
  \bibAnnoteFile{NoStop}{potthoff2001b}%
\bibitem{bulla2000}%
  \BibitemOpen
  \bibfield{author}{%
  \bibinfo {author} {\bibfnamefont{R.}~\bibnamefont{Bulla}}\ and\ \bibinfo
  {author} {\bibfnamefont{M.}~\bibnamefont{Potthoff}},\ }%
  \bibfield{journal}{%
  \Doi{10.1007/s100510050030}{\bibinfo {journal} {European Physical Journal
  B}}\ }%
  \textbf{\bibinfo {volume} {13}},\ \bibinfo {pages} {257} (\bibinfo {year}
  {2000}),\
  \Eprint{http://arxiv.org/abs/arXiv:cond-mat/9905075}{arXiv:cond-mat/9905075}%
  \bibAnnoteFile{NoStop}{bulla2000}%
\bibitem{lichtenstein2000}%
  \BibitemOpen
  \bibfield{author}{%
  \bibinfo {author} {\bibfnamefont{A.~I.}\ \bibnamefont{Lichtenstein}}\ and\
  \bibinfo {author} {\bibfnamefont{M.~I.}\ \bibnamefont{Katsnelson}},\ }%
  \bibfield{journal}{%
  \Doi{10.1103/PhysRevB.62.R9283}{\bibinfo {journal} {Phys. Rev. B}}\ }%
  \textbf{\bibinfo {volume} {62}},\ \bibinfo {pages} {R9283} (\bibinfo {year}
  {2000}),\
  \Eprint{http://arxiv.org/abs/arXiv:cond-mat/9911320}{arXiv:cond-mat/9911320}%
  \bibAnnoteFile{NoStop}{lichtenstein2000}%
\bibitem{gros1993}%
  \BibitemOpen
  \bibfield{author}{%
  \bibinfo {author} {\bibfnamefont{C.}~\bibnamefont{Gros}}\ and\ \bibinfo
  {author} {\bibfnamefont{R.}~\bibnamefont{Valent\'i}},\ }%
  \bibfield{journal}{%
  \Doi{10.1103/PhysRevB.48.418}{\bibinfo {journal} {Phys. Rev. B}}\ }%
  \textbf{\bibinfo {volume} {48}},\ \bibinfo {pages} {418} (\bibinfo {year}
  {1993})%
  \bibAnnoteFile{NoStop}{gros1993}%
\bibitem{senechal2000}%
  \BibitemOpen
  \bibfield{author}{%
  \bibinfo {author} {\bibfnamefont{D.}~\bibnamefont{S\'en\'echal}}, \bibinfo
  {author} {\bibfnamefont{D.}~\bibnamefont{Perez}},\ and\ \bibinfo {author}
  {\bibfnamefont{M.}~\bibnamefont{Pioro-Ladri\`ere}},\ }%
  \bibfield{journal}{%
  \Doi{10.1103/PhysRevLett.84.522}{\bibinfo {journal} {Phys. Rev. Lett.}}\ }%
  \textbf{\bibinfo {volume} {84}},\ \bibinfo {pages} {522} (\bibinfo {year}
  {2000}),\
  \Eprint{http://arxiv.org/abs/arXiv:cond-mat/9908045}{arXiv:cond-mat/9908045}%
  \bibAnnoteFile{NoStop}{senechal2000}%
\bibitem{dahnken2004}%
  \BibitemOpen
  \bibfield{author}{%
  \bibinfo {author} {\bibfnamefont{C.}~\bibnamefont{Dahnken}}, \bibinfo
  {author} {\bibfnamefont{M.}~\bibnamefont{Aichhorn}}, \bibinfo {author}
  {\bibfnamefont{W.}~\bibnamefont{Hanke}}, \bibinfo {author}
  {\bibfnamefont{E.}~\bibnamefont{Arrigoni}},\ and\ \bibinfo {author}
  {\bibfnamefont{M.}~\bibnamefont{Potthoff}},\ }%
  \bibfield{journal}{%
  \Doi{10.1103/PhysRevB.70.245110}{\bibinfo {journal} {Phys. Rev. B}}\ }%
  \textbf{\bibinfo {volume} {70}},\ \bibinfo {pages} {245110} (\bibinfo {year}
  {2004}),\
  \Eprint{http://arxiv.org/abs/arXiv:cond-mat/0309407}{arXiv:cond-mat/0309407}%
  \bibAnnoteFile{NoStop}{dahnken2004}%
\bibitem{tong2005}%
  \BibitemOpen
  \bibfield{author}{%
  \bibinfo {author} {\bibfnamefont{N.-H.}\ \bibnamefont{Tong}},\ }%
  \bibfield{journal}{%
  \Doi{10.1103/PhysRevB.72.115104}{\bibinfo {journal} {Phys. Rev. B}}\ }%
  \textbf{\bibinfo {volume} {72}},\ \bibinfo {eid} {115104} (\bibinfo {year}
  {2005}),\
  \Eprint{http://arxiv.org/abs/arXiv:cond-mat/0504778}{arXiv:cond-mat/0504778}%
  \bibAnnoteFile{NoStop}{tong2005}%
\bibitem{koller2006}%
  \BibitemOpen
  \bibfield{author}{%
  \bibinfo {author} {\bibfnamefont{W.}~\bibnamefont{Koller}}\ and\ \bibinfo
  {author} {\bibfnamefont{N.}~\bibnamefont{Dupuis}},\ }%
  \bibfield{journal}{%
  \Doi{10.1088/0953-8984/18/41/019}{\bibinfo {journal} {J. Phys.: Condens.
  Matter}}\ }%
  \textbf{\bibinfo {volume} {18}},\ \bibinfo {pages} {9525} (\bibinfo {year}
  {2006}),\
  \Eprint{http://arxiv.org/abs/arXiv:cond-mat/0511294}{arXiv:cond-mat/0511294}%
  \bibAnnoteFile{NoStop}{koller2006}%
\bibitem{arrigoni2011}%
  \BibitemOpen
  \bibfield{author}{%
  \bibinfo {author} {\bibfnamefont{E.}~\bibnamefont{Arrigoni}}, \bibinfo
  {author} {\bibfnamefont{M.}~\bibnamefont{Knap}},\ and\ \bibinfo {author}
  {\bibfnamefont{W.}~\bibnamefont{von~der Linden}},\ }%
  \bibfield{journal}{%
  \Doi{10.1103/PhysRevB.84.014535}{\bibinfo {journal} {Phys. Rev. B}}\ }%
  \textbf{\bibinfo {volume} {84}},\ \bibinfo {eid} {014535} (\bibinfo {year}
  {2011}),\ \Eprint{http://arxiv.org/abs/1103.3664}{arXiv:1103.3664
  [cond-mat.quant-gas]}%
  \bibAnnoteFile{NoStop}{arrigoni2011}%
\bibitem{abrikosov1975}%
  \BibitemOpen
  \bibfield{author}{%
  \bibinfo {author} {\bibfnamefont{A.~A.}\ \bibnamefont{Abrikosov}}, \bibinfo
  {author} {\bibfnamefont{L.~P.}\ \bibnamefont{Gorkov}},\ and\ \bibinfo
  {author} {\bibfnamefont{I.~E.}\ \bibnamefont{Dzyaloshinski}},\ }%
  \emph{\bibinfo {title} {Methods of Quantum Field Theory in Statistical
  Mechanics}}\ (\bibinfo {publisher} {Dover Publications},\ \bibinfo {year}
  {1975})%
  \bibAnnoteFile{NoStop}{abrikosov1975}%
\end{thebibliography}
\end{document}